\documentclass[11pt]{article}

\usepackage[margin=1in]{geometry}

\usepackage{amsmath, amsthm, amssymb, microtype, enumerate, stackrel, mathrsfs, tikz}
\usepackage[colorlinks, citecolor=blue, linkcolor=blue, bookmarks=true]{hyperref}
\usepackage[nameinlink, noabbrev, capitalize]{cleveref}

\usepackage{bbm}
\usepackage{ulem} 

\usepackage{cleveref}
\usepackage[ruled,vlined]{algorithm2e}

\usepackage{float}
\usetikzlibrary{positioning}

\def\E{\mathbb{E}}

\newcommand{\INW}{\mathsf{INW}}

\newtheorem{lemma}{Lemma}[section]
\newtheorem{theorem}[lemma]{Theorem}

\newtheorem{corollary}[lemma]{Corollary}
\newtheorem{definition}[lemma]{Definition}
\newtheorem{claim}[lemma]{Claim}

\newtheorem*{remark}{Remark}

\newcommand{\R}{\mathbb{R}}

\newcommand{\eps}{\varepsilon}

\newcommand{\Ext}{\mathrm{E\textsc{xt}}}

\newcommand{\Samp}{\mathrm{Samp}}

\newcommand{\Rot}{\mathsf{Rot}}

\newcommand{\NZ}{\mathsf{NZ}}
\newcommand{\PRG}{\mathsf{PRG}}

\DeclareMathOperator{\poly}{\mathrm{poly}}

\newcommand{\sv}{\overset{sv}{\approx}}
\newcommand{\ds}{\textcircled{p}}
\newcommand{\WE}{\widetilde{\mathbb{E}}}

\newcommand{\BS}{\mathsf{BS}}
\newcommand{\M}{\mathbf{M}}

\newcommand{\reduct}[1]{\mathsf{#1}}

\newcommand{\knote}[1]{{\color{cyan} (Kuan: #1)}}
\newcommand{\wnote}[1]{{\color{magenta} (Wu: #1)}}

\begin{document}

\title{

Weighted Pseudorandom Generators for Read-Once Branching Programs via Weighted Pseudorandom Reductions


}

\author{ 
Kuan Cheng \footnote{CFCS, School of Computer Science, Peking University. ckkcdh@pku.edu.cn.}
\and 
Ruiyang Wu \footnote{CFCS, School of Computer Science, Peking University. wuruiyang@stu.pku.edu.cn.}
}

\date{}
\maketitle

\begin{abstract}

We study weighted pseudorandom generators (WPRGs) and derandomizations for read-once branching programs (ROBPs). 
Denote $n$ and $w$ as the length and the width of a ROBP.
We have the following results.

For standard ROBPs, we give an explicit $\varepsilon$-WPRG with seed length
    $$  O\left(\frac{\log n\log (nw)}{\max\left\{1,\log\log w-\log\log n\right\}}+\log w \left(\log\log\log w-\log\log\max\left\{2,\frac{\log w}{\log \frac{n}{\varepsilon}}\right\}\right)+\log\frac{1}{\varepsilon}\right).$$
When $n = w^{o(1)}$, this is better than the WPRGs of \cite{hozaBetterPseudodistributionsDerandomization2021, cohenErrorReductionWeighted2021, pynePseudodistributionsThatBeat2021, chattopadhyayOptimalErrorPseudodistributions2020}.
Further as a direct application, we attain a WPRG for regular ROBPs with a better seed length 
than that of \cite{chenWeightedPseudorandomGenerators2023, chattopadhyayRecursiveErrorReduction2023}.

For permutation ROBPs with unbounded widths and single accept nodes, we give an explicit $\varepsilon$-WPRG with seed length
    $$     O\left( \log n\left( \log\log n + \sqrt{\log(1/\varepsilon)} \right)+\log(1/\varepsilon)\right), $$
    improving \cite{chenWeightedPseudorandomGenerators2023}.
A key difference to \cite{chenWeightedPseudorandomGenerators2023} is that this implies a WPRG with optimal seed length for short-wide ROBPs with multiple accept nodes. Specifically, after switching to multiple accept nodes in a standard way by replacing $\eps$ with $ \eps/w $, this gives a WPRG with optimal seed length $O(\log w)$ for $n = 2^{O(\sqrt{\log w})}$, and error $1/\poly w$. 
The only previous work attaining optimal seed lengths are Nisan-Zuckerman style PRGs \cite{nisanRandomnessLinearSpace1996, armoniDerandomizationSpaceBoundedComputations1998} but they are only optimal for $n= \poly\log w, \varepsilon = 2^{-\log^{0.9} w}$.

We also give a new Nisan-Zuckerman style derandomization for regular ROBPs with width $w$, length $n = 2^{O(\sqrt{\log w})}$, and multiple accept nodes. We attain optimal space complexity $O(\log w)$ for arbitrary approximation error $\varepsilon = 1/\poly w$. 
When requiring the derandomization to be in $\mathbf{L}$, again the only previous result is by Nisan-Zuckerman style PRGs \cite{nisanRandomnessLinearSpace1996, armoniDerandomizationSpaceBoundedComputations1998}, which are only optimal for $n= \poly\log w, \varepsilon = 2^{-\log^{0.9} w}$.
Also, if compared to \cite{ahmadinejadHighprecisionEstimationRandom2022, chenWeightedPseudorandomGenerators2023, chattopadhyayRecursiveErrorReduction2023}, which can be viewed as Saks-Zhou style derandomizations, then  for $n =2^{O(\sqrt{\log w})}$ our derandomization not only improves the space complexity to optimal, but also substantially improves the time complexity from super polynomial to standard polynomial in $w$. 
Note that derandomizations of \cite{ahmadinejadHighprecisionEstimationRandom2022, chenWeightedPseudorandomGenerators2023, chattopadhyayRecursiveErrorReduction2023} has space complexity $ S= O(\log(nw)\log\log(nw/\varepsilon)) $ and time complexity exponential in $S$.

All our results are based on iterative weighted pseudorandom reductions, which can iteratively reduce fooling long ROBPs to fooling short ones.

\end{abstract}

\section{Introduction.}

Randomness is a fundamental resource in computation, but is it essential? A key conjecture in space-bounded computation is that randomized algorithms in the complexity class $\mathbf{BPL}$ can be efficiently simulated by deterministic logspace algorithms, i.e. $\mathbf{BPL}=\mathbf{L}$. A central approach toward addressing this conjecture is the derandomization of standard-order read-once branching programs (ROBPs), which is usually defined as the following.

\begin{definition}[Read-once branching programs (ROBP)]
    A read-once branching program $f$ of length $n$, width $w$ and alphabet size $|\Sigma|=2^s$ is a directed acyclic graph with $n+1$ layers $V_0,\ldots,V_n$. For any layer $V_i$ except $V_n$, each node $v\in V_i$ has $2^s$ outgoing edges to nodes in $V_{ i+1}$. These edges are labeled by distinct symbols in $\Sigma$. There exists a unique start node $v_{start} \in V_0$ and a set of accept nodes $V_{accept}\subset V_n$. Given an input $x\in \Sigma^n$, the computation of $f(x)$ is defined as: 
    $f(x) = 1$, if there exists a unique path $v_{start},v_1,\ldots,v_n$ such that the edge between $v_i$ and $v_{i+1}$ is labeled by $x_i$, and $v_n\in V_{accept}$;  $f(x) = 0$ otherwise.



\end{definition}
Every problem in $\mathbf{BPL}$ can be reduced to approximating $\mathbb{E}f$ for a corresponding ROBP $f$. A classical method for derandomizing ROBPs is constructing pseudorandom generators (PRGs).

\begin{definition}[PRG]
    Let $\mathcal{F}$ be a class of ROBPs $f:(\{0,1\}^s)^n\to \{0,1\}$.
    An $\varepsilon$-PRG for $\mathcal{F}$ is a function  $G:\{0,1\}^d\to (\{0,1\}^s)^n$
    such that for every $f\in \mathcal{F}$, we have
    \begin{equation*}
        \left|\E_{x\in (\{0,1\}^s)^n}f(x)-\E_{r\in \{0,1\}^d}f(G(r))\right|\leq \varepsilon.
    \end{equation*}
    The input length $d$ is called the seed length of the PRG. We say that $G$ is \textbf{explicit} if it can be computed in space $O(d)$ and time $\poly(d, n)$.

\end{definition}


By the probabilistic method, one can show the existence of a non-uniform $\varepsilon$-PRG for standard-order ROBPs of length $n$, width $w$, and alphabet size $2^s$, with an optimal seed length of $O(s+\log(nw/\varepsilon))$. 
However, constructing explicit PRGs that have short seed lengths turns out to be an exceptional challenge. In a seminal work, Nisan \cite{nisanPseudorandomGeneratorsSpacebounded1992} constructed an explicit $\varepsilon$-PRG for ROBPs of length $n$, width $w$, and binary alphabet $\{0,1\}$, with a seed length of $O(\log n\log(nw/\varepsilon))$. Building on this, Saks and Zhou \cite{saksBPHSPACEDSPACES31999} developed a celebrated algorithm to derandomize $\mathbf{BPL}$, within $O(\log^{3/2}n)$ space deterministically.
Nisan \cite{nisan1992rl} also used \cite{nisanPseudorandomGeneratorsSpacebounded1992} to show that $\mathbf{BPL \subseteq \mathbf{SC}}$.
Impagliazzo, Nisan, and Wigderson \cite{impagliazzoPseudorandomnessNetworkAlgorithms1994} generalized the construction of \cite{nisanPseudorandomGeneratorsSpacebounded1992} by using expanders, to fool more general models in network communication.
In the meantime, for short ROBPs, Nisan and Zuckerman \cite{nisanRandomnessLinearSpace1996} gave another remarkable PRG that has an optimal seed length for wide but short ROBPs, i.e. with seed length $O(\log w)$ when $n = \poly \log w, \eps = 2^{-\log^{0.99} w}$.
Armoni \cite{armoniDerandomizationSpaceBoundedComputations1998}  further extended \cite{nisanPseudorandomGeneratorsSpacebounded1992, nisanRandomnessLinearSpace1996} to construct an improved PRG with seed length\footnote{\cite{armoniDerandomizationSpaceBoundedComputations1998} needs to use an extractor with seed length optimal up to constant factors which is discovered later than \cite{armoniDerandomizationSpaceBoundedComputations1998}, e.g. \cite{guruswamiUnbalancedExpandersRandomness2009}.
} $O\left(\frac{\log n\log (nw/\varepsilon)}{\max\{1, \log\log w-\log\log (n/\varepsilon)\}} \right)$.

\subsection{Weighted pseudorandom generators.}

Despite years of research, the challenging problem of constructing better PRGs for general ROBPs remains open. 
However, PRGs are not the only black-box method for derandomizations. 
A remarkable work by Braverman, Cohen, and Garg \cite{braverman2019pseudorandom} improves the seed length by introducing and constructing WPRGs.
\begin{definition}[WPRG]
    Let $\mathcal{F}$ be a class of ROBPs $f:(\{0,1\}^s)^n\to \{0,1\}$. A $W$-bounded $\varepsilon$-WPRG for $\mathcal{F}$ is a function $(G,\sigma):\{0,1\}^d\to(\{0,1\}^s)^n\times \mathbb{R}$ such that for every $f\in \mathcal{F}$, we have
    \begin{align*}
        &\left|\E_{x\in(\{0,1\}^s)^n}f(x)-\sum_{r\in \{0,1\}^d}\left[\frac{1}{2^d}\sigma(r)\cdot  f(G(r))\right]\right|\leq \varepsilon,\\
        &\forall r, |\sigma(r)|\leq W.
    \end{align*}
    The input length $d$ is called the seed length of the WPRG. We say that $(G,w)$ is \textbf{explicit} if it can be computed in space $O(d)$ and time $\poly(d, n)$.
\end{definition}
As shown in \cite{braverman2019pseudorandom}, under this notion, the seed length can be significantly improved to $\widetilde{O}(\log n  \log (nw) + \log \frac{1}{\eps})$, i.e. for $\varepsilon$ there is only an isolated addend. 
Chattopadhyay and Liao \cite{chattopadhyayOptimalErrorPseudodistributions2020} further improved this construction to attain seed length  $O(\log n  \log (nw) \log \log(nw) + \log \frac{1}{\eps})$.
Then Cohen, Doron, Renard, Renard, Sberlo, and Ta-Shma \cite{cohenErrorReductionWeighted2021}, and also Pyne and Vadhan \cite{pynePseudodistributionsThatBeat2021} gave black-box error reductions from large error PRGs to small error WPRGs.
They both are based on a preconditioned Richardson Iteration, which was previously developed by Ahmadinejad, Kelner, Murtagh, Peebles, Sidford, and Vadhan \cite{ahmadinejadHighprecisionEstimationRandom2022} for high precision derandomization of random walks on Eulerian graphs.
Hoza \cite{hozaBetterPseudodistributionsDerandomization2021} further improved the error reduction based WPRG constructions to attain seed length $O(\log n \log (nw) + \log (1/\varepsilon))$.

Following these new error reduction methods, there are several new progress on derandomizations. Hoza \cite{hozaBetterPseudodistributionsDerandomization2021} improved the derandomization of Saks and Zhou \cite{saksBPHSPACEDSPACES31999} to be $\mathbf{BPL}\subseteq \mathbf{DSPACE}\left(\frac{\log^{3/2}n}{\sqrt{\log\log n}}\right)$. 
Cohen, Doron, Sberlo, and Ta-Shma \cite{cohenApproximatingIteratedMultiplication2023}, also Pyne and Putterman \cite{puttermanOptimalDerandomizationMediumWidth2022},  showed that ROBPs with medium width $w = 2^{O(\sqrt{\log n})}$ can be derandomized in $\widetilde{O}(\log n)$ space.
Cheng and Wang \cite{cheng2024bpl} showed that $\mathbf{BPL} \subseteq \mbox{logspace-uniform } \mathbf{AC}^1$.

WPRG is stronger than Hitting Set Generator (HSG), while HSG is already a powerful tool in derandomization.
\begin{definition}[HSG]
    Let $\mathcal{F}$ be a class of ROBPs $f:(\{0,1\}^s)^n\to \{0,1\}$. An $\varepsilon$-HSG for $\mathcal{F}$ is a function $H:\{0,1\}^d\to (\{0,1\}^s)^n$ such that for every $f\in \mathcal{F}$, if $\E_{x\in (\{0,1\}^s)^n}f(x)\geq \varepsilon$, then $\exists r\in \{0,1\}^d, f(H(r))=1$.

The input length $d$ is called the seed length of the HSG. We say that $H$ is \textbf{explicit} if it can be computed in space $O(d)$ and time $\poly(d, n)$.
\end{definition}
One can use HSG to find an accepting path when the acceptance probability is significant.
Actually, HSG is more powerful than this. 
Cheng and Hoza \cite{chengHittingSetsGive2020} showed how to approximate the acceptance probabilities of ROBPs by HSGs for larger-size ROBPs. 
Pyne, Raz, and Zhan \cite{pyne2023certified} further extended the method to give a deterministic sampler for such tasks.
Constructing better HSGs is also a challenging task. 
Hoza and Zuckerman gave a HSG with seed length $O\left( \frac{\log n \log (nw)}{\max\{1, \log\log w- \log \log n\}} + \log(1/\varepsilon) \right)$ which is optimal when $n = \poly\log w$.

\subsection{WPRG for short-wide standard ROBPs.}
Nisan-Zuckerman PRG and Armoni's generator are better in seed length than Nisan's PRG or INW PRG for short-wide cases when the error is large. 
It is a natural question whether one can transform these PRGs to WPRGs to attain better seed lengths for smaller errors.
A specific interesting case is $n= \poly \log w$, for which NZ PRG \cite{nisanRandomnessLinearSpace1996} and Armoni's PRG \cite{armoniDerandomizationSpaceBoundedComputations1998} have optimal seed lengths of $O(\log w)$ when $\varepsilon=2^{-\log^{1-v}w}$, where $v$ is any constant in $(0,1)$. However, when the error is smaller, such as $\varepsilon=1/\poly(w)$, the seed lengths of the WPRGs of \cite{cohenErrorReductionWeighted2021, hozaBetterPseudodistributionsDerandomization2021, pynePseudodistributionsThatBeat2021} deteriorate to $O(\log w\log\log w)$, which has no advantage over those of the PRGs \cite{nisanPseudorandomGeneratorsSpacebounded1992, impagliazzoPseudorandomnessNetworkAlgorithms1994, armoniDerandomizationSpaceBoundedComputations1998}. 


In this paper, we make progress for answering the question by giving a new WPRG construction that has a better seed length for short-wide ROBPs. Specifically, we have the following result.
\begin{theorem}\label{thm:intro armoni improved} 
    For every $n, w\in \mathbb{N}$, $\varepsilon\in (0,1)$, with $n \ge \log w$, there exists an explicit $\varepsilon$-WPRG with seed length
    \begin{equation*}
        O\left(\frac{\log n\log (nw)}{\max\left\{1,\log\log w-\log\log n\right\}}+\log w \left(\log\log\log w-\log\log\max\left\{2,\frac{\log w}{\log (n/\varepsilon)}\right\}\right)+\log(1/\varepsilon)\right)
    \end{equation*}
    for the class of ROBPs of width $w$, length $n$ and alphabet $\{0,1\}$.
\end{theorem}
The seed length is better than \cite{hozaBetterPseudodistributionsDerandomization2021, cohenErrorReductionWeighted2021, pynePseudodistributionsThatBeat2021, chattopadhyayOptimalErrorPseudodistributions2020}, when $  n \ll w$.
For the Nisan-Zuckerman regime i.e. $n = \poly \log w$, $\varepsilon = 2^{-\log^{1-\alpha} n}$ for any constant $\alpha >0$, the seed length of our \cref{thm:intro armoni improved} is $O(\log w)$, since the subtraction of the two triple-log terms becomes a constant. 
This matches the seed length of the Nisan-Zuckerman PRG \cite{nisanRandomnessLinearSpace1996}.
Further when $n$ is larger or $\eps$ is smaller, our seed length is strictly better.
\cref{tab:armoni comparison} and \cref{tab:nz comparison} summarize the seed length of $\varepsilon$-PRGs and $\varepsilon$-WPRGs for short-wide general ROBPs. 
\begin{table}[H]
    \centering
    \begin{tabular}{c|c|c}
        Seed length&Type&Reference\\
        \hline
        $O\left(\frac{\log n\log (nw/\varepsilon)}{\log\log w-\log\log (n/\varepsilon)}\right)$&PRG&\cite{armoniDerandomizationSpaceBoundedComputations1998,kaneRevisitingNormEstimation2009}\\
         $\widetilde{O}(\log n  \log (nw) + \log \frac{1}{\eps})$ &WPRG &\cite{braverman2019pseudorandom}\\
          $O(\log n  \log (nw) \log \log(nw) + \log \frac{1}{\eps})$ &WPRG & \cite{chattopadhyayOptimalErrorPseudodistributions2020}\\
        $O\left( \log n\log (nw) +\log (w/\varepsilon)\log\log_n(1/\varepsilon)\right)$&WPRG&\cite{cohenErrorReductionWeighted2021}  \\
         $O(\log n\log(nw)+\log(1/\varepsilon))$&WPRG&\cite{hozaBetterPseudodistributionsDerandomization2021}\\
        $O\left(\frac{\log n\log (nw)}{\log\log w-\log\log n}+\log w \log\log\log w+\log(1/\varepsilon)\right)$&WPRG& \cref{thm:intro armoni improved}\\
    \end{tabular}
    \caption{Comparison of seed length of $\varepsilon$-WPRGs for general ROBPs of width $w$, length $n < w $, and alphabet $\{0,1\}$.}
    \label{tab:armoni comparison}
\end{table}

\begin{table}[H]
    \centering
    \begin{tabular}{c|c|c|c}

        Seed length&Type&Reference&Note\\
        \hline
        $O(\log w)$&PRG&\cite{nisanRandomnessLinearSpace1996}&$\varepsilon=2^{-\log^{0.99}w}$\\
        $O(\log(w/\varepsilon)\log\log w)$&PRG&\cite{armoniDerandomizationSpaceBoundedComputations1998, impagliazzoPseudorandomnessNetworkAlgorithms1994, nisanPseudorandomGeneratorsSpacebounded1992}&$\varepsilon\leq 1/\poly(w)$\\
        $O(\log w\log\log w+\log(1/\varepsilon))$&WPRG&\cite{hozaBetterPseudodistributionsDerandomization2021, cohenErrorReductionWeighted2021}&$\varepsilon\leq 1/\poly(w)$\\
        $O(\log w)$&WPRG&\cref{thm:intro armoni improved}&$\varepsilon=2^{-\log^{0.99}w}$\\
        $O(\log w\log\log\log w+\log(1/\varepsilon))$&WPRG&\cref{thm:intro armoni improved}&$\varepsilon\leq 1/\poly(w)$\\
        \hline
        $O(\log w+\log(1/\varepsilon))$&PRG&folklore& Optimal; non-explicit\\
    \end{tabular}
    \caption{Comparison of seed length of   $\varepsilon$-WPRGs for short-wide ROBPs of width $w$, length $n=\poly(\log w)$, and alphabet $\{0,1\}$.}
    \label{tab:nz comparison}
\end{table}
Another interesting point is that our construction deploys a new framework which iteratively applies weighted pseudorandom reductions. We will describe this in detail in later sections. 

As a direct application of our WPRG, we also attain a better WPRG for regular ROBPs. 
\begin{definition}[Regular ROBP]
    A regular ROBP is a standard-order ROBP $f$, where for every $i\in [n]$, the bipartite graph induced by the nodes in layers $V_{i-1}$ and $V_i$ is a regular graph.
\end{definition}
\begin{theorem}\label{thm:WPRG for regular0}

For every $w, n  \in \mathbb{N}$, and every $\varepsilon>0$, there exists an explicit $\varepsilon$-WPRG with seed length
\begin{align*}
    O\left( \log n\left(\sqrt{\log(1/\varepsilon)}+\log w+\log\log n\right)+\log(1/\varepsilon)\right).
\end{align*}
for regular ROBPs  of width $w$, length $n$ and binary alphabet.
\end{theorem}
This is better than that of Chen, Hoza, Lyu, Tal, and Wu \cite{chenWeightedPseudorandomGenerators2023} which has a seed length $\widetilde{O}\left(\log n\left(\sqrt{\log(1/\varepsilon)}+\log w +\log\log n\right)+\log (1/\varepsilon)\right)$, and matches an independent work of Chen and Ta-Shma \cite{chen25better} for the binary case.
For larger alphabets, we can actually attain a seed length with a better dependence on the alphabet size than \cite{chen25better}. See our \cref{thm:WPRG for regular} for details.
The construction of \cite{chen25better} combines the methods of \cite{chenWeightedPseudorandomGenerators2023} and Hoza \cite{hozaBetterPseudodistributionsDerandomization2021}. 
Our construction is using our main technical theorem \cref{thm:final WPRG} to replace the INW PRG generating correlated seeds in \cite{chenWeightedPseudorandomGenerators2023}, with an argument based on weighted pseudorandom reductions.

\subsection{WPRG for unbounded-width permutation ROBPs with a single accept node.}
Our framework works even better for constructing WPRGs against permutation ROBPs.
Permutation ROBPs are interesting special ROBPs, where the transition functions between layers are permutations.
\begin{definition}[permutation ROBP]
    A (standard-order) permutation ROBP is a standard-order ROBP $f$, where for every $i\in [n]$ and $x\in \{0,1\}^s$, the transition matrix from $V_i$ to $V_{i+1}$ through edges labeled by $x$, is a permutation matrix in $\mathbb{R}^{w\times w}$.
\end{definition}

Early work on PRGs for permutation ROBPs \cite{brody2010coin, de2011pseudorandomness, koucky2011pseudorandom, steinke2012pseudorandomness, reingold2013pseudorandomness, chattopadhyay2019pseudorandom} focus on constant width cases. 
A remarkable line of recent studies develops PRG/WPRG/HSGs for unbounded-width permutation ROBPs with single accepting nodes \cite{hozaPseudorandomGeneratorsUnboundedWidth2020,pynePseudodistributionsThatBeat2021, bogdanov2022hitting,chenWeightedPseudorandomGenerators2023}. 
Hoza, Pyne and Vadhan \cite{hozaPseudorandomGeneratorsUnboundedWidth2020} showed an $\varepsilon$-PRG for unbounded-width permutation ROBPs with seed length $\widetilde{O}(\log n\log(1/\varepsilon))$. 
They also proved that any PRG for this class must have seed length $\widetilde{\Omega}(\log n\log(1/\varepsilon))$. 
For the WPRG case, Pyne and Vadhan \cite{pynePseudodistributionsThatBeat2021} showed a $\varepsilon$-WPRG with seed length $\widetilde{O}(\log n\sqrt{\log(n/\varepsilon)}+\log(1/\varepsilon))$.
This was improved by Chen, Hoza, Lyu, Tal, and Wu \cite{chenWeightedPseudorandomGenerators2023} to  $O\left(\log n\sqrt{\log(1/\varepsilon)}\sqrt{\log\log(n/\varepsilon)}+\log(1/\varepsilon)\log\log(n/\varepsilon)\right)$.

We give an improved WPRG against permutation ROBPs as the following.
\begin{theorem}\label{thm:permWPRG}
    For every $n,s\in \mathbb{N}$ and $\varepsilon\in (0,1)$, there exists an explicit $\varepsilon$-WPRG with seed length
    \begin{equation*}
        O\left(s+\log n\left( \log\log n + \sqrt{\log(1/\varepsilon)} \right)+\log(1/\varepsilon)\right)
    \end{equation*}
    for the class of permutation ROBPs of length $n$ and alphabet $\{0,1\}^s$ with a single accept node.
\end{theorem}
The comparison of our result with the previous results is shown in \cref{tab:permutation comparison}.
\begin{table}[H]
    \centering
    \begin{tabular}{c|c|c|c}
        Seed length&Type&Reference&Note\\
        \hline
        $\widetilde{O}(\log n\log(1/\varepsilon))$&PRG&\cite{hozaPseudorandomGeneratorsUnboundedWidth2020}&\\
        $\widetilde{O}(\log n\sqrt{\log(n/\varepsilon)}+\log(1/\varepsilon))$&WPRG&\cite{pynePseudodistributionsThatBeat2021}&\\
        $O\left(\log n\sqrt{\log(1/\varepsilon)}\sqrt{\log\log(n/\varepsilon)}+\log(1/\varepsilon)\log\log(n/\varepsilon)\right)$&WPRG&\cite{chenWeightedPseudorandomGenerators2023}&\\
        $O\left(\log n\left( \log\log n + \sqrt{\log(1/\varepsilon)} \right) +\log(1/\varepsilon)\right)$&WPRG&This work&\\
        \hline
        $\Omega(\log n\log(1/\varepsilon))$&PRG&\cite{hozaPseudorandomGeneratorsUnboundedWidth2020}&lower bound\\
    \end{tabular}
    \caption{Comparison of seed length of $\varepsilon$-PRGs and $\varepsilon$-WPRGs for unbounded-width permutation ROBPs of length $n$ and alphabet $\{0,1\}$ with one accepting node.}
    \label{tab:permutation comparison}
\end{table}
Note that by replacing $\eps$ with $\eps/w$, one can attain a WPRG for multiple accept nodes. In this way our result gives a WPRG with seed length $   O\left(s+\log n\left( \log\log n + \sqrt{\log(w/\varepsilon)} \right)+\log(w/\varepsilon)\right)$.
This reveals a substantial difference between our result and  \cite{chenWeightedPseudorandomGenerators2023}.
Notice that our seed length is optimal $O(\log w)$ for $n =2^{O(\sqrt{\log w})}, \varepsilon = 1/\poly w$, i.e. it implies a Nisan-Zuckerman style PRG for short-wide permutation ROBPs. 
The only prior work attaining optimal seed length for such ROBPs are \cite{nisanRandomnessLinearSpace1996, armoniDerandomizationSpaceBoundedComputations1998} but only work for $n= \poly\log w, \varepsilon = 2^{-\log^{0.9} w}$.
It is not clear how to use \cite{chenWeightedPseudorandomGenerators2023} to achieve optimal seed length even for constant error, since when switching to multiple accept nodes, the only known way is to replace $\eps$ with $\eps/w$.

We also remark that one may want to use the sampler technique of \cite{hozaBetterPseudodistributionsDerandomization2021} on the WPRG of \cite{chenWeightedPseudorandomGenerators2023} to attain similar parameters to our \cref{thm:permWPRG}. However, one barrier is that it is unclear if it is possible to attain samplers supporting the sv-approximation used in the analysis of \cite{chenWeightedPseudorandomGenerators2023}, with desired parameters.

\subsection{Derandomization for short-wide regular ROBPs.}
Our framework also provides a new derandomization for regular ROBPs.
There is an extensive body of work on constructing HSGs, WPRGs, and PRGs for regular ROBPs \cite{bravermanPseudorandomGeneratorsRegular2014,bogdanov2022hitting,de2011pseudorandomness,reingold2013pseudorandomness,chattopadhyayRecursiveErrorReduction2023,chenWeightedPseudorandomGenerators2023}.  
For derandomization, dedicated derandomization algorithms achieve better performance. 
Ahmadinejad, Kelner, Murtagh, Peebles, Sidford, and Vadhan \cite{ahmadinejadHighprecisionEstimationRandom2022} derandomized regular ROBPs of length $n$, width $w$ and alphabet $\{0,1\}$ within error $\varepsilon$, using space $ O(\log(nw)\log\log(nw/\varepsilon))$.
Subsequently,  Chen, Hoza, Lyu, Tal, and Wu \cite{chenWeightedPseudorandomGenerators2023} achieved the same result with a simplified algorithm. 
Chattopadhyay and Liao \cite{chattopadhyayRecursiveErrorReduction2023} constructed another alternate algorithm with the same space complexity.
These derandomizations can be viewed as Sak-Zhou style derandomizations since they can work for full length with small space. However their time complexity is exponential in their space complexity, hence is not polynomial.

Our new derandomization, stated as the following, can be viewed as a Nisan-Zuckerman style derandomization, since it has optimal logspace complexity for short-wide ROBPs.
\begin{theorem}\label{thm:intro regular}
    There exists an algorithm that for any input regular ROBP $f$ of width $w$, length $n = 2^{O(\log^{1/2} w)}$, alphabet $\{0, 1\}^s$, and for any input parameter $\eps \ge 1/\poly w$, outputs an approximation for $\E[f]$ with addtive error $\eps$.
    The algorithm uses $O(s + \log w)$ bits of space.
\end{theorem}
Notice that the space complexity is optimal up to constant factors, placing the derandomization precisely in $\mathbf{L}$, as long as the alphabet bit-length $s = O(\log w)$. Also notice that this implies our algorithm is in time $\poly(w)$. 
For  derandomizing regular ROBPs within $\mathbf{L}$, the only prior feasible methods are Nisan-Zuckerman generator \cite{nisanRandomnessLinearSpace1996} and Armoni's generator \cite{armoniDerandomizationSpaceBoundedComputations1998}. Our result significantly improves the length of the regular ROBPs that can be derandomized in $\mathbf{L}$ from $\poly\log w $ to $2^{O(\log^{1/2} w)} $ and also improves the precision from $2^{-\log^{0.99} w}$ to arbitrary $1/\poly w$. 

If compared to the Saks-Zhou style derandomizations of \cite{ahmadinejadHighprecisionEstimationRandom2022, chenWeightedPseudorandomGenerators2023, chattopadhyayRecursiveErrorReduction2023}, then our algorithm not only attains optimal space complexity $O(\log w)$, but also has a substantial improvement for time complexity from super-polynomial to standard polynomial, as long as $n = 2^{O(\log^{1/2} w)}$. 
Note that derandomizations of \cite{ahmadinejadHighprecisionEstimationRandom2022, chenWeightedPseudorandomGenerators2023, chattopadhyayRecursiveErrorReduction2023} are not in polynomial time even if $n = \poly(\log w)$ and $\eps = O(1)$.

\subsection{Technical Overview}

\paragraph{WPRGs for standard-order ROBPs} We start by reviewing the Richardson iteration based error reduction \cite{ahmadinejadHighprecisionEstimationRandom2022, cohenErrorReductionWeighted2021, pynePseudodistributionsThatBeat2021} that attains low-error WPRGs from large-error PRGs.

The Richardson iteration based error reduction can be viewed as a procedure that produces high-precision approximations for iterated matrices multiplications, given low-precision approximations.
Let $\{A_i\}_{i=1}^n \subseteq \mathbb{R}^{w\times w}$ be transition matrices, and $\{B_{i,j}\}_{0 \leq j < i \leq n} \subseteq \mathbb{R}^{w\times w}$ be approximations satisfying $\|B_{i,j} - A_i \cdots A_{j+1}\| \leq \varepsilon_0/(2(n+1))$. For any integer $k$, a standard method, e.g. \cite{cohenErrorReductionWeighted2021}, indicates there is a weighted sum $P = \sum_{i=1}^K \sigma_i B_{n_{i,1},n_{i,2}} \cdots B_{n_{i,k-1},n_{i,k}}$ of $K = n^{O(k)}$ terms such that  
\[
\left\| A_n \cdots A_1 -  P \right\| \leq \varepsilon_0^{k/2} (n+1),
\]  
where $\sigma_i \in \{-1,0,1\}$ and $\forall i,$ the sequence $n_{i, j}  ,j \in [k]$ is an increasing sequence of indices in $[n]$.\footnote{See \cref{appendix:iterationproof} for details.}  
To apply this to an ROBP $f$, one can instantiate the matrices as
\begin{itemize}
    \item $A_i$: the stochastic matrix of $f$ from layer $i-1$ to $i$.  
    \item $B_{i,j} := \mathbb{E}_{x \sim \{0,1\}^s} \left[ f^{j\to i}(\PRG_{j\to i}(x)) \right]$, where $f^{j\to i}$ is the transition of $f$ from layer $j$ to layer $i$, and $\PRG_{j\to i}$ is an $ \frac{\varepsilon_0}{2(n+1)} $-PRG with seed length $s$ and outputting $i-j$ symbols. 
\end{itemize}  
This immediately yields a high-precision approximation of the expectation of the ROBP:  
\[
\left| \mathbb{E}_U[f(U)] - \sum_{i=1}^K \sigma_i \cdot \mathbb{E}_{x_1,\dots,x_k \sim \{0,1\}^s} f\left( \PRG_{0\to n_{i,1}}(x_1), \dots, \PRG_{n_{i,k-1}\to n_{i,k}}(x_k) \right) \right| \leq \varepsilon_0^{k/2} (n+1).
\]  
Notice that this also immediately gives a WPRG for $f$ with error $\varepsilon_0^{k/2}(n+1)$ by definition. However the $k$ independent PRG callings cost too much randomness. 
\cite{cohenErrorReductionWeighted2021} further reduces the randomness by using an INW generator to generate the seeds for the $k$ independent PRG callings. But since INW does not have optimal seed length, this takes $O(\log k \log \frac{kw}{\eps} +s)$ randomness which has an extra $O(\log k)$ factor in the main term. Using the sampler technique of \cite{hozaBetterPseudodistributionsDerandomization2021}, one can avoid this factor, but only when the starting error $\eps_0$ is already $1/\poly w$. If the large error PRG has $\eps_0 \gg 1/\poly w$, then this does not work as desired. 
For example, if the large-error PRG is the NZ PRG, then it is unclear how to use this to derive an optimal seed length for $\eps = 1/\poly w$.


To address the problem, we introduce a new strategy.
For each $i\in [K]$, one can view
\[
f_i(y_1, \dots, y_k) := f\left( \PRG_{0\to n_{i,1}}(y_1), \PRG_{n_{i,1}\to n_{i,2}}(y_2), \dots, \PRG_{n_{i,k-1}\to n_{i,k}}(y_k) \right),
\]  
as a new ROBP with a shorter length $k$ and a large alphabet, while the width is still $w$.
To fool $f$, one only need to fool these shorter ROBPs.
Based on this observation, our high-level idea is to iteratively apply the above procedure to reduce the ROBPs to even shorter ones until the length become a constant such that the reduced ROBPs can be fooled trivially.

We use the notion \textit{weighted pseudorandom reduction} to describe the construction in details.
\begin{definition}[Weighted Pseudorandom Reduction]
    Let $\mathcal{B}_{n_0,s_0,w}$ be the class of length-$n_0$, width-$w$ ROBPs over alphabet $\Sigma_0=\{0,1\}^{s_0}$. Let $\mathcal{B}_{n_1,s_1,w}$ be the class of length-$n_1$, width-$w$ ROBPs over alphabet $\Sigma_1=\{0,1\}^{s_1}$.  A $(d,K,\varepsilon)$-reduction from  $\mathcal{B}_{n_0,s_0,w}$ to $\mathcal{B}_{n_1,s_1,w}$ is a tuple $(\reduct{R},\sigma)$, in which $\reduct{R}:\left(\{0,1\}^{s_1}\right)^{n_1}\times \{0,1\}^d \rightarrow \left(\{0,1\}^{s_0}\right)^{n_0}$ and $\sigma:\{0,1\}^d\rightarrow \mathbb{R}$. Furthermore, for every $f\in \mathcal{B}_{n_0,s_0,w}$, we have:
    \begin{align*}
        &\left|\E_U f(U)-\frac{1}{2^d}\sum_{i\in \{0,1\} ^d} \sigma(i)\E_{U'} [f(\reduct{R}(U',i))]\right|\leq \varepsilon,\\
        &\forall i, |\sigma(i)|\leq K,\\
        &\forall i, f(\reduct{R}(\cdot,i))\in \mathcal{B}_{n_1,s_1,w}. 
    \end{align*} 
    We call $\reduct{R}$ the reduction function and $\sigma$ the weight function. 
\end{definition}

Let the original length-$n$, width-$w$ ROBP be $f$ over alphabet $\Sigma$. For simplicity, now let the target error $\eps = 1/\poly w$. 
We plan to construct a sequence of length reductions as the following.
Let $n_0=n$.
The $i$-th length reduction, which is instantiated by the Richardson iteration based error reduction with a base $\frac{\varepsilon_i}{2(n_{i-1}+1)}$-PRG, is a $(d_i,K_i,\varepsilon)$-reduction  from  $\mathcal{B}_{n_{i-1},s_{i-1},w}$ to $\mathcal{B}_{n_i,s_i,w}$, where we take 
\[
n_i = O\left( \frac{\log(n_{i-1}/\varepsilon)}{\log(1/\varepsilon_i)} \right).
\]  
This may introduce error
\[
\varepsilon_i^{n_i/2} (n_{i-1}+1) \le \varepsilon,
\] 
as it is instantiated by the Richardson iteration based error reduction.
Notice that as $i$ goes up, one can decrease $\eps_i$ such that \(n_i\) can be smaller and smaller.
One may doubt that a smaller \(\varepsilon_i\) may require a more precise base PRG, which may require a longer seed.  
 However, the seed length depends on $n_{i-1}, w, \varepsilon_i$. As $n_{i-1}$ is already decreased by the previous reduction, one can pick a smaller $\varepsilon_i$ without increasing the seed length.
Another issue is that the alphabet is accumulating quickly as the seed length of the base PRG will become the alphabet bit-length for the next reduction. 
Our solution is to apply an alpahbet reduction after each length reduction to retain a small alphabet. 
That is, we construct our reduction from $\mathcal{B}_{n_{i-1},s_{i-1},w}$ to $\mathcal{B}_{n_i,s_i,w}$ by composing a length reduction (from $\mathcal{B}_{n_{i-1},s_{i-1},w}$ to $\mathcal{B}_{n_i,s'_i,w}$) with a subsequent alphabet reduction (from $\mathcal{B}_{n_i,s'_i,w}$ to $\mathcal{B}_{n_i,s_i,w}$).

More concretely, the $l$ length reductions mentioned above are instantiated by the Richardson iteration based error reduction  
with Armoni's generator \cite{armoniDerandomizationSpaceBoundedComputations1998, kaneRevisitingNormEstimation2009} as base PRGs.  
For a length-$n$, width-$w$ ROBP over $\Sigma = \{0,1\}^s$, we first choose $\varepsilon_0=2^{-\sqrt{ \log w\log n}}$ and apply a $(d_0=\log K_0,K_0,\varepsilon)$-reduction from $\mathcal{B}_{n,s,w}$ to $\mathcal{B}_{n_1,s'_1,w}$, in which $n_1 = O\left(\sqrt{\frac{\log w}{\log n}}\right)$, $K_0=2^{O\left(\sqrt{\log n\log w}\right)}$ and $s'_1=O\left( s + \frac{\log n \log (nw / \varepsilon_0)}{\log\log w - \log\log (n / \varepsilon_0)} \right)$. 
For the subsequent $l-1$ reductions i.e. for $i =  1,2 \ldots, l-1$, we setup a $(d_{i} = \log K_i, K_i, \eps_i)$ reduction from $\mathcal{B}_{n_i,s_i,w}$ to $\mathcal{B}_{n_{i+1},s'_{i+1},w}$, where we assume $s_i = c\log w$ with $c$ being a universal constant.
We choose $\varepsilon_i = 2^{-(\log w) / n_i^{1/3}}$, and consequently $n_{i+1} = n_i^{1/3}$, $K_i=O\left(2^{n_i^{1/3}\log n_i}\right)=o\left(\log ^{1/2 }w\right)$ and $s'_{i+1}=O(s_i+\log w)$.  
We iterate for $l$ steps until we reach $n_l = O(1)$.
One can see that $l \le O(\log\log \log w)$. 

As mentioned before, for every $i$, after the $i$th length reduction, we apply an alphabet reduction to retain a small alphabet.
Each alphabet reduction is realized by a one-level NZ generator. 
Assume we have a ROBP $f'$ with length $m = n_i$ and alphabet bit-length  $s'_i$.
The one-level NZ generator which, given a source \(X\) of \(d = O(s'+\log w)\) bits and a sequence of independent seeds \(Y_1, \ldots, Y_m\), each of \(s_i = O(\log (d/\varepsilon))\) bits, outputs $(\Ext(X,Y_1),\Ext(X,Y_2),\ldots,\Ext(X,Y_m))$ where each $\Ext(X, Y_i)$ is of binary length $s'_i$. 
Now fixing $X=x$, the function 
$f'\left(\Ext(x,y_1),\Ext(x,y_2),\ldots,\Ext(x,y_m)\right)$
can be viewed as an ROBP on input $(y_1, \ldots, y_m)$ with alphabet bit-length $s_i=O\left(\log \frac{s'_i+\log w}{\varepsilon}\right)$. One can check $s_i \le c \log w$ as $\eps = 1/\poly w$, where $c$ is a universal constant.

The seed length of our WPRG is the sum of the randomness of all reductions, together with the true randomness used to fool the final constant length ROBPs. 
For length reductions the seed length is $\sum_{i=0}^l d_i \le  O\left(\log (nw) \right)$.
The first alphabet reduction contributes $O\left(s+\frac{\log n\log (nw)}{\max\left\{1,\log\log w-\log\log n\right\}}\right)$. The remaining alphabet reductions contribute $O(l \cdot \log w)$. 
The true randomness for final constant length ROBPs is $O(\log w)$ since the alphabet bit-length is $O(\log w)$.

Finally for even smaller errors ($\varepsilon \ll 1/\poly(w)$), we apply a variant of Hoza's sampler \cite{hozaBetterPseudodistributionsDerandomization2021} technique, which can also be viewed as an extra level of weighted pseudorandom reduction.

\paragraph{WPRGs for unbounded-width permutation ROBPs with a single accept node}

We start by reviewing the recent WPRG given by \cite{chenWeightedPseudorandomGenerators2023} in the view of weighted pseudorandom reductions. Their WPRG combines the INW generator and a new matrix iteration method. To fool a unbounded-width permutation ROBP of length $n$ and alphabet $\{0,1\}^s$ within error $\eps$, they first prepare a base INW generator $\PRG:\{0,1\}^d\to \left(\{0,1\}^s\right)^n$ with moderate sv-error\footnote{See \cref{def:sv-error}.} $\tau=2^{-O(\log \log n\sqrt{\log (1/\eps)})}$. Define $\PRG_{j\to i}$ as the truncation of $\PRG$ to the first $i-j$ output symbols. Then they construct the $\eps$-WPRG $(G,\sigma)$ in the following form:
\begin{align*}
    &G(j,x_1,\ldots,x_m)=\PRG_{0 \to n_{j,1}}(x_1),\PRG_{n_{j,2}\to n_{j,1}}(x_2),\ldots,\PRG_{n_{j,m}\to n_{j,m-1}}(x_m),\\
    &\sigma(j)=\sigma_j\cdot K,
\end{align*}
where $j\in [K]$, $K=2^{O(m)}$, $m=O\left(\log n\cdot \frac{\log(1/\eps)}{\log(1/\tau)}\right)$, for each $j$  the sequence $n_{j, k}, k\in [m]$  of indices being such that $0\leq n_{j,1}<n_{j,2}<\ldots<n_{j,m}\leq n$, and $\sigma_j \in \{-1, 0, 1\}$. 
The seed length of $\PRG$ is $d=s+O\left(\log n\cdot(\log\log n+\log(1/\tau))\right)$.
Note that for each $j\in[K]$, one can view $f(G(j, x_1,\ldots,x_m))$ as a new permutation ROBP on input $x_1, x_2, \ldots, x_m$. \cite{chenWeightedPseudorandomGenerators2023} applies an $\eps/K$-INW generator to fool such reduced permutation ROBPs, which cost $d+O\left(\log m\cdot (\log\log m+\log (K/\eps)\right)$ bits of randomness. This step introduces some double-logarithmic factors.

To address this problem, we introduce the following new strategy based on weighted pseudorandom reductions.
Let $\mathcal{P}_{n,s}$ denote the class of length-$n$ unbounded-width permutation ROBPs over alphabet $\{0,1\}^s$. Then  $(G, \sigma)$ can be viewed as a $(\log K,K,\varepsilon)$-reduction from $\mathcal{P}_{n,s}$ to $\mathcal{P}_{m,d}$.
Let our target error be $\varepsilon$.
We setup $l$ reductions as the following. The $i$-th reduction is a $(d_i = \log K_i,K_i,\varepsilon_i)$-reduction from  $\mathcal{P}_{n_{i-1},s_{i-1}}$ to $\mathcal{P}_{n_i,s_i}$.  For the first reduction, we set $\tau_1=2^{-O\left(\sqrt{\log(1/\varepsilon)}\right)}$ to obtain a $(\log K_1,K_1,\varepsilon/2)$-reduction from $\mathcal{P}_{n,s}$ to $\mathcal{P}_{n_1,s_1}$. 
One can see $n_1=O\left(\log n\cdot \sqrt{\log(1/\varepsilon)}\right)$, $s_1=s+O\left(\log n\cdot \left(\log\log n+\sqrt{\log(1/\varepsilon)} \right)\right)$  and $K_1=\exp\left( O\left(\log n\cdot \sqrt{\log(1/\varepsilon)}\right)\right)$. For the subsequent $l-1$ reductions, we set $\eps'=(\eps/(2K_1))^2$ and $\tau_i=2^{-\frac{C\log(1/\eps')}{\log^2 n_{i-1}}}$ to obtain a $(\log K_i,K_i,\varepsilon')$-reduction from $\mathcal{P}_{n_{i-1},s_{i-1}}$ to $\mathcal{P}_{n_i,s_i}$, where $C>0$ is a universal constant. One can deduce that $n_{i}=O\left(\log n_{i-1}\frac{\log(1/\eps')}{\log(1/\tau_i)}\right)=\log^3 n_{i-1}$, $s_{i}=s_{i-1}+O(\log n_{i-1}\cdot (\log\log n_{i-1}+\log(1/\tau_i)))=s_{i-1}+O\left(\frac{\log(1/\eps')}{\log n_{i-1}}\right)$ and  $K_i=2^{O(\log^3 n_i)}$.
We do the iteration until $n_{l}= O(1)$. So $l = o(\log \log n)$.

We emphasize that the alphabet bit-length increment from $s_i$ to $s_{i+1}$ is not significant. In fact $s_l=s_1+O(\log(1/\eps'))\cdot\sum_{i=1}^{l-1}\frac{1}{\log n_i}=s_1+O(\log(1/\eps'))$.
Also notice that the overall weight $K_1 K_2K_3\ldots K_l$. 
Here $ K_2K_3\ldots K_l $ is bound by $  2^{O(\log^3 n_1 )}$ since $n_i$ decreases quickly. 
This is negligible compared to  $\eps/K_1$, which means that $\eps'=(\eps/(2K_1))^2$ is small enough to deduce an $\eps$-WPRG. 
Finally, we calculate the seed length. The overall random bits used in our WPRG have two parts, the bits for the final ROBP of length $n_l$ and alphabet $\{0,1\}^{s_l}$ and the randomness used in the weighted reductions. 
For the first part, we need $s_l\cdot n_l$ bits. 
For the second part, we need $\sum_{i} d_i = \log K_1+\sum_{i=2}^l\log K_i=\log K_1+O(\log^3{n_1})$ bits. 
Therefore, one can deduce the overall random bits used in our WPRG is $s_l\cdot n_l + \sum_i d_i = O(s+\log n\cdot(\log\log n+\sqrt{\log(1/\eps)})+\log(1/\eps))$.

\paragraph{Derandomization for short-wide regular ROBPs}
To attain our derandomization, notice that if we only want to derandomize permutation ROBPs with multiple accept nodes then we can directly use our WPRG for such permutation ROBPs.
Recall that the seed length is $   O\left(s+\log n\left( \log\log n + \sqrt{\log(w/\varepsilon)} \right)+\log(w/\varepsilon)\right)$.
Hence it is $O(s + \log w)$ if $n = 2^{O(\sqrt{\log w})}$, $\eps = 1/\poly w$.
In fact, for regular ROBPs, we essentially do the same thing, except that we apply some white-box techniques such that the framework of our WPRG for permutation ROBPs can also work for regular ROBPs. 

To derandomize regular ROBPs with binary alphabet, we show that there is a logspace algorithm that can transform a regular ROBP into a permutation ROBP which has the same expectation (though it probably computes a different function). Then we approximation this expectation with our WPRG for permutation ROBPs.

The hard case is to derandomize regular ROBPs with larger alphabets. 
Let $f$ be a regular ROBP with length $n$, width $w$, alphabet bit-length $s$.
We apply $l$ reductions as the following.
For every \(p\in [l]\), let \(f_{i_1,\dots,i_{p-1}}\) be a reduced regular ROBP of length $n_{p-1}$, with its stochastic matrices \({{\M}_{l\ldots r}}\) from its layer \(l\) to layer \(r\), $\forall l, r\in [n_p]$. 
 \(f_{i_1,\dots,i_{p-1}}\) is $f$ if $p-1=0$.
For every $0\leq l<r\leq n_{p-1}$ one can construct a \(\tau_p\)-sv-approximation \(\widetilde{{\M}}_{l\ldots r}\) of \({{\M}_{l\ldots r}}\) by the derandomizing square method of \cite{ahmadinejadHighprecisionEstimationRandom2022, hozaPseudorandomGeneratorsUnboundedWidth2020, chenWeightedPseudorandomGenerators2023}, if \(f_{i_1,\dots,i_{p-1}}\) has a two-way labeling.
A crucial point is that \(\widetilde{{\M}}_{l\ldots r}\) corresponds to the stochastic matrix of a regular bipartite graph \(\widetilde{G}_{l\ldots r}\) also with a two-way labeling, as long as \(f_{i_1,\dots,i_{p-1}}\) has a two-way labeling.  
Recall that a two-way labeling for an ROBP, defined by \cite{reingold2000entropy, rozenmanDerandomizedSquaringGraphs2005}, requires that every edge has two labels, each for one direction, such that for every vertex all its out-going labels are distinct, and all its in-coming labels are also distinct.
And a two-way labeling naturally supports a rotation technique which, whenever the pseudorandom walk arrives at a new node $v$ from $u$ through an edge $e$, switches the label of $e$ from the out-going label of $u$ to the incoming label of $v$. In this way, though the program is only regular, it can still be fooled by a pseudorandom walk from \cite{ahmadinejadHighprecisionEstimationRandom2022, chenWeightedPseudorandomGenerators2023} with such rotations. 
Using an error reduction polynomial of \cite{chenWeightedPseudorandomGenerators2023}, one can decompose \({{\M}_{0\ldots n_{p-1}}}\) into a signed sum:
\[
{{\M}_{0\ldots n_{p-1}}} \approx \sum_{i_p \in [K_p]} \sigma^{(p)}_{i_p} \cdot \prod_{t=1}^{n_p} \widetilde{{\M}}_{m^{(p)}_{i_p,t-1}\ldots m^{(p)}_{i_p,t}},
\]
where \(0 = m^{(p)}_{i_p,0} < m^{(p)}_{i_p,1} < \cdots < m^{(p)}_{i_p,n_p} = n_{p-1}\) are indices partitioning \(f_{i_1,\dots,i_{p-1}}\) into \(n_p\) segments, \(K_p\) denotes the number of terms, and \(\sigma^{(p)}_{i_p} \in \{-1,0,1\}\) are signs.  
Each product in the sum corresponds to a {concatenation of regular bigraphs} such that the product can also be viewed as a regular ROBP \(f_{i_1,\dots,i_p}\) of length \(n_p\):
\[
f_{i_1,\dots,i_p} = \widetilde{G}_{m^{(p)}_{i_p,0}\ldots m^{(p)}_{i_p,1}} \circ \widetilde{G}_{m^{(p)}_{i_p,1}\ldots m^{(p)}_{i_p,2}} \circ \cdots \circ \widetilde{G}_{m^{(p)}_{i_p,n_p-1}\ldots m^{(p)}_{i_p,n_p}}.
\]
$f_{i_1,\dots,i_p}$ again has a two-way labeling since it is the concatenation of bigraphs with two-way labelings.
If consider all the reductions, then finally we can approximate \(\mathbb{E}[f]\) with a signed sum of expectations of those reduced ROBPs:
\[
\mathbb{E}[f] \approx \sum_{i_1 \in [K_1], \dots, i_l \in [K_l]} \sigma^{(1)}_{i_1} \cdots \sigma^{(l)}_{i_l} \cdot \mathbb{E}[f_{i_1,\dots,i_l}],
\]
where $\forall p\in [l], \forall i_p, \sigma^{(p)}_{i_p}$ is a corresponding sign from the $p$-th reduction.
It turns out all relevant parameters (e.g. the length and alphabet of the reduced ROBPs, the signs, the seed length) can be essentially the same as those parameters of our WPRG for permutation ROBPs. 
Also one can see that since $f$ has a two-way labeling then $\forall p\in [l]$, every reduced program $f_{i_1,\dots,i_p}$ has a two-way labeling by the property mentioned above.
So finally for each $f_{i_1,\dots,i_l}$, whose length is $n_l = O(1)$, one can apply a true random walk using rotations supported by the two-way labeling, and the expectation can be approximated by a weighted sum of the results of these random walks. 
By a more detailed analysis, we show that these rotations are again space efficient so that the overall space complexity is only linear of the seed length.

\section{Preliminaries.}

\subsection{Some notations.}
For convenience of description, we denote $\mathcal{B}_{(n,s,w)}$ as the class of ROBPs with length $n$, alphabet size $2^s$ and width $w$.
We denote $\mathcal{P}_{(n,s)}$ as the class of permutation ROBPs with unbounded width and only one accept node, where $n$ is the length $s$ is the bit-length of the alphabet.
We denote $\mathcal{R}_{(n,s, w )}$ as the class of regular ROBPs with length $n$, alphabet size $2^s$ and width $w$.

Given an ROBP $f$, we denote $f^{[i,j]}:(\{0,1\}^s)^{j-i}\to \mathbb{R}^{w\times w}$ as the matrix representing the transition function of $f$ between layers $V_i$ and $V_j$. Specifically, the entry $[f^{[i,j]   }(x)]_{u,v}=1$ if there exists a path from node $u$ in $V_i$ to node $v$ in $V_j$ that is labeled by the string $x=x_1\ldots x_{j-i}$. Otherwise, $[f^{[i,j]}(x)]_{u,v}=0$.

We use $\circ$ to denote the composition of functions, i.e. for any two functions $f, g$ where any output of $g$ can be an input of $f$, we have that $f\circ g (x) := f(g(x))$.

\subsection{Weighted pseudorandom reductions.}
Here we state the definition of weighted pseudorandom reduction.
\begin{definition}[Weighted Pseudorandom Reduction]\label{def:weighted pseudorandom reduction}
    Let $\mathcal{F}_0$ be a class of functions $\left(\{0,1\}^{s_0}\right)^{n_0}\rightarrow \mathbb{R}$. Let $\mathcal{F}_{simp}$ be a class of functions $\left(\{0,1\}^{s_1}\right)^{n_1}\rightarrow \mathbb{R}$.  A $(d,K,\varepsilon)$-weighted pseudorandom reduction from $\mathcal{F}_0$ to $\mathcal{F}_{simp}$ is a tuple $(\reduct{R},\sigma)$, in which $\reduct{R}:\left(\{0,1\}^{s_1}\right)^{n_1}\times \{0,1\}^d \rightarrow \left(\{0,1\}^{s_0}\right)^{n_0}$ and $\sigma:\{0,1\}^d\rightarrow \mathbb{R}$. Furthermore, for every $f\in \mathcal{F}_0$, we have:
    \begin{align*}
        &\left|\E_U f(U)-\frac{1}{2^d}\sum_{i\in \{0,1\} ^d} \sigma(i)\E_{U'} [f(\reduct{R}(U',i))]\right|\leq \varepsilon,\\
        &\forall i, |\sigma(i)|\leq K,\\
        &\forall i, f(\reduct{R}(\cdot,i))\in \mathcal{F}_{simp}. 
    \end{align*} 
    We call $\reduct{R}$ the reduction function and $\sigma$ the weight function. In some cases, we use the notation $\reduct{R}_i(\cdot):=\reduct{R}(\cdot,i)$  for convenience.
\end{definition}
We emphasize that throughout this work, both $\mathcal{F}_0$ and $\mathcal{F}_{simp}$ are some classes of ROBPs, i.e. all our reductions keep this read-once property which is crucial in our proof.

We will frequently use compositions of reductions.
\begin{lemma}[Composition Lemma]\label{lem:composition}
    Let $\mathcal{F}_0,\mathcal{F}_1,\mathcal{F}_2$ be classes of boolean functions within $\{0,1\}^{s_0}\to \mathbb{R},\{0,1\}^{s_1}\to \mathbb{R},\{0,1\}^{s_2}\to \mathbb{R}$ respectively.
    Let $(\reduct{R}^{(1)},\sigma^{(1)})$ be an explicit $(d_1,K_1,\varepsilon_1)$-weighted pseudorandom reduction from $\mathcal{F}_0$ to $\mathcal{F}_{1}$ and $(\reduct{R}^{(2)},\sigma^{(2)})$ be an explicit $(d_2,K_2,\varepsilon_2)$-weighted pseudorandom reduction from $\mathcal{F}_{1}$ to $\mathcal{F}_{2}$. Then the composition $(\reduct{R}^{(1)}\circ\reduct{R}^{(2)},\sigma^{(1)}\cdot \sigma^{(2)}):$
    \begin{align*}
        &(\reduct{R}^{(1)}\circ\reduct{R}^{(2)})_{(i_1,i_2)}(x):=\reduct{R}^{(1)}_{i_1}(\reduct{R}^{(2)}_{i_2}(x)),\\
        &\sigma^{(1)}\cdot \sigma^{(2)}(i_1,i_2):=\sigma^{(1)}(i_1)\cdot \sigma^{(2)}(i_2),
    \end{align*}
    is an explicit $(d_1+d_2,K_1K_2,\varepsilon_1+K_1\varepsilon_2)$-weighted pseudorandom reduction from $\mathcal{F}_0$ to $\mathcal{F}_{2}$.

\end{lemma}

\begin{proof}
    Let $\reduct{R}=\reduct{R}^{(1)}\circ\reduct{R}^{(2)}$ and $\sigma=\sigma^{(1)}\cdot \sigma^{(2)}$.
    For any $f\in \mathcal{F}_0$, we have:
    \begin{align*}
        &\left|\E_U f(U)-\frac{1}{2^{d_1+d_2}}\sum_{i_1\in \{0,1\} ^{d_1},i_2\in \{0,1\} ^{d_2}} \sigma(i_1,i_2)\E_{U_{s_2}} \left[f(\reduct{R}_{(i_1,i_2)}(U_{s_2}))\right]\right|\\
        &\leq \left|\E_U f(U)-\frac{1}{2^{d_1}}\sum_{i_1\in \{0,1\} ^{d_1}} \sigma^{(1)}(i_1)\E_{U_{s_1}} \left[f(\reduct{R}^{(1)}_{i_1}(U_{s_1}))\right]\right|\\
        &+\frac{1}{2^{d_1}}\sum_{i_1\in \{0,1\} ^{d_1}} |\sigma^{(1)}(i_1)| \left|\E_{U_{s_1}} \left[f(\reduct{R}^{(1)}_{i_1}(U_{s_1}))\right]-\frac{1}{2^{d_2}}\sum_{i_2\in \{0,1\} ^{d_2}} \sigma^{(2)}(i_2)\E_{U_{s_2}} \left[f(\reduct{R}^{(2)}_{i_2}(U_{s_2}))\right]\right|\\
        &\leq \varepsilon_1+K_1\varepsilon_2.
    \end{align*}
    
    Furthermore, $|\sigma(i_1,i_2)|\leq |\sigma^{(1)}(i_1)|\cdot |\sigma^{(2)}(i_2)|\leq K_1K_2$. For any $i_1,i_2$, $x\mapsto f(\reduct{R}^{(1)}_{i_1}(x))\in \mathcal{F}_{1}$, sp we can apply $\reduct{R}^{(2)}_{i_2}$ to this mapping, therefore $x\mapsto f(\reduct{R}_{(i_1,i_2)}(x))\in \mathcal{F}_{2}$. The seed length is $d_1+d_2$. The function is explicit since both components are explicit.
\end{proof}

\begin{lemma}[Composition Lemma for multiple reductions]\label{lem:multi composition}
    For any positive integer $k$,
   let $\mathcal{F}_0,\mathcal{F}_1,\ldots,\mathcal{F}_k$ be classes of boolean functions within $\{0,1\}^{s_0}\to \mathbb{R},\{0,1\}^{s_1}\to \mathbb{R},\ldots,\{0,1\}^{s_k}\to \mathbb{R}$ respectively.
    Let $(\reduct{R}^{(1)},\sigma^{(1)}),\ldots,(\reduct{R}^{(k)},\sigma^{(k)})$ be explicit $(d_1,K_1,\varepsilon_1),\ldots,(d_k,K_k,\varepsilon_k)$-weighted pseudorandom reductions from $\mathcal{F}_0$ to $\mathcal{F}_{1},\ldots,\mathcal{F}_{k}$ respectively. Then the composition $(\reduct{R}^{(1)}\circ\ldots\circ\reduct{R}^{(k)},\sigma^{(1)}\cdot \ldots\cdot \sigma^{(k)}):$
    \begin{align*}
        &(\reduct{R}^{(1)}\circ\ldots\circ\reduct{R}^{(k)})_{(i_1,\ldots,i_k)}(x)=\reduct{R}^{(1)}_{i_1}(\ldots \reduct{R}^{(k)}_{i_k}(x)\ldots),\\
        &\sigma^{(1)}\cdot \ldots\cdot \sigma^{(k)}(i_1,\ldots,i_k)=\sigma^{(1)}(i_1)\cdot \ldots\cdot \sigma^{(k)}(i_k),
    \end{align*}
    is an explicit $(\sum_{i=1}^k d_i,\prod_{i=1}^k K_i,\sum_{i=1}^k \left(\prod_{j=1}^{i-1}K_j\right)\varepsilon_i)$-weighted pseudorandom reduction from $\mathcal{F}_0$ to $\mathcal{F}_{k}$.
\end{lemma}

\begin{proof}
    We prove this by induction on $k$. The base case $k=2$ is shown by \cref{lem:composition}. Suppose the statement holds for $k-1$, then $(\reduct{R}^{(1)}\circ\ldots\circ\reduct{R}^{(k-1)},\sigma^{(1)}\cdot \ldots\cdot \sigma^{(k-1)})$ is an explicit $(\sum_{i=1}^{k-1} d_i,\prod_{i=1}^{k-1} K_i,\sum_{i=1}^{k-1} \left(\prod_{j=1}^{i-1}K_j\right)\varepsilon_i)$-weighted pseudorandom reduction from $\mathcal{F}_0$ to $\mathcal{F}_{k-1}$. Then we can apply Lemma \ref{lem:composition} on $(\reduct{R}^{(1)}\circ\ldots\circ\reduct{R}^{(k-1)},\sigma^{(1)}\cdot \ldots\cdot \sigma^{(k-1)})$ and $(\reduct{R}^{(k)},\sigma^{(k)})$, which gives the desired result.
\end{proof}

\section{WPRG for standard ROBPs.}
\label{sec:WPRGforShortWide}

In this section, we provide a new construction of Weighted Pseudorandom Generator(WPRG) for read-once branching programs(ROBPs). The main idea is iteratively approximating the expectation of a long ROBP by a weighted sum of the expectations of much shorter ROBPs. Let $f$ be a long ROBP, our reduction will follow the paradigm:
\begin{align*}
    \left|\E f(U)-\frac{1}{2^d}\sum_{i\in 2^{d}} \sigma^{(i)}\E \left[f(\reduct{R}^{(i)}(U))\right]\right|<\varepsilon,
\end{align*}
where $f\circ \reduct{R}^{(i)}$ could be computed by a much shorter ROBP for any fixed $i$ and $f$.

We mainly use two types of weighted pseudorandom reductions: alphabet reduction and length reduction. 

\textbf{Alphabet Reduction}: In an alphabet reduction $(\reduct{R},\sigma)$, each $\reduct{R}^{(i)}$ is a $(\{0,1\}^{s'})^n\to (\{0,1\}^{s})^n$ function that maps $n$ short symbols to $n$ long symbols. For any $f\in \mathcal{B}_{(n,s,w)}$, $f\circ \reduct{R}^{(i)}$ can be computed in $\mathcal{B}_{(n,s',w)}$, where the length and width are unchanged but the alphabet bit-length is reduced.

\textbf{Length Reduction}: In a length reduction $(\reduct{R},\sigma)$, each $\reduct{R}^{(i)}$ is a $(\{0,1\}^{s'})^k\to (\{0,1\}^{s})^n$ function that maps $k$ long symbols to $n\gg k$ short symbols. For any $f\in \mathcal{B}_{(n,s,w)}$, $f\circ \reduct{R}^{(i)}$ can be computed in $\mathcal{B}_{(k,s',w)}$, where the length is reduced from $n$ to $k$ but the alphabet bit-length is increased to $s'$.

For the rest of this section, we are going to show the following lemma. The general strategy is to apply the two reductions alternately to reduce the length and maintain the width and alphabet bit-length. 
\begin{lemma}\label{lem:reduction WPRG}
    For every $\varepsilon\geq 1/\poly (w)$, there exists an explicit $\varepsilon$-WPRG for $\mathcal{B}_{(n,s,w)}$ with seed length 
    $$O\left(s+\frac{\log n\log (nw)}{\max\left\{1,\log\log w-\log\log n\right\}}+\log w \left(\log\log\min\{n,\log w\}-\log\log\max\left\{\frac{\log w}{\log \frac{n}{\varepsilon} }, 2\right\}\right)\right)$$ 
    and weight $(8n)^{2\sqrt{\frac{\log w}{\log n}}}$.
\end{lemma}

We note that for smaller target errors, one can further use the sampler trick \cite{hozaBetterPseudodistributionsDerandomization2021} with \cref{lem:reduction WPRG} to reduce the error from $1/\poly(w)$ to an arbitrary $\varepsilon>0$, attaining the following theorem. 
\begin{theorem}
\label{thm:final WPRG}
\footnote{\cref{thm:final WPRG} in a prior version of this paper stated that the seed length dependence on $s$ is $O(s)$, but the proof here only supports the current version. We thank Ben Chen, Gil Cohen, Dean Doron, Yuval Khaskelberg, and Amnon Ta-Shma for pointing out this error. We give a revised construction in \cref{sec:reducing revised} to attain the optimal dependence on $s$. }
    For all integer $n,s,w$, there exists an explicit construction of a $\varepsilon$-WPRG for $\mathcal{B}_{(n,s,w)}$ with seed length
    $$s'+ O\left(\frac{\log n\log (nw)}{\max\{1,\log\log w-\log\log n\}}+\log w \left(\log\log\min\{n,\log w\}-\log\log\max\left\{\frac{\log w}{\log \frac{n}{\varepsilon} }, 2\right\}\right)+\log(\frac{1}{\varepsilon})\right),$$
    and weight $\poly (nw/\varepsilon)$. Here $s'=O\left(s\cdot\frac{\log(\frac{n}{\varepsilon})}{\log (nw)}\right)$.
\end{theorem}
Note that our main theorem \cref{thm:intro armoni improved} is a direct corollary of \cref{thm:final WPRG}.

\subsection{Alphabet reduction.}

The alphabet reduction reduces the alphabet bit-length $s$ to a much smaller $O(\log w)$ and keeps the length $n$ and width $w$ unchanged. The alphabet reduction is achieved by using the Nisan-Zuckerman (NZ) PRG, we start with the construction of the PRG and its main ingredient, extractors.
\begin{definition}[Extractor]
    A $(k,\varepsilon)$-extractor is a function $\Ext:\{0,1\}^n\times \{0,1\}^d\to \{0,1\}^k$ such that for any $X\in \{0,1\}^n$  with $H(X)\geq k$, the distribution of $\Ext(X,U_d)$ is $\varepsilon$-close to the uniform distribution on $\{0,1\}^k$. Here $H(X)=\max_{x\in \{0,1\}^n} -\log \Pr[X=x]$ is the min-entropy of $X$, and $U_d$ is the uniform distribution on $\{0,1\}^d$. 
\end{definition}
\begin{theorem}[Explicit Extractor from \cite{guruswamiUnbalancedExpandersRandomness2009}]\label{thm:GUV}
    There exists a universal constant $C_{GUV}$ that for all positive integer $s$ and positive real $\varepsilon$, there is an explicit construction of a $(2s,\varepsilon/3)$-extractor $\Ext:\{0,1\}^{3s}\times \{0,1\}^{d}\to \{0,1\}^{s}$ with seed length $d=C_{GUV}\cdot \log \frac{ns}{\varepsilon}$.
\end{theorem}

We mainly use the one-level version of NZ PRG.
\begin{lemma}[One-level NZ PRG\cite{nisanRandomnessLinearSpace1996} with a large alphabet]\label{thm:NZonelayer}
    Let $s\geq \log w$. Assume there exists a $(2s, \frac{\varepsilon}{3n})$-extractor $\Ext:\{0,1\}^{3s}\times \{0,1\}^{d}\to \{0,1\}^{s}$.  Let $X$ and $Y_1,\ldots,Y_n$ be independent uniform random variables. Then the following construction
    \begin{align*}
        \NZ(X,Y)=\Ext(X,Y_1), \ldots, \Ext(X,Y_n)
    \end{align*}
    fools any $f\in \mathcal{B}_{(n,s,w)}$ with error at most $\varepsilon$.
\end{lemma}
The original proof of \cite{nisanRandomnessLinearSpace1996} only considers the binary case. So we include a proof in \cref{sec:NZproof} for completeness.

Now we construct the alphabet reduction. Given a ROBP $f\in \mathcal{B}_{(n,s,w)}$, we use $\E f(\NZ(X,Y))$ to approximate $\E f$. By fixing $X$, the computation of $f(\NZ(X,Y))$ can be done by a program of much smaller alphabet, that gives the following reduction.


\begin{lemma}[Alphabet Reduction]\label{lem:alphabet reduction}
    For all  positive integers $w,n,s$ with $s\geq \log w$, there exists an explicit  $(3s,1,\varepsilon)$-weighted pseudorandom reduction from $\mathcal{B}_{(n,s,w)}$ to $\mathcal{B}_{(n,C_{GUV} \log \frac{ns}{\varepsilon},w)}$.
\end{lemma}

\begin{proof}

    Let $\Ext:\{0, 1\}^{3s} \times \{0,1\}^d \rightarrow \{0, 1\}^{s}$ be a $(2s,\frac{\varepsilon}{3n})$-extractor with seed length $d=C_{GUV}\cdot \log \frac{ns}{\varepsilon}$ from Theorem~\ref{thm:GUV}. 
    Let $\NZ$ be defined as in \cref{thm:NZonelayer}. We define the reduction $(\reduct{R},\sigma)$:
    \begin{align*}
        &\reduct{R}_x(y):=\NZ(x,y)\\
        &\sigma_x:= 1.
    \end{align*}
    Now consider that we fix $x$ and let $y_i$'s be free.
    We need to prove that the reduction $(\reduct{R},\sigma)$ is a $(3s,1,\varepsilon)$-weighted pseudorandom reduction.

    Let $f\in \mathcal{B}_{(n,s,w)}$, then by \cref{thm:NZonelayer},
    $$|\E f-2^{-3s}\sum_{x\in \{0,1\}^{3s}}\E f(\NZ(x,*))|=|\E f-\E f(\NZ(X,Y))|\leq \varepsilon.$$
    Therefore, $(\reduct{R},\sigma)$ approximates $f$ with error at most $\varepsilon$.

    To prove that $f(\NZ(x,*))\in \mathcal{B}_{(n,d,w)}$ for all $f\in \mathcal{B}_{(n,s,w)}$ and $x\in \{0,1\}^{3s}$, $y\in \{0,1\}^{d}$. We construct the ROBP $g_x:=f(\NZ(x,*))$ as follows:
    
    Let $V_0,\ldots,V_n$ be the layers of $f$, each consisting of $w$ nodes. The ROBP $g_x$ is also defined on $V_0,\ldots,V_n$. The labeled edge set of $g_x$ is defined as follows: 
    \begin{align*}
        E_{g_x}=\{(u,v,y):u\in V_{i-1},v\in V_i, y\in \{0,1\}^{d}, \exists \text{ an edge from $u$ to $v$ labeled by $\Ext(x,y)$ in $f$}\}
    \end{align*}
    The start node and accept nodes of $G$ are the same as those in $f$. 
    Then $g_x$ is in $\mathcal{B}_{(n,d,w)}$ and $g_x(y_1,\ldots,y_n)=f(\Ext(x,y_1),\ldots,\Ext(x,y_n))=f(\NZ(x,y))$.
\end{proof}

\subsection{Length reduction framework from Richardson Iteration.}
\label{subsec:lengthReducFrame}

A length reduction reduces the length $n$ to a much smaller $k$ but may increase the alphabet bit-length. 
In this subsection, we construct a length reduction using the error reduction given by \cite{ahmadinejadHighprecisionEstimationRandom2022, cohenErrorReductionWeighted2021, pynePseudodistributionsThatBeat2021}.
\begin{lemma}[framework of the length reduction]\label{lem:framework}
    For any positive integers $n,s,w$, positive odd integer $k$ and positive real $\varepsilon$, assume there exists an explicit $\frac{\varepsilon}{(n+1)^2}$-PRG for $\mathcal{B}_{(n,s,w)}$ with seed length $d = d(n, s, w)$. Then there exists an explicit $(\log K, K ,\varepsilon^\frac{k+1}{2}\cdot(n+1))$-weighted pseudorandom reduction from $\mathcal{B}_{(n,s,w)}$ to $\mathcal{B}_{(k,d,w)}$, where $K = (8n)^{k+1}$.

\end{lemma}

To prove \cref{lem:framework}, we need the following error reduction.
\begin{theorem}[Error Reduction based on Richardson Iteration  \cite{ahmadinejadHighprecisionEstimationRandom2022}\cite{cohenErrorReductionWeighted2021}\cite{pynePseudodistributionsThatBeat2021}]\label{thm:richardson iteration} 
    Let $\{A_i\}_{i=1}^n\subset \mathbb{R}^{w\times w}$ be a sequence of matrices. Let $\{B_{i,j}\}_{i,j=0}^n \subset \mathbb{R}^{w\times w}$ be a family of matrices such that for every $i+1< j$, $\|B_{i,j}-A_{i+1}\ldots A_j\|\leq \varepsilon/(2(n+1))$ for some submultiplicative norm $\|\cdot\|$, $\|A_i\|\leq 1$ for all $i$ and also $B_{i-1,i}=A_i$ for all $i$. Then for any odd $k\in \mathbb{N}$, there exists a $K=(8n)^{k+1}$, a set of indices $\{n_{i,j}\}_{i\in [K], j\in [k]}$ with $0\leq n_{i,1}\leq \ldots\leq n_{i,k}= n$, and  signs $\sigma_i\in\{-1,0,1\} , i\in [K]$ such that (We set $B_{i,i}=I$ for all $i$):

    \begin{align*}
        \left\lVert A-\sum_{i\in [K]}\sigma_i\cdot B_{0,n_{i,1}}B_{n_{i,1},n_{i,2}}\ldots B_{n_{i,k-1},n_{i,k}}\right\rVert\leq \varepsilon^{(k+1)/2}\cdot(n+1).
    \end{align*}

\end{theorem}
A proof of \cref{thm:richardson iteration} is in \cref{appendix:iterationproof}.

\begin{proof}[Proof of Lemma~\ref{lem:framework}]
    Let $k,K,n_{i,j},\sigma'_i$ be as in Theorem~\ref{thm:richardson iteration}, Let $\PRG$ be a $\varepsilon/(2(n+1))$-PRG for $\mathcal{B}_{(n,s,w)}$ with seed length $d$ in the assumption. We define the reduction $(\reduct{R},\sigma)$ as follows:
    \begin{align*}
        &\reduct{R}_i(x_1,\ldots,x_k)=\PRG(x_1)_{n_{i,1}}, \PRG(x_2)_{n_{i,2}-n_{i,1}}, \ldots, \PRG(x_k)_{n_{i,k}-n_{i,k-1}}\\
        &\sigma_i=\sigma'_i K.
    \end{align*}
    Here $\PRG(x_i)_{c}$ denotes the first $c$ symbols of $\PRG(x_i)$ and each symbol is in $\{0,1\}^s$.
    
    We will show that $(\reduct{R},\sigma)$ is a $( \log K, K,\varepsilon^{\frac{k+1}{2}}\cdot(n+1))$-weighted pseudorandom reduction from $\mathcal{B}_{(n,s,w)}$ to $\mathcal{B}_{(k,d,w)}$.

    Let $f\in \mathcal{B}_{(n,s,w)}$. Recall that $f^{[i,j]}(x)$ denotes the transition matrix of $f$ from layer $i$ to layer $j$ with input $x=(x_1,\ldots,x_{j-i})$. Define $A_i=\E_{x\in \{0,1\}^s}f^{[i-1,i]}[x]$ and $B_{i,j}=\E_{x\in \{0,1\}^d}f^{[i,j]}[\PRG(x)_{j-i}]$. By the definition of the PRG, we have $\|B_{i,j}-A_{i+1}\ldots A_j\|_\infty\leq \varepsilon/(n+1)$.

    Therefore, by \cref{thm:richardson iteration},
    \begin{align*}
        &\left| \E f-\frac{1}{K}\sum_{i=1}^K \sigma(i)\E_{X} f(\reduct{R}_i(X))\right|\\
        =& \left\lVert \E f^{[0,n]} - \sum_{i=1}^K \sigma_i \E f^{[0,n]}(\PRG(X_1)_{n_{i,1}}, \PRG(X_2)_{n_{i,2}-n_{i,1}},\ldots,\PRG(X_k)_{n_{i,k}-n_{i,k-1}})\right\rVert_\infty\\
        \leq& \left\lVert \E f^{[0,n]}- \sum_{i=1}^K \sigma_i \E f^{[0,n_{i,1}]}(\PRG(X_1)_{n_{i,1}})\E f^{[n_{i,1},n_{i,2}]}(\PRG(X_2)_{n_{i,2}-n_{i,1}})\cdots \right.\\
        &\qquad\qquad\qquad\left. \vphantom{\sum_{i=1}^K} \cdots\E f^{[n_{i,k-1},n_{i,k}]}(\PRG(X_k)_{n_{i,k}-n_{i,k-1}})\right\rVert_\infty\\
        =& \left\lVert A_1\ldots A_n - \sum_{i=1}^K \sigma_i B_{0,n_{i,1}}B_{n_{i,1},n_{i,2}}\ldots B_{n_{i,k-1},n_{i,k}}\right\rVert_\infty\\
        \leq& \varepsilon^{\frac{k+1}{2}}\cdot(n+1).
    \end{align*}


    The weight is $K=(8n)^{k+1}$ by  \cref{thm:richardson iteration}, and so the seed length is $\log K=O(k\log n)$.

    Finally, we need to show that $f\circ \reduct{R}_i \in \mathcal{B}_{(k,d,w)}$ for all $f\in \mathcal{B}_{(n,s,w)}$ and $i \in [K]$. 
    We construct the ROBP $g:=f\circ \reduct{R}_i$ as follows: 
    
    Let $V_0,\ldots,V_n$ be the layers of $f$, each consisting of $w$ nodes. The ROBP $g$ is defined on $V_0,V_{n_{i,1}},\ldots,V_{n_{i,k}}$. The labelled edge set of $g$ is defined as follows:
    \begin{align*}
        E_g= \{ & (u,v,x): j\in [n], u\in V_{n_{i,j-1}}, v\in V_{n_{i,j}}, x\in \{0,1\}^{d}, \\
                & \text{there exists a path from $u$ to $v$ in $f$ through $\PRG(x)_{n_{i,j}-n_{i,j-1}}$ } \}
    \end{align*}
    The start node and accept nodes of $g$ are the same as that in $f$. Then $g$ is in $\mathcal{B}_{(k,d,w)}$ and $g(x_1,\ldots,x_n)=f(\PRG(x_1)_{n_{i,1}}, \PRG(x_2)_{n_{i,2}-n_{i,1}}, \ldots,\PRG(x_k)_{n_{i,k}-n_{i,k-1}})=f(\reduct{R}_i(x_1,\ldots,x_k))$.
\end{proof}

\subsection{Length reduction instantiated by Armoni's PRG.}

We use Armoni's PRG for $\mathcal{B}_{(n,s,w)}$ to instantiate the framework of \cref{subsec:lengthReducFrame}.

\begin{theorem}[Armoni's PRG\cite{armoniDerandomizationSpaceBoundedComputations1998, kaneRevisitingNormEstimation2009}]\label{thm:armoni prg}
    For all positive integer $n,s,w$ and positive real $\varepsilon$, there exists an explicit construction of a $\varepsilon$-PRG for $\mathcal{B}_{(n,s,w)}$ with seed length $O(s+\frac{\log n\log (nw/\varepsilon)}{\log\log w-\log\log (n/\varepsilon)})$.
    Here the big-O of the seed length hides a universal constant.
\end{theorem}

We have two instantiations that have different parameters.
\begin{lemma}[Length reduction 1]\label{lem:length reduction}
    For any constant $a,c,C\in \mathbb{N}$, for all integer $n,w$, if $n\le \log^c w  $, then there exists an explicit $(\log K,K,1/w^a)$-weighted pseudorandom reduction from $\mathcal{B}_{(n,C \log w,w)}$ to $\mathcal{B}_{(n^{1/c}, C'\log w,w)}$, where $K\leq (8n)^{n^{1/c}+1}$, $C'$ is a constant depending on $c, C, a$.
\end{lemma}

\begin{proof}
Fix $c,C$ to be constants. We use Armoni's PRG\footnote{Readers can notice that for this setting, the Armoni's PRG that we use, should automatically become the Nisan-Zuckerman PRG.}
for $\mathcal{B}_{(n,C\log w,w)}$ with error $\varepsilon_0=\frac{2^{-\frac{a\log w}{n^{1/c}}}}{(n+1)^2}$.
By \cref{thm:armoni prg}, the seed length is $O\left(C\log w+\frac{a\log w\log \log w}{\log\log w/c+O(1)}\right)=C' \log w $, where $C'$ depends on $c, C, a$.
Then we can set $k=2 n^{1/c} \ge \frac{a\log w+\log n}{1/2\cdot\log(1/\varepsilon_0)}$ in Lemma~\ref{lem:framework} and invoke it to attain this lemma.
\end{proof}

\begin{lemma}[Length reduction 2]\label{lem:armoni reduction}
    For any constant $a\in \mathbb{N}$, for all integer $n,s,w$, there exists an explicit $(\log K,K, 1/w^a)$-weighted pseudorandom reduction from $\mathcal{B}_{(n,s,w)}$ to $\mathcal{B}_{\left(\log^{\frac{1}{2}} w,O\left(s+\frac{\log n\log (nw)}{\log\log w-\log\log n}\right),w\right)}$, where $K=(8n)^{\sqrt{\frac{\log w}{\log n}}+1}$.

\end{lemma}

\begin{proof}
 We use Armoni's PRG for $\mathcal{B}_{(n,s,w)}$  with error  $\varepsilon_0=\frac{2^{-2a\sqrt{\log n\log w}}}{(n+1)}$. 
By \cref{thm:armoni prg}, the seed length is $O\left(s+\frac{\log n\log (nw)}{\log\log w-\log\log n}\right)$. 
Then we can set $k=\frac{a\log w+\log n}{1/2\cdot \log(1/\varepsilon_0)}\leq \sqrt{\frac{\log w}{\log n}}\leq \log^{1/2} w$ in Lemma~\ref{lem:framework} and invoke it to attain this lemma.
\end{proof}

\subsection{Combining the reductions.}
\label{sec:combine reduction}

We combine the reductions from Lemma~\ref{lem:alphabet reduction}, Lemma~\ref{lem:length reduction} and Lemma~\ref{lem:armoni reduction} to give the following reduction. 

\begin{lemma}[Main reduction] \label{lem:final reduction}
 For all integer $n,s,w$, for all $\varepsilon\geq 1/\poly(w)$, there exists an explicit   $\left(O\left(s+\frac{\log n\log (nw)}{\log\log w-\log\log n}+\log w (\log\log\min\{n,\log w\}-\log\log\max\left\{\frac{\log w}{\log \frac{n}{\varepsilon} }, 2\right\})\right),(8n)^{2\sqrt{\frac{\log w}{\log n}}}, \varepsilon \right)$-weighted pseudorandom reduction from $\mathcal{B}_{(n,s,w)}$ to $\mathcal{B}_{\left(O(1),O(\log w),w\right)}$.
\end{lemma}

\begin{figure}[H]
    \begin{tikzpicture}[shorten >=1pt, auto]

    \node (q0)   {$\mathcal{B}(n,s,w)$};
    \node[right=7cm of q0] (q1) {$\mathcal{B}(n_0,s_0,w)$};
    \node[below=4cm of q1] (q2) {$\mathcal{B}(n_i,s_i,w)$};
    \node[below=4cm of q0] (q3) {$\mathcal{B}(n'_i,s'_i,w)$};

    \path[->,align=center]
        (q0) edge node[above] { Length reduction 2:\\
            $n_0 = \log^{1/3}w$ ,  $s_0 = O\left(s+\frac{\log n\log (nw)}{\log\log w-\log\log n}\right)$,\\
                        $d_0 = O(\log w)$,    $K_0 = (8n)^{2\sqrt{\frac{\log w}{\log n}}}$, $\varepsilon_0 = \varepsilon/4$.}(q1);
        (q1);
    \path[->,align=center]
        (q1) edge node[right] {Alphabet reduction:\\
         $n_1 = n_0= \log^{1/3}w$, \\  $s_1 =  2C_{GUV}\log \frac{1}{\varepsilon'_i} $,\\
                        $d_1=O(s_1) =  O\left(s+\frac{\log n\log (nw)}{\log\log w-\log\log n}\right)$,\\    $K_1 = 1$\\ $\varepsilon_1 = \varepsilon/(4K_0)$.                    
                        }(q2);
    \path[->,align=center,bend right=10]
        (q2) edge node[above] {Length reduction 1:\\ $n'_i = n^{1/3}_i$ ,  $s'_i = C_{large} \log w$\\
                        $d_i=O(n_i)$,    $K_i = O(2^{n_i})$, $\varepsilon_i = (\varepsilon/(4K_0))^2$.}(q3);
    \path[->,align=center,bend right=10]
        (q3) edge node[below] {Alphabet reduction:\\ $n_{i+1} = n'_i$ ,  $s_{i+1} = 2C_{GUV}\log \frac{1}{\varepsilon'_i}$\\
                        $d'_{i}=O(\log w)$,    $K'_{i} = 1 $,$\varepsilon'_i = (\varepsilon/(4K_0))^2$.}(q2);
    \end{tikzpicture}
    \caption{The recursion we use to iteratively reduce the ROBP in the construction of the WPRG~\cref{lem:reduction WPRG}.  The arrows represent the reduction from an ROBP to simpler ROBPs. 
    }\label{fig:ROBP reduction}
\end{figure}
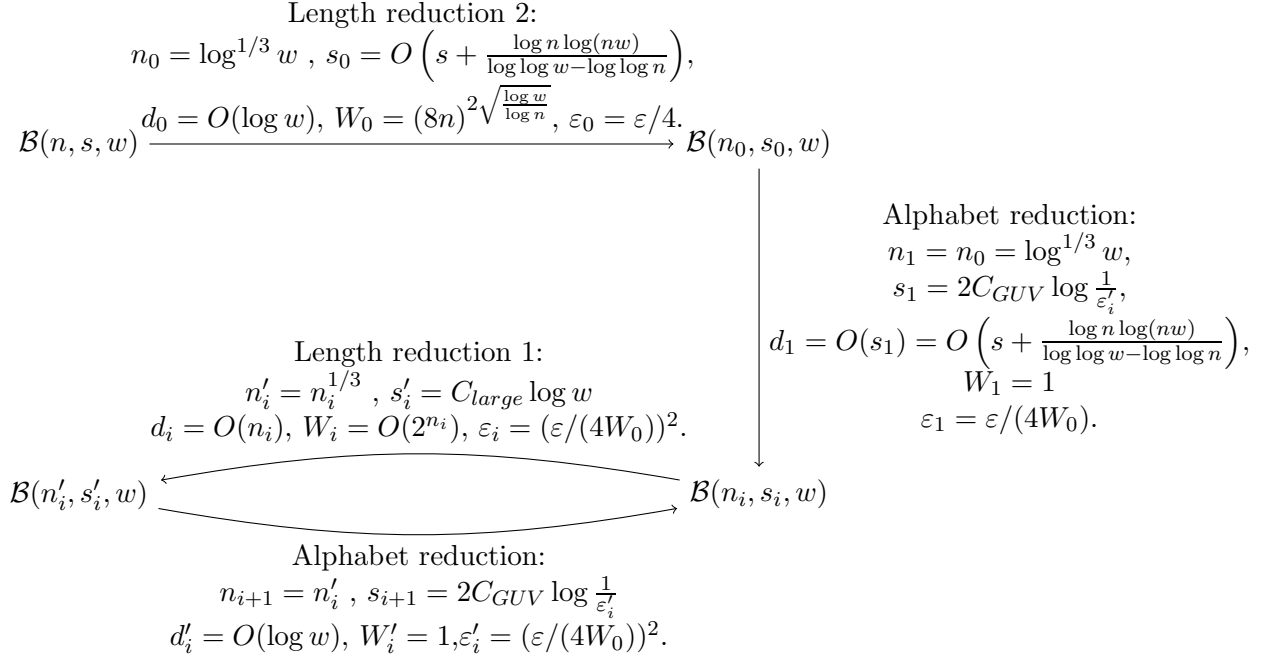

\begin{proof}
    We set up the reductions as the following, which is also illustrated in \Cref{fig:ROBP reduction}.

    We have the following reductions:

    Let $(\reduct{R}^{(1)},\sigma^{(1)})$ be a $(d_1,K_1, \varepsilon/4)$-weighted pseudorandom reduction from $\mathcal{B}_{(n,s,w)}$ to $\mathcal{B}_{(n_1,s_1,w)}$ given by Lemma~\ref{lem:armoni reduction}, where
    \begin{itemize}
        \item $d_1=O(\log w)$,
        \item $n_1=\min\{n,\log^{1/2} w\}$,
        \item $s_1=O(s+\frac{\log n\log (nw)}{\log\log w-\log\log n})$,
        \item $K_1=(8n)^{\sqrt{\frac{\log w}{\log n}}+1}$.
    \end{itemize}

    Let $(\reduct{R}^{(2)},\sigma^{(2)})$ be a $(s_2,K_2, \varepsilon/(4K_1))$-weighted pseudorandom reduction from $\mathcal{B}_{(n_1,s_1,w)}$ to $\mathcal{B}_{(n_2,s_2,w)}$ given by Lemma~\ref{lem:alphabet reduction}, where
    \begin{itemize}
        \item $d_2=O(s+\frac{\log n\log (nw)}{\log\log w-\log\log n})$,
        \item $n_2=n_1$,
        \item $s_2\leq 2C_{GUV}\log (1/\varepsilon)$,
        \item $K_2=1$.
    \end{itemize}
    
    Using \cref{lem:composition}, $(\reduct{R}^{(1)}\circ \reduct{R}^{(2)},\sigma^{(1)}\cdot \sigma^{(2)})$ is a $(d_1+d_2,K_1,\varepsilon/2)$ weighted pseudorandom reduction from $\mathcal{B}_{(n,s,w)}$ to $\mathcal{B}_{(\log^{1/2} w, 2C_{GUV}\log (1/\varepsilon),w)}$. We set $\varepsilon'=(\varepsilon/K_1)^2$ and continue the reduction with the following parameters:

    $n_3=\log^{1/2} w,n_{i+1}=n_i^{1/3}$ for $i=3,\ldots,l-1$. $n_l=\frac{\log w}{\log 1/\varepsilon'}+2=\frac{\log w}{\log 1/\varepsilon+\sqrt{\log n\log w}}+2$.
    
    It is clear that  
    $$l=\log\log\min\{n,\log w\}-\log\log\left(\frac{\log w}{\log n/\varepsilon}+2\right)+O(1).$$

    For $i=3,\ldots,l-1$, let $(\reduct{R}^{(i)},\sigma^{(i)})$ be a $(d_i,K_i, \varepsilon')$-weighted pseudorandom reduction from $\mathcal{B}_{(n_i,2C_{GUV}\log (1/\varepsilon'),w)}$ to $\mathcal{B}_{(n_{i+1},C_{large}\log w,w)}$ given by Lemma~\ref{lem:length reduction}, setting constants $c =3 $, $ C = \frac{2C_{GUV}\log (1/\varepsilon')}{\log w}$, $a= \log_w (1/\varepsilon')$ in \cref{lem:length reduction}, where
    \begin{itemize}
        \item $d_i= \log K_i = O(n_i^{1/2}\log n_i)$,
        \item $K_i=\exp(O(n_i^{1/2}\log n_i))$.
    \end{itemize}
    Notice that since $n_3=\log^{1/2} w$ and $n_i$ is monotonously decreasing, the condition $n_i\leq \log^3 w$ is always satisfied, so the condition in \cref{lem:length reduction} is always met.


    For $i=3,\ldots,l-2$, let $(\widehat{\reduct{R}}^{(i)},\widehat{\sigma}^{(i)})$ be a $(\widehat{d}_i,\widehat{K}_i, \varepsilon')$-weighted pseudorandom reduction from $\mathcal{B}_{(n_{i+1},C_{large}\log w,w)}$ to $\mathcal{B}_{(n_{i+1}, 2C_{GUV}\log (1/\varepsilon'),w)}$ given by Lemma~\ref{lem:alphabet reduction}, where
    \begin{itemize}
        \item $\widehat{d}_i=O(\log w)$,
        \item $\widehat{K}_i=1$.
    \end{itemize}

    We invoke \cref{lem:multi composition} to compose the $2l-7$ reductions, which gives $(\reduct{R}^{(3)}\circ \widehat{\reduct{R}}^{(3)}\circ \reduct{R}^{(4)}\circ \widehat{\reduct{R}}^{(4)}\circ \ldots \circ \reduct{R}^{(l-1)},\sigma^{(3)}\cdot \widehat{\sigma}^{(3)}\cdot \sigma^{(4)}\cdot \widehat{\sigma}^{(4)}\cdot \cdots \cdot \sigma^{(l-1)})$. 
    Also by \cref{lem:multi composition}, we know the reduction is a $(d^*,K^*,\varepsilon^*)$-weighted pseudorandom reduction from $\mathcal{B}_{(n,s,w)}$ to $\mathcal{B}_{(n_l, 2C_{GUV}\log (1/\varepsilon'),w)}$. The latter class is contained in $\mathcal{B}_{(O(1),C_{GUV}\log w,w)}$.The parameters $(d^*,K^*,\varepsilon^*)$ is shown as following:
    \begin{itemize}
        \item $d^*=\sum_{i=3}^{l-1}d_i=O(l\log w)$,
        \item $K^*=\prod_{i=3}^{l-1}K_i\cdot \widehat{K}_i=\exp(\sum_{i=3}^{l-1}n_i^{1/2}\log n_i)\leq \exp(l\log^{1/3} w)<\exp(\log^{1/2} w)$,
        \item $\varepsilon^*=\sum_{i=3}^{l-1}\varepsilon'\cdot \prod_{j=3}^{i-1}K_j\cdot \widehat{K}_j\leq l\cdot \varepsilon\cdot K<\varepsilon'\cdot \exp(\log^{1/2} w)<\varepsilon/(2K_1)$.
    \end{itemize}

    We invoke \cref{lem:composition} again and compose the $(d_1+d_2,K_1,\varepsilon/2)$-weighted pseudorandom reduction from $\mathcal{B}_{(n,s,w)}$ to $\mathcal{B}_{(\log^{1/3} w,2C_{GUV}\log (1/\varepsilon),w)}$ with the $(d^*,K^*,\varepsilon^*)$-weighted pseudorandom reduction from $\mathcal{B}_{(\log^{1/3} w,2C_{GUV}\log (1/\varepsilon),w)}$ to $\mathcal{B}_{(O(1),C_{GUV}\log w,w)}$. We get the desired reduction with the following parameters:
    \begin{itemize}
        \item Seed length: $s_1+s_2+s^*=O(s+\frac{\log n\log (nw)}{\log\log w-\log\log n}+\log w (\log\log\log w-\log\log\frac{\log w}{\log (n/\varepsilon)})$,
        \item Weight: $K_1\cdot K^*\leq (8n)^{2\sqrt{\frac{\log w}{\log n}}}$,
        \item Error: $\varepsilon/2+K_1\cdot \varepsilon/(2K_1)=\varepsilon$.
    \end{itemize}

    The lemma follows.    
\end{proof}

The reduction naturally gives a WPRG, which is the PRG in \Cref{lem:reduction WPRG}

\begin{lemma}[Lemma~\ref{lem:reduction WPRG} restated]
\label{lem:lemma reduction WPRG}
    For all $C>0,\varepsilon>1/\poly (w)$, there exists an explicit $\varepsilon$-WPRG for $\mathcal{B}_{(n,C\log w,w)}$ with seed length $O\left(\frac{\log n\log (nw)}{\log\log w-\log\log n}+\log w \left(\log\log\log w-\log\log\max\left\{\frac{\log w}{\log \frac{n}{\varepsilon} }, 2\right\}\right)\right)$ and weight $(8n)^{2\sqrt{\frac{\log w}{\log n}}}$.
\end{lemma}

\begin{proof}
    Let $(\reduct{R},\sigma)$ be the reduction from Lemma~\ref{lem:final reduction}. We construct the WPRG $(G,w)$ with the same weight function $w$ and $G$ is defined as $G(x,i)=\reduct{R}(x,i)$. 
    By \cref{lem:final reduction}, to fool the original ROBP we only need to provide randomness for the seed of $(\reduct{R},\sigma)$ and also provide a random string $x$ which can fool an arbitrary ROBP in  $\mathcal{B}_{(O(1),C_{GUV}\log w,w)}$.
    We use true randomness for $x$ and this only takes $O(\log w)$ random bits.
    Also the randomness used by $(\reduct{R},\sigma)$ is  $O\left(\frac{\log n\log (nw)}{\log\log w-\log\log n}+\log w (\log\log\log w-\log\log\frac{\log w}{\log (n/\varepsilon)})\right)$.
    So the total seed length is as stated.
    The weight also directly follows from \cref{lem:final reduction}.
\end{proof}

\subsection{Reducing error beyond $1/\poly(w)$.}

To further reduce the error from $1/\poly(w)$ an arbitrary small $\varepsilon$, we use the sampler trick from~\cite{hozaBetterPseudodistributionsDerandomization2021}. The original technique is applied on PRGs to get a WPRG with a smaller error. Here we also use the technique but we apply it on a WPRG and get a WPRG with a smaller error.

Recall the definition of averaging samplers:
\begin{definition}[Averaging Sampler]\label{def:Sampler}
    A $(\alpha,\gamma)$-averaging sampler is
    a function $\Samp:\{0,1\}^r \times \{0,1\}^p\to \{0,1\}^q$ such that for every function $f:\{0,1\}^q\to [-1,1]$, it holds that:
    \begin{align*}
        \Pr_{x \in \{0,1\}^r }\left[\left|2^{-p}\sum_{y\in \{0,1\}^p}  f(\Samp(x,y)) - \E [f] \right| \ge \alpha\right]\leq \gamma.
    \end{align*}
    
\end{definition}

\begin{lemma}[\cite{chattopadhyayOptimalErrorPseudodistributions2020} Appendix B, \cite{goldreich2011sample, reingold2000entropy}]\label{lem:good sampler}
    For all $\alpha>0,\gamma>0$, there exists an explicit $(\alpha,\gamma)$-averaging sampler with seed length $r=q+O(\log(1/\alpha)+\log(1/\gamma))$ and $p=O(\log(1/\alpha)+\log\log(1/\gamma))$.
\end{lemma}

We now give the construction of a WPRG for class $\mathcal{B}_{(n,s,w)}$ with small error $\varepsilon$. First, let $(G_0:\{0,1\}^d\to(\{0,1\}^s)^n,\sigma_0:\{0,1\}^d\to\R)$ be a WPRG for $\mathcal{B}_{(n,s,w)}$ , error $1/(2w\cdot (n+1)^2)$ and weight $W$, which is constructed from Lemma~\ref{lem:reduction WPRG}. We then set the parameters as follows:
\begin{itemize}
    \item $k=\frac{\log (n/\varepsilon)}{\log (nw)}$,
    \item $\alpha=1/(W\cdot w^2\cdot (n+1)^2)$,
    \item $\gamma= \varepsilon/(2(2n)^k\cdot (W)^{k+1}\cdot w^2)$.
\end{itemize}
Let $\Samp:\{0,1\}^r \times \{0,1\}^p\to \{0,1\}^d$ be a $(\alpha,\gamma)$-averaging sampler via \cref{lem:good sampler}. Let $K,n_{i,j},\sigma_{i}$ be as the corresponding parameters in Lemma~\ref{thm:richardson iteration}, which depend on our $k, n$ here. 
Our WPRG is constructed as follows:
\begin{align*}
    \begin{cases}
        &G(x,y_1,\ldots,y_{k},i)=Z(x,y_1,i),Z(x,y_2,i),\ldots,Z(x,y_k,i),\\
        &\forall i\in [K],j\in[k], Z(x,y_j,i)=\begin{cases}
        &G_0(\Samp(x,{y_j}))_{n_{i,j}-n_{i,j-1}} \qquad if\ n_{i,j}-n_{i,j-1}>1,     \\
        &(y_j)_s, \qquad if \ n_{i,j}-n_{i,j-1}=1,     \\
        \end{cases}\\
        &\sigma(x,y_1,\ldots,y_{k},i)=\sigma_i K\cdot\prod_{j=1}^{k}\sigma_0(\Samp(x,y_j)),\\
    \end{cases}
\end{align*}
where $x\in\{0,1\}^r$, each $y_j\in\{0,1\}^p\cup\{0,1\}^s$. $(y_j)_s$ is $y_j$ truncated to the first $s$ bits. 
Notice that in the Richardson-iteration error-reduction
polynomial, each monomial alternates between approximate multi-step transition
matrices and exact one-step transition matrices. The approximated transitions are attained by using the outputs of $G_0$, and the exact transitions are attained by using those  $(y_j)_s$.

Now we show that the construction serves as a good WPRG for $\mathcal{B}_{(n,s,w)}$ with error $\varepsilon$. 
\begin{lemma}\label{lem:sampler for small error}
    For all $n,s,w$, assume there exists a $W$-bounded $1/(2w\cdot (n+1)^2)$-WPRG $(G_0,\sigma_0)$  for $\mathcal{B}_{(n,s,w)}$ with seed length $d$, and a $(\alpha,\gamma)$-averaging sampler $\Samp:\{0,1\}^r \times \{0,1\}^p\to \{0,1\}^d$ with $\alpha,\gamma$ as defined above. Then there exists a $\varepsilon$-WPRG for $\mathcal{B}_{(n,s,w)}$ with seed length $r+kp+ks$ and weight $\sigma^{\log(n/\varepsilon)/\log (nw)}\cdot \poly(nw/\varepsilon)$.
\end{lemma}
The proof of \cref{lem:sampler for small error} is deferred to \cref{append:smallererr}.

Combining the lemma with Lemma~\ref{lem:reduction WPRG}, we prove the main theorem of the section:

\begin{theorem}
    For all integer $n,s,w$, there exists an explicit construction of a $\varepsilon$-WPRG for $\mathcal{B}_{(n,s,w)}$ with seed length 
    $$O\left(s\cdot\frac{\log(n/\varepsilon)}{\log (nw)}+\frac{\log n\log (nw)}{\log\log w-\log\log n}+\log w (\log\log\min\{n,\log w\}-\log\log\frac{\log w}{\log (n/\varepsilon)})+\log(1/\varepsilon)\right)$$ 
    and weight $\poly (nw/\varepsilon)$.
\end{theorem}

\begin{proof}
    Let $(G_0,\sigma_0)$ be a $1/(2w\cdot (n+1)^2)$-WPRG for $\mathcal{B}_{(n,s,w)}$ with seed length 
    $$d=O\left(s+\frac{\log n\log (nw)}{\log\log w-\log\log n}+\log w (\log\log\min\{n,\log w\}-\log\log\frac{\log w}{\log (n/\varepsilon)})\right)$$ and weight $K=\poly (nw/\varepsilon)$ given by Lemma~\ref{lem:reduction WPRG}. 
    
    Let $\Samp$ be a $(1/(2W\cdot w^2\cdot (n+1)^2),\varepsilon/(2(2n)^k\cdot \sigma^{k}))$-averaging sampler with seed length $r=d+O(\log(nw/\varepsilon))$ and $p=O(\log(nw)+\log\log(1/\varepsilon))$ given by Lemma~\ref{lem:good sampler}. 

    By Lemma~\ref{lem:sampler for small error}, we get a $\varepsilon$-WPRG for $\mathcal{B}_{(n,s,w)}$ with seed length 
    \begin{align*}
        &r+(2k+1)p+(2k+1)s\\
        = &O\left(s\cdot\frac{\log(n/\varepsilon)}{\log(nw)}+\frac{\log n\log (nw)}{\log\log w-\log\log n}+\log w (\log\log\min\{n,\log w\}-\log\log\frac{\log w}{\log (n/\varepsilon)})+\log(1/\varepsilon)\right)
     \end{align*}
    and weight $\poly (nw/\varepsilon)$.

\end{proof}

\subsection{Application: a WPRG for regular branching program with a better seed length.}
\label{subsec:appforregularROBP}

One application of our WPRG is to improve the WPRG for regular ROBPs by Chen, Hoza, Lyu, Tal, and Wu\cite{chenWeightedPseudorandomGenerators2023}. Their construction can be viewed as a weighted pseudorandom reduction from long regular ROBPs to short standard ROBPs with large alphabets. Then they fool the short ROBPs with an INW generator. We show that if one replaces this INW generator with our WPRG \cref{thm:final WPRG revised} (or alternatively, by the WPRG of \cite{chenImprovedErrorReduction2026}) instead, then we can slightly improve the seed length.

First we recall the ``Weight'' of an ROBP defined in \cite{bravermanPseudorandomGeneratorsRegular2014}, which is different from the weight of a WPRG.
\begin{definition}[Weight of ROBPs]
Let $f\in \mathcal{B}_{n,s,w}$ be an ROBP. For every $v\in [n]\times [w]$, let $f^{v\to}$ denote the subsection of $f$ starting at $v$ and has the same accept nodes as $f$. Let $E$ be the edges of $f$. We define the weight of $f$ as:
\[
    \mathsf{W}(f)=\sum_{(u,v)\in E}\left|\E[f^{v\to}]-\E[f^{u\to}]\right|.
\]
 \end{definition}

 Recall that we denote the class of all regular ROBPs of length $n$, width $w$ and alphabet $\Sigma=\{0,1\}^{s}$ as $\mathcal{R}_{n,s,w}$. Our result can be stated as the following.
\begin{theorem}\label{thm:WPRG for regular}
For every $w, n, s \in \mathbb{N}$ 
, and every $\varepsilon>0$, there exists an explicit $\varepsilon$-WPRG for $\mathcal{B}_{n,s,w}$ with seed length
\begin{align*}
    O\left(s+\log n\left(\sqrt{\log(1/\varepsilon)}+\log w +\log(\mathsf{W}(f))+\log\log n\right) +\log(1/\varepsilon)\right).
\end{align*}
Furthermore, the weight of this WPRG is bound by $\poly(n^{{1+\sqrt{\log( 1/\varepsilon)}}},w)$.
\end{theorem}
One main result of \cite{bravermanPseudorandomGeneratorsRegular2014} states that the weights of regular ROBPs over a binary alphabet are bounded by $O(w^2)$.
\begin{lemma}[Lemma 6 of \cite{bravermanPseudorandomGeneratorsRegular2014}]\label{lem:regular bounded weight}
    For every $n,w\in \mathbb{N}$ and $f\in \mathcal{R}_{n,1,w}$, $\mathsf{W}(f)\leq O(w^2)$.
\end{lemma}
Combining \cref{lem:regular bounded weight} with \cref{thm:WPRG for regular} immediately gives \cref{thm:WPRG for regular0}.

Before we prove \cref{thm:WPRG for regular}, we recall the following from \cite{chenWeightedPseudorandomGenerators2023, chattopadhyayRecursiveErrorReduction2023}.
In Section 6.3 of \cite{chenWeightedPseudorandomGenerators2023},  the WPRG $(G,w):[K]\times(\{0,1\}^s)^r\to\{0,1\}^n\times\mathbb{R}$ for $\mathcal{R}_{n,1,w}$ 
, has the form
\begin{align*}
    G(i,y_1,y_2,\ldots y_r)&=\left(G_{i,1}(y_1),G_{i,2}(y_2),\ldots,G_{i,r}(y_r)\right)\\
    \sigma(i,y_1,y_2,\ldots y_r)&=K\cdot \gamma_i,
\end{align*}
where $i\in [K]$ and $\forall j\in[r], j\in \{0,1\}^s$. The parameters satisfy $K=n^{O(m)}, r=O(m\log n),\gamma_i\in \{-1,+1\}$, where $m=\Theta\left(\frac{\log(w/\varepsilon)}{\sqrt{\log(1/\varepsilon)}+\log w}\right)=O\left(\sqrt{\log(1/\varepsilon)}\right)$. $G_{i,j}$ is the PRG by \cite{bravermanPseudorandomGeneratorsRegular2014} for $\mathcal{R}_{n,1,w}$ within target error $\tau=\min\left\{2^{-\sqrt{\log (1/\varepsilon)}},1/\poly(w,\mathsf{W}(f),\log n)\right\}$. Therefore, the length of each $y$ is  
\begin{align*}
    s=O\left(\log n\left(\sqrt{\log(1/\varepsilon)}+\log w +\log(\mathsf{W}(f))+\log\log n\right)\right).
\end{align*}
We view the WPRG as a weighted pseudorandom reduction from regular ROBPs to standard ROBPs by fixing $i$ and let $y_1,\dots, y_r$ be free. Moreover, this WPRG can naturally extend to large alphabet cases although not explicitly stated in  \cite{chenWeightedPseudorandomGenerators2023}. \cite{chattopadhyayRecursiveErrorReduction2023} gives a similar result. We restate both of their results as the following theorem, and we include a short proof in \cref{appendix:regular analysis} for completeness. 
\begin{theorem}[Section 6.3 of \cite{chenWeightedPseudorandomGenerators2023}, also Section 3 of \cite{chattopadhyayRecursiveErrorReduction2023}]\label{thm:CHLTW for regular}
For all $n,w\in \mathbb{N}$ and $\varepsilon>0$, there exists a $(\log K, K, \varepsilon)$-weighted pseudorandom reduction from $\mathcal{B}_{n,s,w}$ to $\mathcal{B}_{n_1,n_1,w}$, where $K=n^{O(\sqrt{\log(1/\varepsilon})}, n_1=O(\log n\sqrt{\log(1/\varepsilon})$ and
\begin{align*}
    s_1=O\left(s+\log n\left(\sqrt{\log(1/\varepsilon)}+\log w +\log(\mathsf{W}(f))+\log\log n\right)\right).
\end{align*}

\end{theorem}

Now prove \cref{thm:WPRG for regular}.
\begin{proof}[Proof of \cref{thm:WPRG for regular}]

   In the case of $w>n$, \cref{thm:final WPRG revised} already gives a suitable WPRG. So we assume $w\leq n$.
   Denote $f\in\mathcal{B}_{n,s,w}$ as the original regular ROBP. Let $(\reduct{R},\sigma)=\{(\reduct{R}_i,\sigma_i)\}_{i\in [K]}$ be the $(\log K, K, \varepsilon/2)$ -reduction from $\mathcal{R}_{n,s,w}$ to $\mathcal{B}_{n_1,s_1,w}$, given by \cref{thm:CHLTW for regular}, where $K=n^{O(\sqrt{\log(1/\varepsilon})}, n_1=O(\log n\sqrt{\log(1/\varepsilon)})$, and
    $$s_1=O\left(s+\log n\left(\sqrt{\log(1/\varepsilon)}+\log w + \log(\mathsf{W}(f))+\log\log n\right)\right).$$
   Then we have
   \begin{align*}
   &\left|\E[f]-\sum_{i\in[K]}\frac{\sigma_i}{K}\cdot \E_{x\in (\{0,1\}^{s_1})^{n_1}}\left[f(\reduct{R}_i(x)\right]\right|\leq \varepsilon/2,
   \end{align*}
   in which $|\sigma_i|\leq K$.
  Let $(G,\sigma'):\{0,1\}^d\to (\{0,1\}^{s_1})^{n_1}\times \mathbb{R}$
be the $\varepsilon/(2K)$-WPRG given by \cref{thm:final WPRG revised}.
   We have
   
   \begin{align*}
   &\left|\E[f]-\sum_{i\in [K]}\frac{\sigma_i}{K}\cdot \sum_{y\in\{0,1\}^d}\frac{\sigma'(y)}{2^d} \left[f(\reduct{R}_i(G(y))\right]\right|\\
   \leq& \left|\E[f]-\sum_{i\in[K]}\frac{\sigma_i}{K}\cdot \E_{x\in (\{0,1\}^{s_1})^{n_1}}\left[f(\reduct{R}_i(x)\right]\right|
   +\sum_{i\in[K]}\frac{\sigma_i}{K}\left| \E_{x\in (\{0,1\}^s)^r}\left[f(\reduct{R}_i(x)\right]-\sum_{y\in\{0,1\}^d}\frac{\sigma'(y)}{2^d} \left[f(\reduct{R}_i(G(y))\right] \right|\\
   \leq& \varepsilon/2+K\cdot \varepsilon/(2K)=\varepsilon.
   \end{align*}
 Therefore $(\reduct{R}\circ G, \sigma\cdot \sigma')$ is a $\varepsilon$-WPRG for $f$ with seed length $d_{total}=\log K+d$.

 By \cref{thm:final WPRG revised}, the $\varepsilon/(2K)$-WPRG requires
\begin{align*}
    d=&O\left(s_1+\frac{\log n_1\log (n_1w)}{\max\left\{1,\log\log w-\log\log n_1\right\}}+\log w \log\log\min\{n_1,\log w\}+\log (K/\varepsilon)\right)
\end{align*}
We use the assumption that $w\leq n$
, organize the terms and find
\begin{align*}
    d_{total}=&O\left(s+\log n\left(\sqrt{\log(1/\varepsilon)}+\log w  +\log(\mathsf{W}(f)) +\log\log n\right)+ \log n_1\log n+\log(K/\varepsilon)\right)\\
    =& O\left(s+\log n\left(\sqrt{\log(1/\varepsilon)}+\log w +\log(\mathsf{W}(f))+\log\log n\right)+\log(1/\varepsilon)\right),
\end{align*}
since $ \log n_1\log (n_1 w) = O( \log n_1 \log n),  \log w \log\log\min\{n_1,\log w\} = O(\log n \log n_1)$.


Now we compute the weight, which is the upper bound of $|\sigma_i|\cdot|\sigma'(x)|$ over $i\in[K],x\in\{0,1\}^d$. By \cref{thm:final WPRG revised}, $|\sigma'(x)|\leq \poly(r,w,\varepsilon/2K)=\poly(n^{{1+\sqrt{\log( 1/\varepsilon)}}},w)$.  $|\sigma_i|\leq K=n^{O(1+\sqrt{\log( 1/\varepsilon)})}$. Therefore, the total weight is bound by $\poly(n^{{1+\sqrt{\log( 1/\varepsilon)}}},w)$.

\end{proof}

\section{WPRG for Permutation Read-once Branching Programs.}
\label{sec:PROBPWPRG}

In this section we focus on WPRGs against permutation ROBPs. First we recall the definition of permutation ROBPs with only one accept node.
\begin{definition}[Permutation ROBPs]
    A ROBP $f\in \mathcal{B}_{(n,s,w)}$ is a permutation ROBP if for every $i\in [n]$, $x\in \{0,1\}^s$, the matrix $f^{[i-1,i]}(x)$ is a permutation matrix.

    We denote the class of permutation ROBPs with unbounded width and only one accept node by $\mathcal{P}_{(n,s)}$, where $n$ is the length $s$ is the bit-length of the alphabet.
\end{definition}
Now we give our WPRG for  $\mathcal{P}_{(n,s)}$.
\begin{theorem}\label{thm:permutation-reduction-main}[Restatement of \cref{thm:permWPRG}]
    There exists an $\varepsilon$-WPRG for $\mathcal{P}_{(n,s)}$ with seed length $s+O(\log n(\log\log n+\sqrt{\log(1/\varepsilon)})+\log(1/\varepsilon))$ and weight $2^{O(\log n\sqrt{\log(1/\varepsilon)})}$.
\end{theorem}
Before we prove the theorem, we introduce some more definitions and lemmas.
\begin{definition}[PSD norm]
    Let $A$ be a $w\times w$ positive semi-definite matrix. The PSD norm on $\mathbb{R}^w$ with respect to $A$ is defined as $\|x\|_A=\sqrt{x^TAx}$.
\end{definition}

\begin{definition}[sv-approximation \cite{ahmadinejad2023singular}]
    Let $\widetilde{W}$ and $W$ be two $w\times w$ doubly stochastic matrices. We say that $\widetilde{W}$ is a $\varepsilon$-singular-value approximation of $W$, denoted by $\widetilde{W} \sv_{\varepsilon}W$, if for all $x,y\in \mathbb{R}^w$,
    \begin{align*}
        \left|y^T(\widetilde{W}-W)x\right|\leq \frac{\varepsilon}{4}\left( \|x\|^2_{I-W^TW}+\|y\|^2_{I-WW^T}\right).
    \end{align*}
    
\end{definition}

\begin{definition}[fooling with sv-error]\label{def:sv-error}
    Let $(G,\sigma):\{0,1\}^s \rightarrow \{0,1\}^n\times \mathbb{R}$ be a WPRG. Let $\mathcal{C}$ be a family of matrix valued functions $\mathbf{B}: \{0,1\}^s \rightarrow \mathbb{R}^{w\times w}$. We say that $G$ fools $\mathcal{C}$ with $\varepsilon$-sv-error if for all $\mathbf{B}\in \mathcal{C}$, $\sum_{z\in \{0,1\}^s}[\frac{1}{2^s}\mathbf{B}(G(z))\cdot \sigma(x)]\sv_{\varepsilon}\E_{x \in \{0,1\}^n}[\mathbf{B}(x)]$.
\end{definition}

We describe the PRG against permutation ROBPs from \cite{hoza2021pseudorandom}. 
In the following descriptions we regard ROBPs of $\mathcal{P}_{(n,s)}$ as matrix-valued functions.
\begin{lemma}[Lemma 4.1 in \cite{chenWeightedPseudorandomGenerators2023}, originally by  \cite{hoza2021pseudorandom}]\label{lem:sv-fool}
    For all $s,n,\varepsilon$ there exists a PRG (actually the INW generator) that fools $\mathcal{P}_{(n,s)}$ with $\varepsilon$-sv-error. The seed length is
    \begin{align*}
        s+O(\log n(\log\log n+\log(1/\varepsilon))).
    \end{align*} 
\end{lemma}
\begin{remark}
    We note that Lemma 4.1 in \cite{chenWeightedPseudorandomGenerators2023} only proves the case for $s=2$, but their proof can naturally be generalized to handle arbitrary $s$.  See \cref{sec:sv-fool sketch}.
\end{remark}

Next we describe the key ingredient in previous error reductions for WPRGs against permutation ROBPs. 

\begin{lemma}[Claim 9.3 in \cite{chenWeightedPseudorandomGenerators2023}]\label{lem:sv-error-reduction-polynomials}
    Let $n \in \mathbb{N}$, let $\tau \in \left(0, \frac{1}{64 \log^2 n}\right)$, and let $G:\{0,1\}^d\to (\{0,1\}^s)^n$ be a PRG that fools  $\mathcal{P}_{(n,s)}$ with sv-error $\tau$. Let $k \in \mathbb{N}$, and let 
    \[
    \alpha = \left(4 \cdot \sqrt{\tau} \cdot \log n \right)^{k+1}.
    \]
    Then there exists a WPRG $G^{(k)}, \sigma^{(k)}$ that can be written in the form of 
    \begin{align*}
        &G^{(k)}(i, x_1, \ldots, x_r)=G(x_1)_{n_{i,1}}, G(x_2)_{n_{i,2}}, \ldots,G(x_r)_{n_{i,r}}, \\
        &\sigma^{(k)}(i) = \gamma_i\cdot K.
    \end{align*}
    where $K=n^{O(k)},i\in [K],r=O(k\log n)$, $\gamma_i\in \{-1,0,1\}$ 
    and $0\leq n_{i,j}\leq n$, such that $G^{(k)}$ fools $\mathcal{P}_{(n,s)}$ with entrywise error $\alpha$.

\end{lemma}

\subsection{One level of the reduction.}
  We show that the previous lemma already gives a reduction:
\begin{lemma}\label{lem:onelevelWPRPerm}
    For any integer $n,k,s$ and any $\varepsilon>0$, there exists $\tau=\Omega(\varepsilon^{\frac{2}{k+1}}/\log^2 n)$ such that there exists a $(O(k\log n), n^{O(k)},\varepsilon)$-weighted pseudorandom reduction $(\reduct{R},\sigma)$ from $\mathcal{P}_{(n,s)}$ to $\mathcal{P}_{(O(k\log n),s+O(\log n(\log\log n+\log 1/\tau)))}$.
\end{lemma}

\begin{proof}
    Let $G$ be the INW PRG that fools $\mathcal{P}_{(n,s)}$ with $\tau$-sv-error, such that $\tau=\min\{\frac{\varepsilon^{2/(k+1)}}{16\log^2 n},\frac{1}{64\log^2 n}\}$. Then the seed length is $d = s+O(\log n(\log\log n+\log 1/\tau))$ by Lemma \ref{lem:sv-fool}.

    We invoke \cref{lem:sv-error-reduction-polynomials} with $G$ and $k$, which gives $(G^{(k)},\sigma^{(k)})$. This WPRG fools $\mathcal{P}_{(n,s)}$ with entrywise error $\left(4 \cdot \sqrt{\tau} \cdot \log n \right)^{k+1}<\varepsilon$. We define the reduction $(\reduct{R},\sigma): \reduct{R}_i(x_1,...,x_r)=G^{(k)}(i,x_1,...,x_r),w_i=\sigma^{(k)}(i)$.
    By \cref{lem:sv-error-reduction-polynomials}, the weight of this reduction is bounded by $K=n^{O(k)}$. 
    The seed length of the reduction is $\log K=O(k\log n)$, since we need $\log K=O(k\log n)$ bits to choose $i$.

    Given any $f\in \mathcal{P}_{(n,s)}$, notice that each $f\circ \reduct{R}_i$ is a permutation ROBP of length $r=O(k\log n)$ and alphabet size $2^d=2^{s+O(\log n(\log\log n+\log 1/\tau))}$. The reason is that the transition matrix of $f\circ \reduct{R}_i$ from the $j-1$-th layer to the $j$-th layer on input $x\in \{0,1\}^{d}$ can be expressed as $(f\circ \reduct{R}_i)^{[j-1,j]}(x)$. As $(f\circ \reduct{R})^{[j-1,j]}(x)=f^{[n_{i,j-1},n_{i,j}]}(G(x)_{n_{i,j}-n_{i,j-1}})$ and the latter is a product of $n_{i,j}-n_{i,j-1}$ permutation matrices, the transition matrix of $f\circ \reduct{R}_i$ is also a permutation matrix. Hence $f\circ \reduct{R}_i$ is a permutation ROBP.
    Also, note that each $f\circ \reduct{R}_i$ only has one accept node since its accept node is the same as that of $ f$.
    
    Therefore, the reduction is a $(O(k\log n), O(n^k),\varepsilon)$-weighted pseudorandom reduction.
    
\end{proof}

Setting $k=\sqrt {\log (1/\varepsilon)}$ in \cref{lem:onelevelWPRPerm}, we immediately have the following lemma:
\begin{lemma}\label{cor:permutation-reduction-1}
    For any integer $n,s$ and any $\varepsilon>0$, there exists a $(O(\log n\sqrt{\log (1/\varepsilon)}),2^{O(\log n\sqrt{\log (1/\varepsilon)})},\varepsilon)$-weighted pseudorandom reduction from $\mathcal{P}_{(n,s)}$ to $\mathcal{P}_{(O(\log n\sqrt{\log (1/\varepsilon)}),s+O(\log n(\sqrt{\log (1/\varepsilon)}+\log \log n)))}$.
\end{lemma}

Setting $k=O(\log^2 n)$ in \cref{lem:onelevelWPRPerm}, we immediately have the following lemma:
\begin{lemma}\label{cor:permutation-reduction-2}
    For any integer $n,s$ and $\varepsilon>0$, there exists a $(O(\log^3 n),2^{O(\log^3 n)},\varepsilon)$-weighted pseudorandom reduction from $\mathcal{P}_{(n,s)}$ to  $\mathcal{P}_{(\log^3 n,s+O(\frac{\log (1/\varepsilon)}{\log n}+\log n\log \log n))}$.
\end{lemma}

\subsection{Recursion of reductions.}
We give a multi-level recursive procedure which  gradually reduces the program from $\mathcal{P}_{(n,s)}$ to $\mathcal{P}_{(O(1),s+O(\log n(\log\log n+\sqrt{\log(1/\varepsilon)})))}$ with error $\varepsilon$. 

We start by reducing the length of the permutation ROBP to $O(\log n\sqrt{\log(1/\varepsilon)})$ and the weight to $2^{O(\log n\sqrt{\log(1/\varepsilon)})}$ with error $\varepsilon/2$. Previous works use an INW PRG to fool the reduced ROBP, which leads to an extra double logarithmic factor in the seed length. Here we use iterative reductions instead to avoid this extra factor.

\begin{lemma}[Reduction for permutation ROBP]\label{lem:permutation-reduction-main}
    For any integer $n,s$ and any $\varepsilon>0$, there exists a $\left( O\left(\log n\sqrt{\log(1/\varepsilon)} \right),2^{\log n\sqrt{\log(1/\varepsilon)}},O(\varepsilon) \right)$-weighted pseudorandom reduction from $\mathcal{P}_{(n,s)}$ to $\mathcal{P}_{\left(O(1),s+O\left(\log n \left(\log\log n+\sqrt{\log(1/\varepsilon)}\right)+\log(1/\varepsilon)\right) \right)}$.
\end{lemma}

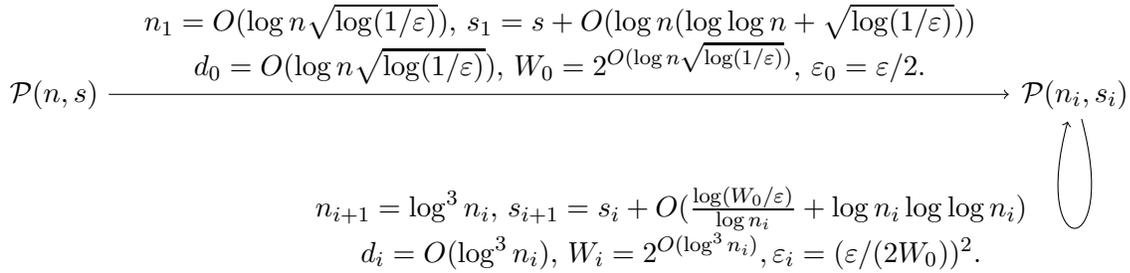
\begin{figure}[H]\label{fig:WPRGforPerm}
    \begin{tikzpicture}[shorten >=1pt, auto]

    \node (q0)   {$\mathcal{P}(n,s)$};
    \node[right=12cm of q0] (q1) {$\mathcal{P}(n_i,s_i)$};

    \path[->,align=center]
        (q0) edge node[above] {
            $n_1=O(\log n \sqrt{\log(1/\varepsilon)})$, $s_1 = s+O(\log n (\log\log n+\sqrt{\log(1/\varepsilon)}))$\\
                        $d_0 = O(\log n \sqrt{\log(1/\varepsilon)})$,    $K_0 = 2^{O(\log n \sqrt{\log(1/\varepsilon)})}$, $\varepsilon_0 = \varepsilon/2$.}(q1);
        (q1);
    \path[->,loop below,align=center, distance=2cm]
        (q1) edge node[left=0.5cm] 
         {
            $n_{i+1}=\log^3 n_i$, $s_{i+1} = s_i+O(\frac{\log(W_0/\varepsilon)}{\log n_i}+\log n_i\log\log n_i)$\\
                        $d_i = O(\log^3 n_i)$,    $W_i = 2^{O(\log^3 n_i)}, \varepsilon_i = (\varepsilon/(2W_0))^2$.                    
                        }(q1);

    \end{tikzpicture}
    \caption{Reduction for permutation ROBPs in~\cref{lem:permutation-reduction-main}. 
    }
\end{figure}

\begin{proof}

    Let $(\reduct{R}^{(0)},\sigma^{(0)})$ be the $(d_0,K_0,\varepsilon/2)$-weighted pseudorandom reduction from $\mathcal{P}_{(n,s)}$ to $\mathcal{P}_{(n_1,s_1)}$  such that $n_1=O(\log n\sqrt{\log(1/\varepsilon)})$ and $s_1=s+O(\log n(\log\log n+\sqrt{\log(1/\varepsilon)}))$. This reduction is guaranteed by Lemma \ref{cor:permutation-reduction-1} and have the following parameters: 
    \begin{enumerate}
        \item $d_0=O(\log n\sqrt{\log(1/\varepsilon)})$,
        \item $K_0=2^{O(\log n\sqrt{\log(1/\varepsilon)})}$.
    \end{enumerate}

    Now we define a new parameter $\varepsilon'=\varepsilon/2K_0$. Let 
    \[n_0=n,n_i=\log^3 n_{i-1}\]
    for each $i=1,\ldots,l$ and $n_l= 32768$. It is clear that $l\leq \log\log n$.

    For each $i=1,\ldots,l$, let $(\reduct{R}^{(i)},\sigma^{(i)})$ be the $(d_i,K_i,(\varepsilon')^2)$-weighted pseudorandom reduction from $\mathcal{P}_{(n_{i},s_{i})}$ to $\mathcal{P}_{(n_{i+1},s_{i+1})}$, which is guaranteed by Lemma \ref{cor:permutation-reduction-2}. The reduction has the following parameters:
 
    \begin{enumerate}
        \item $d_i=O(\log^3 n_{i})$,
        \item $K_i=2^{\log^3 n_i}$,
        \item $s_{i+1}=s_{i}+O(\frac{\log(1/\varepsilon')}{\log n_{i}}+\log n_{i}\log \log n_{i})=s_{i}+O(\frac{\log(1/\varepsilon)}{\log n_{i}})$.
    \end{enumerate}

    Using \cref{lem:multi composition}, we first composite $(\reduct{R}^{(1)},\sigma^{(1)}),\ldots,(\reduct{R}^{(l)},\sigma^{(l)})$ to get a $(d^*,K^*,\varepsilon^*)$-weighted pseudorandom reduction $(\reduct{R}^{(*)},\sigma^{(*)})$ from $\mathcal{P}_{(n_0,s_0)}$ to $\mathcal{P}_{(n_l,s_l)}$ with $\reduct{R}^{(*)}=\reduct{R}^{(1)}\circ \ldots \circ \reduct{R}^{(l)}$ and $\sigma^{(*)}=\sigma^{(1)}\cdot \ldots \cdot \sigma^{(l)}$. We have the following parameters:

    \begin{enumerate}
        \item $d^*=d_1+\ldots+d_l\leq O(l\log^3 n_1)\leq O(\log^{0.1}(n/\varepsilon))$,
        \item $K^*=K_1\cdot \ldots \cdot K_l\leq 2^{O(l\cdot \log^3n_1)}\leq 2^{O(\log^{0.1}(n/\varepsilon))}$,
        \item $\varepsilon^*=\sum_{i=1}^{l}(\varepsilon')^{2}\cdot \prod_{j=0}^{i-1}K_j\leq(\varepsilon')^{2}\cdot K^*\cdot l\leq \varepsilon'/2$,
        \item $s_l=s_0+\sum_{i=1}^{l}(s_i-s_{i-1})=s_0+\sum_{i=1}^{l}O(\frac{\log(1/\varepsilon')}{\log n_{i}})=s_0+O(\log 1/\varepsilon')$.
    \end{enumerate}
    The first three items are attained by directly plugging in our parameters.
    The last item holds because $n_i$ decreases rapidly. Note that $n_{i-1}= 2^{\sqrt[3]{n_i}}$, and when $n_i\geq 2^{15}=32768$, we have $n_{i-1}\geq n^2_i$. When $n_l\geq 32768$, it holds that $\sum_{i=1}^{l}\frac{\log(1/\varepsilon')}{\log n_{i}}\leq \log 1/\varepsilon'\cdot (1/\log n_l+1/(2\log n_l)+1/(4\log n_l)+\cdots +1/(2^l\log n_l))\leq \log 1/\varepsilon'$.

    Finally, we composite $(\reduct{R}^{(*)},\sigma^{(*)})$ with $(\reduct{R}^{(0)},\sigma^{(0)})$ to get the final reduction $(\reduct{R},\sigma)$ from $\mathcal{P}_{(n,s)}$ to $\mathcal{P}_{(O(1),s_l))}$. 
    By \cref{lem:composition}, the parameters are:
    \begin{enumerate}
        \item $d=d^*+d_0=O(\log n\sqrt{\log(1/\varepsilon)})$,
        \item $K=K^*\cdot K_0\leq 2^{O(\log n\sqrt{\log(1/\varepsilon)})}$,
        \item $\varepsilon=\varepsilon/2+\varepsilon'/2\cdot K_0\leq \varepsilon$,
        \item $s=s_l=s_0+O(\log 1/\varepsilon')=s+O(\log n(\log\log n+\sqrt{\log(1/\varepsilon)})+\log(1/\varepsilon))$.
    \end{enumerate}
\end{proof}
One can also see our reduction strategy for the above lemma through Figure \ref{fig:WPRGforPerm}.

Now we show the main theorem of this section:
\begin{theorem}[Theorem \ref{thm:permutation-reduction-main} restated]
    There exists an $\varepsilon$-WPRG for $\mathcal{P}_{(n, s)}$ with seed length $s+O(\log n(\log\log n+\sqrt{\log(1/\varepsilon)})+\log(1/\varepsilon))$ and weight $2^{O(\log n\sqrt{\log(1/\varepsilon)})}$.
\end{theorem}

\begin{proof}
    Let $(\reduct{R}, \sigma)$ be the reduction from Lemma \ref{lem:permutation-reduction-main}. We construct the WPRG $(G,\sigma) $ with the same weight function $\sigma$, and $G$ is defined as $G(x,i)=\reduct{R}(x,i)$.
    By \cref{lem:permutation-reduction-main},  to fool the original ROBP we only need to provide randomness for the seed of $(\reduct{R},\sigma)$ and also provide a random string $x$ which can fool an arbitrary ROBP in  $\mathcal{P}_{(O(1),s+O(\log n(\log\log n+\sqrt{\log(1/\varepsilon)})+\log(1/\varepsilon)))}$.
    We can use another $\varepsilon$ error INW generator from \cite{hozaPseudorandomGeneratorsUnboundedWidth2020} to generate this constant length $x$ and this only takes $s + O\left(\log n\left(\log\log n+\sqrt{\log(1/\varepsilon)}\right)+\log(1/\varepsilon) \right)$ random bits.
    Also the randomness used by $(\reduct{R},\sigma)$ is $O(\log n\sqrt{\log(1/\varepsilon)})$.
     So the total seed length is as stated.
     The weight also directly follows from \cref{lem:permutation-reduction-main}.

\end{proof}

It is well known that a WPRG can fool a Permutation ROBP with an arbitrary number of accept nodes and error $\varepsilon$ if it can fool the program with only one accept node with error $\varepsilon/w$. Therefore, we immediately have the following corollary:

\begin{corollary}\label{corol:multiaccperm}
    There exists an $\varepsilon$-WPRG for length $n$, width $w$, alphabet size (out degree) $2^s$ Permutation ROBPs having an arbitrary number of accept nodes, with seed length $s+O(\log n(\log\log n+\sqrt{\log(w/\varepsilon)})+\log(w/\varepsilon))$ and weight $2^{O(\log n\sqrt{\log(w/\varepsilon)})}$.
\end{corollary}

\section{Derandomization for Regular Branching Programs.}\label{sec:derand regular}

In this section, we show our derandomization for short-wide regular ROBPs, which indicates that
when $n\leq 2^{\log^{1/2} w}$ and $\varepsilon=1/\poly(w)$, there is a $O(\log w)$ space derandomization, where the space complexity is optimal up to constant factors. 
We achieve this by constructing a more general derandomization algorithm as the following.
\begin{theorem}\label{thm:regularderand}
    There exists an algorithm that for any input regular ROBP $f$ of width $w$, length $n$, alphabet $\{0,1\}^s$, and any input  parameter $\varepsilon>0$ as input, outputs an approximation for $\E[f]$ with addtive error $\varepsilon$. The algorithm uses $O(s+\log n(\log\log n+\sqrt{\log(w/\varepsilon)})+\log(w/\varepsilon))$ bits of space.
\end{theorem}
Notice that our main theorem \cref{thm:intro regular} is a direct corollary of \cref{thm:regularderand}. We remark that one may want to further improve the space dependence on $\varepsilon$ by first using \cref{thm:regularderand} and then calling the Richardson iteration method of \cite{ahmadinejadHighprecisionEstimationRandom2022}. This can be done, but our main purpose here is to get a derandomization in $\mathbf{L}$ exactly for short-wide ROBPs. If applying that additional operation, then the algorithm is not in $\mathbf{L}$ and also not in polynomial time.

Next as a warm-up, we show that binary regular ROBPs can be transformed in logspace to binary permutation ROBPs such that they have the same acceptance probability (but probably do not compute the same function). 
We are unaware of any previous work stating this transformation.
\begin{lemma}
    There is an algorithm which on inputting a regular ROBP $P$ with length $n$ width $w$ and binary edge labels, outputs a permutation ROBP $P'$ with length $n$ width $w$ and binary edge labels, such that $P'$ has exactly the same acceptance probability as $P$ (but may not computing the same function as $P$).
    The algorithm has workspace $O(\log (nw))$.
\end{lemma}
\begin{proof}
$P'$ has the same vertex set as that of $P$.
The algorithm visits each layer $i $ of $P$ in the order $i = 0, 1, 2, \ldots, n$.
For layer $i$, consider the subgraph $S$ between $V_i$ and $V_{i+1}$. The algorithm visits each node $v \in V_i$ in order.
For each $v \in V_i$, it tests if $v$ can be reached by any previously visited node of $V_i$ in $S$. This can be done by revisiting each previous node $v'$ and go through the cycle that $v'$ is in. If none of the previous node can reach $v$, then do the following traverse in $S$. Starting from $v$, choose an edge with label $0$, and then go to the corresponding neighbor. For each next node $u$ visited, denote $\ell$ as the label of the adjacent edge through which we reach $u$. If the other adjacent edge $e'$ of $u$ has a label $\ell'$ such that $ \ell' = \ell$, then flip the label $\ell'$.
After this we traverse through $e'$ to go to the next neighbor. 
Go on doing this until it reaches $v$ again.

Notice that starting from $v$ the traverse can go through every node of $V_i \cup V_{i+1}$ reachable from $v$. Because for each next neighbor $u$, it can only be either $v$ or a node which is not touched before in this traverse. In other words, this traverse is going through the cycle of $S$ containing $v$.
Also note that after this traverse, every node reachable from $v$, because of the flipping, can have different labels for its two adjacent edges.
So after we visited every $v \in V_i$, every node of $S$ reachable by nodes in $V_i$ has its two adjacent edges with different labels because of the flips. Hence the graph with these new labels is a permutation.
The acceptance probability maintains the same since the edges are the same as before except that we flip some edge labels. After flippings, each node still has its two outgoing edges with different labels.

The traverse can be done in logspace since recording the index of a layer takes space $O(\log n)$, recording the index of a node in a layer takes space $O(\log w)$, and going through a cycle of $S$ takes space $O(\log w)$.

\end{proof}
After this transformation, we can run our WPRG of \cref{corol:multiaccperm} to attain the desired derandomization for the binary case.

However for alphabets with size larger than $2$, the problem turns out to be more complicated. 
In the rest of this section, we focus on the case with large alphabets.

\subsection{More preliminaries.}

We recall some standard definitions in the literature of  \cite{reingold2000entropy, rozenmanDerandomizedSquaringGraphs2005, ahmadinejadHighprecisionEstimationRandom2022, chenWeightedPseudorandomGenerators2023, chattopadhyayRecursiveErrorReduction2023}. 

\begin{definition}[Regular bigraph]
    A bigraph is a triple $G=(U,V,E)$, where $U$ and $V$ are two sets of vertices and $E\subseteq U\times V$ is a set of edges going from $U$ to $V$. A bigraph is called $d$-regular if every vertex in $U$ has $d$ outgoing edges and every vertex in $V$ has $d$ incoming edges. 

    The transition matrix of a regular bigraph is a the matrix $\mathbf{M}\in \mathbb{R}^{V\times U}$ such that $\mathbf{M}_{v,u}$ is the fraction of edges going from $u$ that go to $v$.

\end{definition}

\begin{definition}[One-way labeling]
    A one-way labeling of a $d$-regular bigraph $G$ assigns a label $i\in [d]$ to each edge in $G$ such that for every vertex $u\in U$, the labels of the outgoing edges of $u$ are distinct. If $G$ has a one way labeling, we say that $G[u,i]=v$ if the outgoing edge of $u$ that is labeled $i$ goes to $v$.
\end{definition}

\begin{definition}[Two-way labeling]
    A two-way labeling of a $d$-regular bigraph $G$ is a labeling of the edges of $G$ such that:
    \begin{itemize}
        \item Every edge $(u,v)$ has two labels in $[d]$, the `outgoing label' and the `incoming label'.
        \item For every vertex $u\in U$, the outgoing labels of the outgoing edges of $u$ are distinct.
        \item For every vertex $v\in V$, the incoming labels of the incoming edges of $v$ are distinct.
    \end{itemize}
\end{definition}

\begin{definition}[Rotation map]\label{def:rot map}
    Let $G=(U,V,E)$ be a $d$-regular bigraph with a two-way labeling. The rotation map of $G$ is a function $\Rot_G:U\times [d]\to  V\times [d]$ such that $\Rot_G(u,i)=(v,j)$ if there is an edge $(u,v)\in E$ with the outgoing label $i$ and the incoming label $j$.
    
\end{definition}

\begin{definition}[Derandomized Product]
    Let $G_1=(U,V,E_1)$ and $G_2=(V,T,E_2)$ be two $d$-regular bigraphs where $G_1$ has a two-way labeling and $G_2$ has a one-way labeling. Let $H=([d],[d],E_H)$ be a $c$-regular bigraph with one way labeling. The derandomized product of $G_1$, $G_2$ and $H$ is a $(c\cdot d)$-regular bigraph with one-way labeling denoted by $G_1\ds_H G_2$ defined as follows. To compute $(G_1\ds_H G_2)[v_0,(i_0,j_0)]$ for  $v_0\in U$ and $(i_0,j_0)\in [d]\times [c]$:
    \begin{itemize}
        \item Let $(v_1,i_1)=\Rot_{G_1}(v_0,i_0)$.
        \item Let $i_2=H[i_1,j_0]$.
        \item Let $v_2=G_2[v_1,i_2]$.
        \item Output $(G_1\ds_H G_2)[v_0,(i_0,j_0)]=:v_2$.
    \end{itemize}
    
\end{definition}

\begin{definition}[Two-way labeling of derandomized product]
    Let $G_1=(U,V,E_1)$ and $G_2=(V,T,E_2)$ be two $d$-regular bigraphs where both $G_1$ and $G_2$ have two-way labeling. Let $H=([d],[d],E_H)$ be a $c$-regular bigraph with two-way labeling. We define the two-way labeling of the derandomized product $G_1\ds_H G_2$ as follows. To compute the rotation map $\Rot_{G_1\ds_H G_2}$ for any vertex $v_0\in U$ and any pair $(i_0,j_0)\in [d]\times [c]$, we do the following:
    \begin{itemize}
        \item Let $(v_1,i_1)=\Rot_{G_1}(v_0,i_0)$.
        \item Let $(i_2,j_1)=\Rot_H(i_1,j_0)$.
        \item Let $(v_2,i_3)=\Rot_{G_2}(v_1,i_2)$.
        \item Output $\Rot_{G_1\ds_H G_2}(v_0,(i_0,j_0))=(v_2,(i_3,j_1))$.
    \end{itemize}
    
\end{definition}

To ensure that the derandomized product behaves like the direct concatenation of the two bigraphs, we need $H$ to be a spectral expander. 

\begin{definition}[spectral expander]
    Let $H=(U,V,E_H)$ be a $d$-regular bigraph with transition matrix $W_H\in \mathbb{R}^{U\times V}$, where $|U|=|V|$. Define $J\in \mathbb{R}^{U\times V}$ where $J_{i,j}=1/|U|$ for all $i\in U,j\in V$. Then the spectral expansion of $H$ is denoted by $\lambda(H)$ and defined as follows:
    \begin{align*}
        \lambda(H)=\|W_H-J\|_2
    \end{align*}
\end{definition}

We also need the definitions for regular branching programs:

\begin{definition}[Regular Branching Program]
    Let $f$ be a ROBP in $\mathcal{B}_{n,s,w}$. We call $f$ a regular branching program if all vertices in the graph of $f$ except the vertices in the first layer have precises $2^s$ incoming edges.
\end{definition}

A general regular branching program may not have a two-way labeling. If we equip it with a two-way labeling for each step of the transition, we call it a regular ROBP with two-way labeling.

\begin{definition}[Regular ROBP with Two-way labeling]\label{def:regular label}
    A regular ROBP with two-way labeling is a regular ROBP $f$ and every edge in $f$ has two labels, the 'outgoing label' and the 'incoming label'. For every vertex $u$ in $V(f)\backslash V_n$, the outgoing labels of the outgoing edges of $u$ are distinct. For every vertex $v$ in $V(f)\backslash V_0$, the incoming labels of the incoming edges of $v$ are distinct.  We compute $f$ according to the outgoing labels of the edges, i.e., for every $(x_1,\ldots,x_n)\in (\{0,1\}^s)^n$, we find the unique path $(v_0=v_{start},v_1,\ldots,v_n)$ in $f$ such that for every $i\in [n]$, the edge $(v_{i-1},v_i)$ has the outgoing label $x_i$, and we output 1 iff $v_n$ is an accepting vertex.

    We denote the class of all regular ROBPs with two-way labeling $\mathcal{R}^{tw}_{n,s,w}$.

    Like bigraphs, we define $\Rot_f:([w]\times [n])\times \{0,1\}^s\to ([w]\times [n])\times \{0,1\}^s$ such that $\Rot_f(u,\sigma)=(v,\sigma')$ if there is an edge $(u,v)$ in $f$ with the outgoing label $\sigma$ and the incoming label $\sigma'$.
\end{definition}

We recall the derandomization algorithm for regular branching programs in \cite{chenWeightedPseudorandomGenerators2023}.

Let $f$ be a regular ROBP with two-way labeling in $\mathcal{R}^{tw}_{n,s,w}$. For each $i\in [\log n]$, let $H_t=([2^s\cdot c^{t-1}],[2^s\cdot c^t],E_{H_t})$ be a $c$-regular bigraph with $\lambda(H_t)\leq \lambda$. Define $\widetilde{G}_{j\to j+1}:=(V_{t-1},V_t,E_t)$, which is a $2^s$-regular bigraph with two-way labeling. Define $E(SC_n)=\{(j,j+2^t):j\in [n-2^t],t\in [\log n], 2^t \text{ is the largest power of 2 dividing } j\}$. For each $(j,j+2^t)\in E(SC_n)$ recursively define the graph $\widetilde{G}_{j \to j+2^t}$ as follows:
\begin{align*}
    \widetilde{G}_{j\to j+2^t} := \widetilde{G}_{j\to j+2^{t-1}}\ds_{H_i} \widetilde{G}_{j+2^{t-1}\to j+2^t}
\end{align*}

The following lemma shows that the transition matrix of the final graph $\widetilde{G}_{0\to n}$ is close to the product of the transition matrices of the intermediate graphs.

\begin{lemma}[Lemma A.20 of \cite{chenWeightedPseudorandomGenerators2023}] \label{lem:sv approximation algorithm}
    Let $n, d, c \in \mathbb{N}$. For each $t \in \{0, 1, \ldots, \log n\}$ and each $j \in [n-2^t]$, let $\widetilde{G}_{j+2^{t-j}} = \big(V_j, V_{j+2^t}, E_{j+2^{t-j}}\big)$ be a $(d \cdot c^t)$-regular bigraph with a two-way labeling. Furthermore, let $\lambda \in \big(0, \frac{1}{6 \log^2 n}\big)$, and for each $t \in [\log n]$, let $H_t = \big([d \cdot c^{t-1}], [d \cdot c^{t-1}], E_{H_t}\big)$ be a $c$-regular bigraph with a one-way labeling satisfying $\lambda(H_t) \leq \lambda$. Assume also that for each $t \in [\log n]$ and each $j \in [n-2^t]$, we have
\[
\widetilde{G}_{ j \to j+2^t} = \widetilde{G}_{ j \to j+2^{t-1}} \ds_{H_t} \widetilde{G}_{j+2^{t-1}\to j+2^t}, \forall t \ge 1,
\]
and 
\[
\widetilde{G}_{ j \to j+1} = G_{j \to j+1},
\]
(where the equation above merely denotes equality as graphs with one-way labelings). For each $(i, j) \in E(\mathsf{SC}_n)$, let $\widetilde{\mathbf{W}}_{j\gets i}$ be the transition matrix of $\widetilde{G}_{i\to j}$, and let 
\[
\mathbf{W}_{j\gets i} = \widetilde{\mathbf{W}}_{j\gets j-1} \cdots \widetilde{\mathbf{W}}_{i+1\gets i}
\] 
Then 
\[
\widetilde{\mathbf{W}}_{n\gets 0} \sv_{11\lambda \cdot \log n} \mathbf{W}_{n\gets 0}.
\]
    
\end{lemma}

\begin{theorem}[Claim A.23 of \cite{chenWeightedPseudorandomGenerators2023}] \label{thm:derandomization algorithm}
    The following algorithm computes 
    $$\Rot_{\widetilde{G}_{0\to n}}(v_0, (x,e_1, \ldots, e_{\log n}))$$

\begin{enumerate}
    \item For $i = 1$ to $n$:
    \begin{enumerate}
        \item[(a)] Update $(v, x) \gets \Rot_{\widetilde{G}_{i-1\to i}}(v, x)$, so now $v \in V^{(i)}$.
        \item[(b)] If $i < n$:
        \begin{enumerate}
            \item Let $t \in [\log n]$ be the smallest positive integer such that $i$ is not a multiple of $2^t$, i.e., the binary expansion of $i$ has precisely $t-1$ trailing zeroes.
                
            \item Update $(x,e_1, \ldots, e_{t-1}) \gets \Rot_{H_t}((x,e_1, \ldots, e_{t-1}), e_t)$.
        \end{enumerate}
    \end{enumerate}
    \item Output $(v, e)$.
\end{enumerate}

\end{theorem}

We also need the efficiently computable spectral expander $H_t$. The following lemma shows that such a spectral expander exists. The construction of $H_t$ is based on Margulis-Gabber-Galil graphs \cite{margulis1973explicit, gabber1981explicit}.

\begin{lemma}[Space-efficient expanders, Lemma A.22 of \cite{chenWeightedPseudorandomGenerators2023}] \label{lem:space-efficient expanders}
\label{lem:HExpander}
    For every $d \in \mathbb{N}$ that is a power of two, for every $\lambda \in (0, 1)$, there is a bigraph $H = \big([d], [d], E_H\big)$ with a two-way labeling satisfying the following:
    \begin{itemize}
        \item $\lambda(H) \leq \lambda$.
        \item $H$ is $c$-regular where $c$ is a power of two and $c \leq \mathrm{poly}(1 / \lambda)$.
        \item $\mathrm{Rot}_H$ can be evaluated in space that is linear in its input length, i.e., space $O(\log(d / \lambda))$.
    \end{itemize}
\end{lemma}

We use the following notation for the algorithm in \cref{thm:derandomization algorithm} equipped with the  efficiently computable spectral expanders. 

\begin{definition}\label{def:DerandWalk}
    We define the algorithm $\operatorname{DerandWalk}$. Let $n,s,w\in\mathbb{N}$,
    and $0<\tau < \frac{1}{6\log n}$. For every $t\in[\lceil\log n\rceil]$, let $c\in \mathbb{N}$ and $H_t = \big([2^s \cdot c^{t-1}], [2^s \cdot c^{t-1}], E_{H_t}\big)$ be a $c$-regular bigraph with a two-way labeling satisfying $\lambda(H_t) \leq \lambda$ constructed by \cref{lem:HExpander}.
    For any $f \in \mathcal{R}^{tw}_{n,s,w}$, $0\leq l <r \leq n$, $u\in[w]$, $x\in\{0,1\}^s$ and  $e_1, e_2\ldots,e_{\lceil\log n\rceil}\in [c]$, we denote $\operatorname{DerandWalk}(\tau,f,l,r,u,(x,e_1,\ldots,e_{\lceil\log n\rceil}))$ as the algorithm in \cref{thm:derandomization algorithm} instantiated with the $l$-th layer to the $r$-th layer of $f$ and $\{H_t\}$, starting at node $u$ with seed $(x,e_1,\ldots,e_{\lceil\log n\rceil})$, i.e. $\operatorname{DerandWalk}(\tau,f,l,r,u,(x,e_1,\ldots,e_{\lceil\log n\rceil}))=\Rot_{\widetilde{G}_{l \to r}}(u, (x,e_1, \ldots, e_{\log n}))$, where $\widetilde{G}_{l\to r}$ is the bigraph generated by \cref{lem:sv approximation algorithm} with the $l$-th layer to the $r$-th layer of $f$.
\end{definition}

We use the following error reduction polynomial in our algorithm, which gives a good sv-approximation for a sequence of transitions.

\begin{theorem}[Recursion  \cite{chattopadhyayRecursiveErrorReduction2023}]\label{lem:derandErrorRedPoly} 
    Let $\{A_i\}_{i=1}^n\subset \mathbb{R}^{w\times w}$ be a sequence of doubly stochastic matrices. Let $\{B_{i,j}\}_{i,j=0}^n \subset \mathbb{R}^{w\times w}$ be a family of matrices such that $B_{i,j} \sv_{\tau/(10\log n)} A_{i+1}\ldots A_j$ Assuming that $B_{i-1,i}=A_i$ for all $i$.  Then for any $k\in \mathbb{N}$, there exists $K=O((2n)^k)$, $t=O(k\log n)$, a set of indices $m_{i,j}$ for each $i\in [K],j\in [t]$ with $0\leq m_{i,1}\leq \ldots\leq m_{i,t}= n$ and a set of signs $\sigma_i\in\{-1,0,1\}$ for each $i\in [K]$ such that:
    \begin{align*}
        \sum_{i\in [K]}\sigma_i\cdot B_{0,m_{i,1}}B_{m_{i,1},m_{i,2}}\ldots B_{m_{i,t-1},m_{i,t}} \sv_{ \tau^{k}} A_1A_2\ldots A_n .
    \end{align*}

\end{theorem}

\subsection{The derandomization for regular ROBPs with large alphabets.}



\begin{theorem}[Restatement of \cref{thm:regularderand}]\label{thm:regularderandmain}
    There exists an algorithm that takes as input a regular ROBP and a parameter $\varepsilon>0$, and outputs an approximation of the acceptance probability with additive error $\varepsilon$. Furthermore, if the regular ROBP has length $n$, width $w$ and alphabet bit length $s$, then the algorithm uses $O(s+\log n(\log\log n+\sqrt{\log(w/\varepsilon)})+\log(w/\varepsilon))$ bits of space.
\end{theorem}

Let $f\in \mathcal{R}^{tw}_{n,w,s}$ be a regular ROBP with two-way labeling and let $\M_{l\ldots r}$ denote its transition matrix from layer $l$ to layer $r$. For every $0\leq l<r\leq n$, we invoke \cref{lem:sv approximation algorithm}, which gives a regular bigraph $\widetilde{G}_{l\ldots r}$ with two-way labeling such that its transition matrix $\widetilde{\M}_{l\ldots r}$ is a $\tau$-sv-approximation of $\M_{l\ldots r}$. Then by \cref{lem:derandErrorRedPoly}, for any $ k\in \mathbb{N}$, we have $K = O((2n)^k)$ and a partition $0=m_{i,0}<m_{i,1}<\cdots<m_{i,l}=n_0$ for every $i\in [K]$ and $\sigma_{i}$ such that
  \begin{align*}
      \sum_{i\in [K]} \sigma_i \widetilde{\M}_{m_{i,0}\ldots m_{i,1}}\widetilde{\M}_{m_{i,1}\ldots m_{i,2}}\cdots \widetilde{\M}_{m_{i,l-1}\ldots m_{i,l}} \text{ approximates } \M_{0\ldots n_0} \text{ with entry-wise error } \tau^{k},
  \end{align*}

  Given the summation, we can view it as a sum of regular ROBPs with two-way labelings. For every $i\in [K]$, we define $f_{i} \in \mathcal{R}^{tw}_{n_1,w,s_1}$ as the concatenation of the bipartite graphs 
  $$\widetilde{G}_{m_{i,0}\ldots m_{i,1}}, \widetilde{G}_{m_{i,1}\ldots m_{i,2}}, \ldots, \widetilde{G}_{m_{i,l-1}\ldots m_{i,l}},$$
  which still has a two-way labeling.
  We set the starting vertex and accepting vertices of $f_{i}$ to be the same as those of $f$. Then we obtain 
  \begin{align*}
      \left|\E_{x} f(x) - \sum_{i\in [K]} \sigma_i \E_{y} f_{i}(y)\right| \leq \tau^{k} w.
    \end{align*}

    The same procedure can be applied to each regular ROBP $f_{i}$, which gives shorter ROBPs $f_{i_1,i_2}$ for every $i_2\in [K_2]$. We repeate the procedure until we attain regular ROBPs $f_{i_1,\ldots,i_\ell}$ with constant length, which can be fooled by true randomness. We formalize the above into the following detailed proof.

    \begin{proof}[Proof of \cref{thm:regularderandmain}]
        Let $f\in \mathcal{R}^{tw}_{n,w,s}$ be the regular ROBP to derandomize. Let $\ell \in \mathbb{N}$ be the number of iterations we need to perform and let $\tau_i,\varepsilon_i$ be the error parameters of the $i$-th iteration for every $i\in [\ell]$. The exact values of $\ell,\{\tau_i,\varepsilon_i\}_{i\in [\ell]}$ will be specified later. 
      For each iteration $i\in [\ell]$, define the error reduction exponent $k_i = \lceil \log_{1/\tau_i} (2/\varepsilon_i)\rceil$, and we setup $n_i,K_i$ iteratively. Let $n_0=n$. 
      Consider every $i\in [\ell]$. 
      Let $n_i = t, K_i = K$, where $t=O(k_{i}\cdot \log n_{i-1})$ and $K=O((2n_{i-1})^{k_i})$ are the parameters whose existence is guaranteed by \cref{lem:derandErrorRedPoly}, when initiating  \cref{lem:derandErrorRedPoly} with $n = n_{i-1}, k = k_i, \tau = \tau_i$.
       
        We iteratively define $f_{i_1,i_2,\ldots,i_p}\in \mathcal{R}^{tw}_{n_p,w,s_p}$ for every $i_1\in [K_1], i_2\in [K_2],\ldots,i_p\in [K_p]$ and $p\in \{0,1,\ldots,\ell\}$ as follows:
        \begin{enumerate}
            \item When $p=0$, $f_{i_1,i_2,\ldots,i_p}$ is defined to be $f$.
            \item For every $p\in [\ell]$, assume we have defined $f_{i_1,i_2,\ldots,i_{p-1}}$.
            We denote $\M^{(i_1,i_2,\ldots,i_{p-1})}_{l\ldots r}$ to be the stochastic matrix of $f_{i_1,i_2,\ldots,i_{p-1}}$ from layer $l$ to layer $r$. For every $0\leq l<r\leq n_{p-1}$, let $\widetilde{G}^{(i_1,i_2,\ldots,i_{p-1})}_{l\ldots r}$ be a regular bigraph with a two-way labeling obtained by approximating $\M^{(i_1,i_2,\ldots,i_{p-1})}_{l\ldots r}$ by the algorithm of \cref{lem:sv approximation algorithm} with error parameter $\tau_p$ \footnote{we always consider it as a length $2^{\left\lceil \log n_{p-1}\right\rceil}$ ROBP regardless of $r-l$}. 
            Let $\widetilde{\M}^{(i_1,i_2,\ldots,i_{p-1})}_{l\ldots r}$  be the stochastic matrix of  $\widetilde{G}^{(i_1,i_2,\ldots,i_{p-1})}_{l\ldots r}$. By \cref{lem:derandErrorRedPoly}, for every $i_p\in [K_p]$ we have a partition $0=m^{(p)}_{i_p,0}<m^{(p)}_{i_p,1}<\cdots<m^{(p)}_{i_p,n_p}=n_{p-1}$ and weights $\sigma^{(p)}_{i_p}\in\{-1,0,1\}$ such that
            \begin{equation}\label{def:f_i1i2}
                \begin{aligned}
                &\sum_{i_p\in [K_p]} \sigma^{(p)}_{i_p} \widetilde{\M}^{(i_1,i_2,\ldots,i_{p-1})}_{m^{(p)}_{i_p,0}\ldots m^{(p)}_{i_p,1}}\widetilde{\M}^{(i_1,i_2,\ldots,i_{p-1})}_{m^{(p)}_{i_p,1}\ldots m^{(p)}_{i_p,2}}\cdots \widetilde{\M}^{(i_1,i_2,\ldots,i_{p-1})}_{m^{(p)}_{i_p,n_p-1}\ldots m^{(p)}_{i_p,n_p}} \text{ approximates }\\ 
                &\M^{(i_1,i_2,\ldots,i_{p-1})}_{0\ldots n_{p-1}} \text{ with entry-wise error } \tau_p^{k_p}.
            \end{aligned}
            \end{equation}
            
            We define $f_{i_1,i_2,\ldots,i_p}$ to be the concatenation of $\widetilde{G}^{(i_1,i_2,\ldots,i_{p-1})}_{m^{(p)}_{i_p,0}\ldots m^{(p)}_{i_p,1}}, \widetilde{G}^{(i_1,i_2,\ldots,i_{p-1})}_{m^{(p)}_{i_p,1}\ldots m^{(p)}_{i_p,2}}, \ldots, \widetilde{G}^{(i_1,i_2,\ldots,i_{p-1})}_{m^{(p)}_{i_p,n_p-1}\ldots m^{(p)}_{i_p,n_p}}$, which also has a two-way labeling.

        \end{enumerate}

        Having defined the regular ROBPs, we use $|\Sigma_p|=2^{s_p}$ to denote the alphabet size of $f_{i_1,i_2,\ldots,i_p}$. Note that for every $i_1\in [K_1], i_2\in [K_2],\ldots,i_p\in [K_p]$ and $p\in \{0,1,\ldots,\ell\}$, it holds that $f_{i_1,i_2,\ldots,i_p}$ always has the same alphabet size, since all such regular ROBPs are constructed from \cref{thm:derandomization algorithm} with the same set of parameters.

        Our derandomization algorithm computes
        \begin{align*}
            &\sum_{i_1\in [K_1],i_2\in [K_2],\ldots,i_\ell\in [K_\ell]} \sigma^{(1)}_{i_1}\sigma^{(2)}_{i_2}\cdots \sigma^{(\ell)}_{i_\ell} \cdot \E_{x\in (\{0,1\}^{s_\ell})^{n_\ell}} f_{i_1,i_2,\ldots,i_\ell}(x).
        \end{align*}
        The algorithm operates as follows, assuming we can access the ROBP $f_{i_1,i_2,\ldots,i_\ell}$ with two-way labeling:
        \begin{enumerate}
            \item Enumerate all tuples $(i_1,i_2,\ldots,i_\ell)\in [K_1]\times [K_2]\times \cdots \times [K_\ell]$ and every string $(x_1,x_2,\ldots,x_{n_\ell})\in (\Sigma_\ell)^{n_\ell}$. For each tuple $(i_1,i_2,\ldots,i_\ell,x_1,x_2,\ldots,x_{n_\ell})$, we compute the following:
            \begin{enumerate}
                \item Let $u\in [w]$ be the starting vertex of $f_{i_1,i_2,\ldots,i_\ell}$.
                \item For every $t\in [n_\ell]$, we compute the rotation map $u\gets \Rot_{f_{i_1,i_2,\ldots,i_\ell}}(u,x_t)$. ($\Rot$ outputs a new pair $(u,x_t)$, but we discard $x_t$.)
                \item If $u$ is an accepting vertex of $f_{i_1,i_2,\ldots,i_\ell}$, then add $ \sigma^{(1)}_{i_1}\sigma^{(2)}_{i_2}\cdots \sigma^{(\ell)}_{i_\ell}/(2^{n_l\cdot s_l})$ to the result; otherwise, add $0$ to the result.
            \end{enumerate}
            \item Return the result.
        \end{enumerate}
        Since we do not have a direct access to $f_{i_1,i_2,\ldots,i_\ell}$, we need to compute it with a recursive program. For every $p\in [\ell]$ all tuples $(i_1,i_2,\ldots,i_p)\in [K_1]\times [K_2]\times \cdots \times [K_p]$ we compute $(u',x')\gets \Rot_{f_{i_1,i_2,\ldots,i_p}}(u,x)$ with the following recursive program, under the assumption that we have already defined the program to compute $ \Rot_{f_{i_1,i_2,\ldots,i_{p-1}}}(\cdot,\cdot)$:
        \begin{enumerate}
            \item If $p=0$, then we compute $\Rot_f(u,x)$ directly with the original ROBP $f$.
            \item Otherwise, assume $u=(a,b)$ is the $a$-th vertex in the $b$-th layer of the ROBP $f_{i_1,i_2,\ldots,i_p}$. Then we let $b'\gets b+1$ and compute $(a',x')\gets \Rot_{\widetilde{G}^{(i_1,i_2,\ldots,i_{p-1})}_{m^{(p)}_{i_p,b}\ldots m^{(p)}_{i_p,b+1}}}(a,x)$ with $\operatorname{DerandWalk}$ from \cref{def:DerandWalk}:

            \begin{enumerate}
                \item Let $l\gets m^{(p)}_{i_p,b}, r\gets m^{(p)}_{i_p,b+1}$.
                \item Let $(a',x')\gets \operatorname{DerandWalk}(\tau_p,f_{i_1,i_2,\ldots,i_{p-1}},l,r,a,x)$, during which the algorithm calls $\Rot_{f_{i_1,i_2,\ldots,i_{p-1}}}$.
                \item Let $u'\gets (a',b')$.
                \item Output $(u', x')$.
            \end{enumerate}
    
       
          

        \end{enumerate}

        We now examine both the error bound and the complexity of the algorithm. First we bound the error of the algorithm. By the construction of $f_{i_1,i_2,\ldots,i_p}$, we have
        \begin{align*}
            \left|\E_{x} f_{i_1,i_2,\ldots,i_{p-1}}(x) - \sum_{i_p\in [K_p]} \sigma^{(p)}_{i_p} \E_{y} f_{i_1,i_2,\ldots,i_p}(y)\right| &\leq \tau_p^{k_p} w \leq \varepsilon_p w
        \end{align*}
        for every $i_1\in [K_1], i_2\in [K_2],\ldots,i_{p-1}\in [K_{p-1}]$. Using triangular inequalities, we have the following bound:
        \begin{align*}
            &\left|\E_{x} f(x) - \sum_{i_1\in [K_1],i_2\in [K_2],\ldots,i_\ell\in [K_\ell]}  \sigma^{(1)}_{i_1}\sigma^{(2)}_{i_2}\cdots \sigma^{(\ell)}_{i_\ell} \cdot \E_{y} f_{i_1,i_2,\ldots,i_\ell}(y)\right|\\
            &\leq \sum_{p=0}^{\ell-1} \sum_{i_1\in [K_1],i_2\in [K_2],\ldots,i_p\in [K_p]} \left|\E_{x} f_{i_1,i_2,\ldots,i_p}(x) - \sum_{i_{p+1}\in [K_{p+1}]} \sigma^{(p+1)}_{i_{p+1}} \E_{y} f_{i_1,i_2,\ldots,i_p,i_{p+1}}(y)\right|\\
            &\leq \varepsilon_1 w + K_1\cdot \varepsilon_2 w + K_1K_2\cdot \varepsilon_3 w + \cdots + K_1K_2\cdots K_{\ell-1}\cdot \varepsilon_\ell w
        \end{align*}

        For the complexity of the algorithm, notice that the recursive algorithm computing $\Rot_{f_{i_1,i_2,\ldots,i_p}}$ requires $(s_p-s_{p-1})+\log n_p$ more bits than that of the algorithm computing $\Rot_{f_{i_1,i_2,\ldots,i_{p-1}}}$, since it only needs to store $b_l,b_r$ and the current $e_1,e_2,\ldots,e_{\log \left\lceil n_{p-1}\right\rceil}$ during the computation of $((a,b),y) \gets \Rot_{f_{i_1,i_2,\ldots,i_{p-1}}}((a,b),y)$.
        Given that the base case $p=0$ requires $O(\log n_0 + s_0 + \log w)$ bits of space, the algorithm to compute $\Rot_{f_{i_1,i_2,\ldots,i_\ell}}$ requires $O(\sum_{p=0}^{\ell} \log n_p + s_\ell + \log w)$ bits of space. The main algorithm to compute the expectation of $f_{i_1,i_2,\ldots,i_\ell}$ requires $O(s_\ell +\sum_{p=0}^{\ell} \log K_p + \log w)$ bits to store the current tuple $(i_1,i_2,\ldots,i_\ell)$ and the current string $(x_1,x_2,\ldots,x_{n_\ell})$ and maintain the result. Therefore, the total space complexity of the algorithm is
        \begin{align*}
            O\left(\sum_{p=0}^{\ell} \log n_p + s_\ell + \log w + \sum_{p=0}^{\ell} \log K_p\right).
        \end{align*}


        We conclude the proof by setting up the parameters $\ell,\{\tau_i,\varepsilon_i\}_{i\in [\ell]}$. We set $\varepsilon_1=\varepsilon/(2w),k_1=\lceil \log_{1/\tau_1} (2/\varepsilon_1)\rceil=\sqrt{\log(2w/\varepsilon)}$ and $\tau_1=\min\{\frac{\varepsilon_1^{2/(k_1+1)}}{16\log^2 n_0},\frac{1}{64\log^2 n_0}\}$. Then $n_1=O(\log n \sqrt{\log(w/\varepsilon)}),K_1=n^{O(\log n \sqrt{\log(w/\varepsilon)})}$. For every $i\in \{2,\ldots,\ell-1\}$, we set $\varepsilon_i=\varepsilon/(2wK_1)^2, k_i= O(\log^2 n_{i-1})$ and $\tau_i=\min\{\frac{\varepsilon_{i-1}^{2/(k_i+1)}}{16\log^2 n_{i-1}},\frac{1}{64\log^2 n_{i-1}}\}$. $k_i$ is chosen such that $n_i=O(k_i\log n_{i-1})=\log^3 n_{i-1}$ and $K_i=O((2n_{i-1})^{k_i})=n^{O(\log^3 n_{i-1})}$. We set $\ell$ to be the smallest integer such that $n_\ell\leq 32768$. We can easily verify that $\ell =o(\log\log n)$.
        Given the parameters, the construction of \cref{lem:sv approximation algorithm} gives $s_{i}=s_{i-1}+O(\log n_i(\log\log n_i + \log \tau_i))$ for every $i\in [\ell]$. Therefore, $s_1=s+O(\log n (\log\log n + \sqrt{\log(w/\varepsilon)}))$ and $s_i=s_{i-1}+O(\frac{\log(wK_1/\varepsilon)}{\log n_{i-1}})$ for every other $i\in [\ell]$. 

        Plugging the parameters into the error bound, we have
        \begin{align*}
            &\left|\E_{x} f(x) - \sum_{i_1\in [K_1],i_2\in [K_2],\ldots,i_\ell\in [K_\ell]} \sigma_{i_1}\sigma_{i_2}\cdots \sigma_{i_\ell} \cdot \E_{y} f_{i_1,i_2,\ldots,i_\ell}(y)\right|\\
             \leq &\varepsilon_1 w + K_1\cdot \varepsilon_2 w + K_1K_2\cdot \varepsilon_3 w + \cdots + K_1K_2\cdots K_{\ell-1}\cdot \varepsilon_\ell w\\
            \leq &\varepsilon/2 + \ell \cdot K_1K_2\cdots K_{\ell-1} \cdot \frac{\varepsilon}{2wK_1^2} \cdot w
            \leq \varepsilon,
        \end{align*}
        by noticing that $K_1 K_2 \cdots K_{\ell-1} \le K^2_1$.

        Plugging the parameters also gives the space complexity of the algorithm as
        \begin{align*}
             &O\left(\sum_{i=0}^{\ell} \log n_i + s_\ell + \log w + \sum_{i=0}^{\ell} \log K_i\right)\\
            =& O\left( s+\log n \left(\log\log n + \sqrt{\log(w/\varepsilon)} \right) + \log (w/\varepsilon) \right).
        \end{align*}
    \end{proof}

\bibliographystyle{alpha}
\bibliography{ref}
\appendix

\section{A proof for \cref{thm:NZonelayer}.}\label{sec:NZproof}

We reprove the lemma for completeness. 

\begin{definition}[total variation]
Let ${A}$ and ${B}$ be two random variables defined on a common probability space $X$. The total variation distance between ${A}$ and ${B}$ is defined as
\begin{align*}
    d_{TV}({A},{B})=\frac{1}{2}\sum_{x\in X}\left|\Pr[{A}=x]-\Pr[{B}=x]\right|.
\end{align*}
\end{definition}

\begin{definition}[Min-Entropy]
    Let ${X}$ be a random variable. The min-entropy of ${X}$ is defined as
    \begin{align*}
        H_{\infty}({X})=-\log_2\left(\max_{x\in X}\Pr[{X}=x]\right).
    \end{align*}
\end{definition}

\begin{definition}[Conditional Min-Entropy \cite{vadhanPseudorandomness2012} Problem 6.7]
    Let ${X}$ and ${A}$ be two random variables. The conditional min-entropy of ${X}$ given ${Y}$ is defined as
    \begin{align*}
        \tilde{H}_{\infty}({X}|{Y})=-\log_2\left(\underset{y\in Y}{\E}\sup_{x\in X}\Pr[{X}=x|{Y}=y]\right).
    \end{align*}
\end{definition}

We need the following lemmas.

\begin{lemma}[Chain rule of Conditional Min-Entropy \cite{vadhanPseudorandomness2012} Problem 6.7]\label{lem:chainrule}
    If $|\text{supp}(A)|\leq 2^s$, then $\widetilde{H}_{\infty}(X|A)\geq H_{\infty}(X)-s$.
\end{lemma}

\begin{lemma}[Problem 6.8 of \cite{vadhanPseudorandomness2012}]\label{lem:extractor for conditional min-entropy}
    Let $Ext:\{0,1\}^n\times \{0,1\}^d\to \{0,1\}^m$ be a $(k,\varepsilon)$-extractor. If $\tilde{H}_{\infty}({X}|{A})\geq k$, then 
    $$
    d_{TV}\left((Ext({X},U_d),A),(U_m,A)\right)\leq 3\varepsilon.
    $$
    (Here $U_m$ and $U_d$ are independent uniform random variables of length $m$ and $d$ respectively.)
\end{lemma}

\begin{lemma}[Data Processing Inequality]
    Let $X,Y$ be two random variables in the same probability space. Let $A$ be a random variable that is independent of both $X$ and $Y$. Let $f$ be a function. Then 
    \begin{align*}
        d_{TV}(f(X,A),f(Y,A))\leq d_{TV}(X,Y).
    \end{align*}
\end{lemma}

Now we can prove the lemma.
\begin{lemma}[\cref{thm:NZonelayer} restated]
    Let $s\geq \log w$. Assume there exists a $(2s, \frac{\varepsilon}{3n})$-extractor $\Ext:\{0,1\}^{3s}\times \{0,1\}^{d}\to \{0,1\}^{s}$.  Let $X$ and $Y_1,\ldots,Y_n$ be independent uniform random variables. Then the following construction
    \begin{align*}
        \NZ(X,Y)=\Ext(X,Y_1), \ldots, \Ext(X,Y_n)
    \end{align*}
    fools any $f\in \mathcal{B}_{(n,s,w)}$ with error at most $\varepsilon$.
\end{lemma}

\begin{proof}
    Let $f\in \mathcal{B}_{(n,s,w)}$. Let $X$ and $Y_1,\ldots,Y_n$ be independent uniform random variables. We use a hybrid argument.  Define $U_i$ to be independent uniform random variables of length $s$ for each $i\in [n]$. Define $Z_i=\Ext(X,Y_i)$ for each $i\in [n]$. Define $R_i$ to be the random variable over $V_i$, which represents the distribution of the state of $f$ on input $U_1,\ldots,U_i$. Define $\widetilde{R}_i$ to be the random variable over $V_i$ which represents the distribution of the state of $f$ on input $Z_1,\ldots,Z_i$. We will show that $d_{TV}(R_i,\widetilde{R}_i)\leq \frac{i\cdot \varepsilon}{n}$ for all $i\in [n]$.

    The base case $i=0$ is trivial, as both $R_0$ and $\widetilde{R}_0$ represent the initial state of $f$. Assume the statement holds for $i-1$. 
    For the case $i$, notice that $|\text{supp}(\widetilde{R}_{i-1})|\leq w\leq 2^s$. By the chain rule \cref{lem:chainrule}, 
    $$\tilde{H}_{\infty}(X|\widetilde{R}_{i-1})\geq 2s.$$ 
    Therefore \cref{lem:extractor for conditional min-entropy} implies 
    $$
    d_{TV}((Z_i,\widetilde{R}_{i-1}),(U_i,\widetilde{R}_{i-1}))\leq \frac{\varepsilon}{n}.
    $$
    By the data processing inequality, we have
    $$
    d_{TV}((U_i,\widetilde{R}_{i-1}),(U_i,R_{i-1}))\leq d_{TV}(\widetilde{R}_{i-1},R_{i-1})\leq \frac{(i-1)\varepsilon}{n}.
    $$
    Using the triangle inequality, we have
    $$
    d_{TV}((Z_i,\widetilde{R}_{i-1}),(U_i,R_{i-1}))\leq \frac{i\varepsilon}{n}.
    $$
    Denote the transition function of $f$ from $V_{i-1}$ to $V_i$ as $T_i$. Then $\widetilde{R}_i=T_i(Z_i,\widetilde{R}_{i-1})$ and $R_i=T_i(U_i,R_{i-1})$. Using the data processing inequality again, we have
    $$
    d_{TV}(R_i,\widetilde{R}_i)\leq d_{TV}((U_i,R_{i-1}),(Z_i,\widetilde{R}_{i-1}))\leq \frac{i\varepsilon}{n}.
    $$
    Therefore, $d_{TV}(R_n,\widetilde{R}_n)\leq \varepsilon$ and the lemma is proved.
\end{proof}

\section{A proof for \cref{thm:richardson iteration}.}\label{appendix:iterationproof}
We prove \cref{thm:richardson iteration} using the Richardson Iteration for completeness. The Richardson iteration is a method to obtain a finer approximation of $L^{-1}$ from a coarse approximation $B$ of $L^{-1}$ and the invertible matrix $L$ itself. 

\begin{lemma}
    Let $L\in \mathbb{R}^{m\times m}$ be an invertible matrix and $B\in \mathbb{R}^{m\times m}$ such that $\|B-L^{-1}\|\leq \varepsilon_0$ for a submultiplicative norm $\|\cdot\|$. For any non-negative integer $k$, define
    \begin{align*}
        R(B,L,k)=\sum_{i=0}^{k}(I-BL)^{i}B
    \end{align*}
    Then $\|L^{-1}-R(B,L,k)\|\leq \|L^{-1}\|\cdot\|L\|^{k+1}\cdot \varepsilon_0^{k+1}$.
\end{lemma}

\begin{proof}
    \begin{align*}
        \|L^{-1}-R(B,L,k)\|&=\left\lVert \left(I-\sum_{i=0}^{k}(I-BL)^{i}BL\right)L^{-1}\right\rVert\\
        &=\left\lVert(I-BL)^{k+1}L^{-1}\right\rVert\\ 
        &\leq \left\lVert I-BL\right\rVert^{k+1}\cdot \left\lVert L^{-1}\right\rVert\\
        &\leq \left\lVert L^{-1}\right\rVert\cdot \left\lVert L\right\rVert^{k+1}\cdot \varepsilon_0^{k+1}
    \end{align*}
\end{proof}

\begin{theorem}[\cref{thm:richardson iteration} restated]
    Let $\{A_i\}_{i=1}^n\subset \mathbb{R}^{w\times w}$ be a sequence of matrices. Let $\{B_{i,j}\}_{i,j=0}^n \subset \mathbb{R}^{w\times w}$ be a family of matrices such that for every $i+1< j$, $\|B_{i,j}-A_{i+1}\ldots A_j\|\leq \varepsilon/(2(n+1))$ for some submultiplicative norm $\|\cdot\|$, $\|A_i\|\leq 1$ for all $i$ and also $B_{i-1,i}=A_i$ for all $i$. Then for any odd $k\in \mathbb{N}$, there exists a $K=(8n)^{k+1}$, a set of indices $\{n_{i,j}\}_{i\in [K], j\in [k]}$ with $0\leq n_{i,1}\leq \ldots\leq n_{i,k}= n$, and  signs $\sigma_i\in\{-1,0,1\} , i\in [K]$ such that (We set $B_{i,i}=I$ for all $i$):

    \begin{align*}
        \left\lVert A-\sum_{i\in [K]}\sigma_i\cdot B_{0,n_{i,1}}B_{n_{i,1},n_{i,2}}\ldots B_{n_{i,k-1},n_{i,k}}\right\rVert\leq \varepsilon^{(k+1)/2}\cdot(n+1).
    \end{align*}

\end{theorem}
\begin{proof}

Define $L\in \mathbb{R}^{(n+1)w\times (n+1)w}$ and $B\in \mathbb{R}^{(n+1)w\times (n+1)w}$ as follows:

\begin{align*}
    L&=
    \begin{pmatrix}
        I & 0 & 0 & \ldots & 0 & 0\\
        -A_1 & I & 0 & \ldots & 0 & 0\\
        0 & -A_2 & I & \ldots & 0 & 0\\
        \vdots & \vdots & \vdots & \ddots & \vdots & \vdots\\
        0 & 0 & 0 & \ldots & -A_{n} & I\\
    \end{pmatrix},
    B=
    \begin{pmatrix}
        I & 0 & 0 & \ldots & 0 & 0\\
        B_{0,1} & I & 0 & \ldots & 0 & 0\\
        B_{0,2} & B_{1,2} & I & \ldots & 0 & 0\\
        \vdots & \vdots & \vdots & \ddots & \vdots & \vdots\\
        B_{0,n} & B_{1,n} & B_{2,n} & \ldots & B_{n-1,n} & I\\
    \end{pmatrix}
\end{align*}
This means that 
\begin{align*}
    L^{-1}=
    \begin{pmatrix}
        I & 0 & 0 & \ldots & 0 & 0\\
        A_1 & I & 0 & \ldots & 0 & 0\\
        A_1A_2 & A_2 & I & \ldots & 0 & 0\\
        \vdots & \vdots & \vdots & \ddots & \vdots & \vdots\\
        A_1A_2\ldots A_{n} & A_2\ldots A_{n} & A_3\ldots A_{n} & \ldots & A_{n} & I\\
    \end{pmatrix}
\end{align*}

By assumption, there exists a submultiplicative norm $\|\cdot\|$ on $\mathbb{R}^{w\times w}$. We induce a submultiplicative norm on $\|\cdot\|_A$ on $\mathbb{R}^{(n+1)w\times (n+1)w}$: Let $M=(M_{i,j})_{i,j=1}^{n+1}\in \mathbb{R}^{(n+1)w\times (n+1)w}$, where each $M_{i,j}\in \mathbb{R}^{w\times w}$. Define $\|M\|_A=\max_{j\in [n+1]}\sum_{i=1}^{n+1}\|M_{i,j}\|$. It is easy to verify that $\|\cdot\|_A$ is a submultiplicative norm, and $\|M\|_A=\|M\|_1$ when $w=1$.

Beacuse $\|B_{i,j}-A_{i+1}\ldots A_j\|\leq \varepsilon/(n+1)^2$ for all $i+1< j$, we have $\|B-L^{-1}\|_A\leq \varepsilon$. 
Also, Because $\|A_i\|\leq 1$ for all $i$, we have $\|L\|_A\leq 2$ and $\|L^{-1}\|_A\leq n+1$.

Define $M=R(B,L,(k-1)/2)$, then 
\begin{align*}
    \|L^{-1}-M\|_A\leq \|L^{-1}\|_A\cdot\|L\|_A^{\frac{k+1}{2}}\cdot \left(\frac{\varepsilon}{n+1}\right)^{\frac{k+1}{2}}\leq \varepsilon^{(k+1)/2}\cdot(n+1).
\end{align*}

Expand $M=(M_{i,j})_{i,j=1}^{n+1}$, where each $M_{i,j}\in \mathbb{R}^{w\times w}$. We foucus on $M_{1,n+1}$, which is a $w\times w$ matrix and $\|M_{1,n+1}-A_1A_2\ldots A_n\|\leq \|L^{-1}-M\|_A\leq \varepsilon^{(k+1)/2}\cdot(n+1)$.

To express $M_{1,n+1}$, we define
\begin{align*}
    \Delta_{i,j}=\begin{cases}
        B_{i,j-1}A_j-B_{i,j} &  i<j,\\
        0 & i\geq j.
    \end{cases}
\end{align*}

Then we have
\begin{align*}
    M_{1,n+1}=B_{0,n}+\sum_{j=1}^{(k-1)/2}\sum_{0<r_1<\cdots<r_j<n}\Delta_{0,r_1}\Delta_{r_1,r_2}\cdots \Delta_{r_{j-1},r_j}B_{r_j,n}.
\end{align*}

Defining $M^{(0)}_{i,j}=B_{i,j-1}A_j=B_{i,j-1}B_{j-1,j}$ and $M^{(1)}_{i,j}=B_{i,j}$, we have
\begin{align*}
    M_{1,n+1}=B_{0,n}+\sum_{j=1}^{(k-1)/2}\sum_{0<r_1<\cdots<r_j<n}\sum_{t_1,\ldots,t_j\in \{0,1\}}(-1)^{t_1+\cdots+t_j} M^{(t_1)}_{0,r_1}M^{(t_2)}_{r_1,r_2}\ldots M^{(t_j)}_{r_j,n}.
\end{align*}

Encoding $j,r_1,\ldots,r_j,t_1,\ldots,t_j$ into a single index $i\in[K]$, rewrite $M^{(t_1)}_{0,r_1}M^{(t_2)}_{r_1,r_2}\ldots M^{(t_j)}_{r_j,n}$ as $B_{0,n_{i,1}}B_{n_{i,1},n_{i,2}}\ldots B_{n_{i,k-1},n_{i,k}}$ for some $n_{i,1}\leq \ldots\leq n_{i,k}=n$, define $\sigma_i=(-1)^{t_1+\ldots+t_j}$, we have the desired result.

Finally, we bound $K$ by $\frac{k-1}{2}\cdot (n+1)^{(k-1)/2}\cdot 2^{(k-1)/2}\leq (8n)^{k+1}$.

\end{proof}

\section{A proof for \cref{lem:sampler for small error}.}\label{append:smallererr}

In this section we provide a proof for \cref{lem:sampler for small error}, which is similar to that of \cite{hozaBetterPseudodistributionsDerandomization2021}. We start by sampling a random matrix using a sampler, using a union bound.

\begin{lemma}\label{lem:matrix sampler}
    Let $\Samp: \{0, 1\}^p \times \{0, 1\}^r \rightarrow \{0, 1\}^q$ be a $(\alpha,\gamma)$-averaging sampler.  Then for every matrix valued function $f:\{0,1\}^q\to \mathbb{R}^{w\times w}$ such that $\|f\|_{1}\leq C$, we have:
    \begin{align*}
        \Pr_{x \in \{0,1\}^r }\left[\left\|2^{-p}\sum_{y\in \{0,1\}^p}  f(\Samp(x,y)) - \E [f] \right\|_{1} \ge C \alpha w \right]\leq w^2\gamma.
    \end{align*}
\end{lemma}

\begin{proof}
    \begin{align*}
        &\Pr_{x \in \{0,1\}^r }\left[\left\|2^{-p}\sum_{y\in \{0,1\}^p}  f(\Samp(x,y)) - \E [f] \right\|_{1} \ge C \alpha w \right]\\
        =&\Pr_{x \in \{0,1\}^r }\exists_{i\in [w]}\left[\sum_{j\in [w]} \left[2^{-p}\sum_{y\in \{0,1\}^p}  f(\Samp(x,y))_{i,j} - \E [f]_{i,j} \right] \ge C \alpha w \right]\\
        \leq&\Pr_{x \in \{0,1\}^r }\exists_{i\in [w],j\in [w]}\left[\left|2^{-p}\sum_{y\in \{0,1\}^p}  f(\Samp(x,y))_{i,j} - \E [f]_{i,j} \right| \ge C \alpha \right]\\
        \leq&\sum_{i\in [w],j\in [w]}\Pr_{x \in \{0,1\}^r }\left[\left|C^{-1}2^{-p}\sum_{y\in \{0,1\}^p}  f(\Samp(x,y))_{i,j} - C^{-1}\E [f]_{i,j} \right| \ge \alpha \right]\\
        \leq& w^2 \gamma.
    \end{align*}
\end{proof}

\begin{lemma}[\cref{lem:sampler for small error} restated]
    For all $n,s,w$, assume there exists a $W$-bounded $1/(2w\cdot (n+1)^2)$-WPRG $(G_0,w_0)$  for $\mathcal{B}_{(n,s,w)}$ with seed length $d$, and a $(\alpha,\gamma)$-averaging sampler $\Samp:\{0,1\}^r \times \{0,1\}^p\to \{0,1\}^d$ with $\alpha,\gamma$ as defined above. Then there exists a $\varepsilon$-WPRG for $\mathcal{B}_{(n,s,w)}$ with seed length $r+kp+ks$ and weight $W^{log(n/\varepsilon)/\log (nw)}\cdot \poly(nw/\varepsilon)$.
\end{lemma}

\begin{proof}
    Take any $f\in \mathcal{B}_{(n,s,w)}$. Let $A_i=\E f^{[i-1,i]}(G_0(U))$ for $i\in [n]$. Define $B_{i,j}=\frac{1}{2^p}\sum_{z\in \{0,1\}^d} w_0(z)\cdot f^{[i,j]}(G(z))$. For any $x\in \{0,1\}^r$ and $0\leq i<i+1<j\leq n$, we define $B^x_{i,j}=\frac{1}{2^p}\sum_{y\in \{0,1\}^p} w_0(\Samp(x,y))\cdot f^{[i,j]}((G(\Samp(x,y)))_{j-i})$. For $j=i+1$, we define $B^x_{i,i+1}=A_i$ as the true transition matrix.

    We call a seed $x$ to be `good' iff for all $i,j\in [n]$, $\|B_{i,j}-B^x_{i,j}\|_1\leq 1/(2w\cdot (n+1))$. Otherwise, we call $x$ to be `bad'. 
    
    Since $w_0$ is $W$-bounded, $\|w_0(z)\cdot f^{[i,j]}(G(z))\|_1$ is at most $W$. Note that $W\alpha w\leq 1/(2w\cdot (n+1)^2)$. By Lemma~\ref{lem:matrix sampler}, for any $i,j\in [n]$, 
    \begin{align*}
        \Pr_{x\in \{0,1\}^r}[\|B_{i,j}-B^x_{i,j}\|_1>1/(2w\cdot (n+1))]\leq w^2\gamma=\varepsilon/(2(2n)^k\cdot W^{k}).
    \end{align*}

    By a union bound, the probability $\Pr_{x\in \{0,1\}^r}[x \text{ is bad}]$ is at most $\varepsilon/(2(2n)^k\cdot W^{k})$. 

    For a good seed $x$, we have $\|A_i\ldots A_j-B^x_{i,j}\|_1\leq \|A_i\ldots A_j-B_{i,j}\|_1+\|B_{i,j}-B^x_{i,j}\|_1\leq 1/(w\cdot (n+1)^2)$. By Theorem~\ref{thm:richardson iteration}, 
    we have:
    \begin{align*}
        \left\|A_1\cdots A_n-\sum_{i=1}^{k}\sigma_i B^x_{0,i_1}B^x_{i_1,i_2}\ldots B^x_{i_{k-1},i_k}\right\|_1
        \leq ((n+1)w)^{-k}\cdot (n+1)\leq \varepsilon/2.
    \end{align*}

    For a bad seed, we can upper-bound $\|B^x_{i,j}\|_1$ by $K$. Therefore, we have:
    \begin{align*}
         \left\|A_1\cdots A_n-\sum_{i=1}^{K}\sigma_i B^x_{0,i_1}B^x_{i_1,i_2}\ldots B^x_{i_{k-1},i_k}\right\|_1
        \leq 1+K\cdot W^k
    \end{align*}
    
    Combining the two cases, we have the following inequality:
    \begin{align*}
        &\left|\E f- \frac{1}{K\cdot 2^{kp+r}}\sum_{x\in \{0,1\}^r}\sum_{y_1,\ldots,y_{k}\in \{0,1\}^p}w(x,y_1,\ldots,y_{k},i)f(G(x,y_1,\ldots,y_{k}))\right|\\
        \leq&\left\|A_1\cdots A_n-\frac{1}{2^r}\sum_{x\in \{0,1\}^r}\sum_{i=1}^{K}\sigma_i B^x_{0,i_1}B^x_{i_1,i_2}\ldots B^x_{i_{k-1},i_k}\right\|_1\\
        \leq&\frac{1}{2^r}\sum_{\substack{x\in \{0,1\}^r \\ x \text{ is bad}}}\left\|A_1\cdots A_n-\sum_{i=1}^{K}\sigma_i B^x_{0,i_1}B^x_{i_1,i_2}\ldots B^x_{i_{k-1},i_k}\right\|_1\\
        &+\frac{1}{2^r}\sum_{\substack{x\in \{0,1\}^r \\ x \text{ is good}}}\left\|A_1\cdots A_n-\sum_{i=1}^{K}\sigma_i B^x_{0,i_1}B^x_{i_1,i_2}\ldots B^x_{i_{k-1},i_k}\right\|_1\\
        \leq&\varepsilon/2+(1+K\cdot W^k)\cdot\gamma\\
        \leq&\varepsilon.
    \end{align*}

    The resulting generator uses \(r+kp+ks\) random bits in total: \(r\) bits for the sampler seed, \(p\) bits for each of the \(k\) sampler queries, and \(s\) bits for each of the \(k\) one-step transitions.

    The lemma follows.
\end{proof}

\section{Proof sketch for \cref{lem:sv-fool}.}\label{sec:sv-fool sketch}

Though not explicitly mentioned, the analysis of the INW generator in appendix B of \cite{chenWeightedPseudorandomGenerators2023} works for large alphabet ROBPs. Here we slightly modifies their proof to make it explicitly works.

We start with the definition of the INW generator. Some notions are defined in the preliminaries of \cref{sec:derand regular} 

\begin{definition}[INW generator for large alphabets]
Let $n,c,d$ be powers of two. For each $t\in [\log n]$, let $H_t$ be a $c$-regular bigraph $H_t=([d\cdot c^{t-1}],[d\cdot c^t],E_{H_t})$, with a one-way label. Relative to the family $(H_t)_{t\in [\log n]}$, we recursively define the INW generator as follows:

Define $\INW_0:\{0,1\}^{\log d}\to \{0,1\}^{\log d}$ as the trivial PRG. For each $t\in [\log n]$, having defined $\INW_{t-1}:\{0,1\}^{d+(t-1)\log c}\to \{0,1\}^{\log d\cdot 2^{t-1}}$, we define $\INW_t:\{0,1\}^{d+t\log c}\to \{0,1\}^{\log d\cdot 2^t}$ as $\INW_t(x,y)=\INW_{t-1}(x),\INW_{t-1}(H_t[x,y])$. Here the comma denotes concatenation.
    
\end{definition}

The INW generator relates to the derandomized product of permutation ROBPs, since they have consistent one-way labelings.

\begin{definition}[consistent consistent one-way labelings]
    Let $G=(U,V,E)$ be a $d$-regular bigraph. A consistent one-way labeling of $G$ is a one-way labeling of $G$ such that for every $v\in V$, the labels of the incoming edges of $v$ are distinct, i.e., the $G[v,i]=G[u,i]$ implies $u=v$. In this case, we can extend the labeling to a two-way labelings:
    \begin{align*}
        \Rot_G(u,i)=(G[u,i],i),
    \end{align*}
    i.e., the incoming label and the outgoing label of an edge $(u,v)$ are the same.
\end{definition}

\begin{lemma}[Derandomized Product with consistent consistent one-way labelings,\cite{rozenmanDerandomizedSquaringGraphs2005}]
    Let $G_1=(U,V,E_1)$ and $G_2=(V,W,E_2)$ be two $d$-regular bigraphs with consistent one-way labelings. Let $H$ be a $x$-regular bigraph with one-way labeling. Then $G_1\ds_H G_2$ has a consistent one-way labeling.
\end{lemma}

Let $f$ be a permutation ROBP in $\mathcal{P}_{(n,s)}$. Then $f$ is also a regular ROBP with a consistent labeling. It can be sv-approximated with the method of \cref{lem:sv approximation algorithm}. Furthermore, the method corresponds to fooling $f$ with the INW generator.

\begin{lemma}[equivariance between INW generator and \cref{lem:sv approximation algorithm}]
    Let $f$ be a permutation ROBP in $\reduct{P}_{(n,s)}$. Let $t\in [\log n]$, $(j,j+2^t)\in E_{SC_n}$ and $\widetilde{G}_{j,j+2^t}$ be the $(d\cdot c^t)$-regular bigraph with two-way labeling as defined in \cref{lem:sv approximation algorithm}. Then for any $u\in V^{j}$ and $x\in \{0,1\}^{d+t\log c}$, we have
    $$
    \exists x',\widetilde{G}_{j,j+2^t}[u,x]=[v,x'] \Leftrightarrow \left[f^{[j,j+2^t]}(\INW(x))\right]_{u,x}=v,
    $$
\end{lemma}

\begin{proof}
    The proof is by induction on $t$. For the base case $t=0$, both sides mean $u$ has an outgoing edge labeled $x$ going to $v$.
     Assume the statement holds for $t-1$. Recall that 
    \[
\widetilde{G}_{ j \to j+2^t} = \widetilde{G}_{ j \to j+2^{t-1}} \ds_{H_t} \widetilde{G}_{j+2^{t-1}\to j+2^t}, \forall t \ge 1,
\] 
    Then for $t$, the left side can be computed as follows (where $x=(z,y)$):
    \begin{align*}
        &(w,z)=\Rot_{\widetilde{G}_{j,j+2^{t-1}}}(u,z),\\
        &(z',y')=\Rot_{H_t}(z,y),\\
        &(v,w)=\Rot_{\widetilde{G}_{j+2^{t-1},j+2^t}}(z',y').
    \end{align*}
    Where the right side can be computed as follows:
    \begin{align*}
        &f^{[j,j+2^{t-1}]}(\INW_{t-1}(z))=w,\\
        &H_t[z,y]=z',\\
        &f^{[j+2^{t-1},j+2^t]}(\INW_{t-1}(z'))=v.
    \end{align*}
    The induction hypothesis implies that the two sides are equivalent.
\end{proof}

Finally, we can prove the lemma.

\begin{lemma}[\cref{lem:sv-fool} restated]
    For all $s,n,\varepsilon$ there exists a PRG (actually the INW generator) that fools $\mathcal{P}_{(n,s)}$ with $\varepsilon$-sv-error. The seed length is
    \begin{align*}
        s+O(\log n(\log\log n+\log(1/\varepsilon))).
    \end{align*} 
\end{lemma}
\begin{proof}
    Let $\lambda=\min\{\varepsilon/(11\log n),1/6(\log^2 n)\}$. Let $d=2^s$ and for each $t\in [\log n]$, let $H_t$ be a $c$-regular  expanders with $\lambda(H_t)\leq \lambda$ from \cref{lem:space-efficient expanders}. Let $\{\INW_t\}_{t\in [\log n]}$ be the INW generator from the definition. Let $\INW=\INW_{\log n}$. Then $\INW$ is a PRG with seed lengths
    $$ \log d+t\cdot \log c=s+O(\log n(\log\log n+\log(1/\varepsilon))).$$

    For any $f\in \mathcal{P}_{(n,s)}$, the derandomized product of $f$ coincides with the the output of $f\circ INW$, i.e. $\widetilde{\mathbf{W}}_{n\gets 0}=E_X f^{[0,n]}(\INW(X))$. Since $\widetilde{\mathbf{W}}_{n\gets 0}\sv_{11\lambda\log n} \widetilde{\mathbf{W}}_{n\gets n-1}\cdots \widetilde{\mathbf{W}}_{1\gets 0}$ by  \cref{lem:sv approximation algorithm}, $INW$ fools $f$ with $\varepsilon$-sv-error.
\end{proof}

\section{Reduction for regular ROBPs over large alphabets.}\label{appendix:regular analysis}

In this section, we show that the WPRG construction in Section 3 of \cite{chattopadhyayRecursiveErrorReduction2023} naturally extends to the case of regular ROBPs with large alphabets. This will immediately give us \cref{thm:CHLTW for regular}. We start by introducing an equivalent definition of the weight of ROBPs, 

\begin{definition}[An Equivalent Definition of Weight of ROBPs]
Let $f\in \mathcal{B}_{n,s,w}$ be an ROBP. Let $V=\bigcup_{i=0}^{n}V_i$ be its vertices and $E=\bigcup_{i=1}^{n}E_{i-1,i}$ be its edges, where $E_{i-1,i}\subseteq V_{i-1}\times V_i$ For every $0\leq l\leq r\leq n$, Let $\M_{l\ldots r}$ be the stochastic matrix of $f$ from layer $l$ to layer $r$. Let $e_i$ denote the unitary vector at the $i$-th coordinate. For every $y\in \mathbb{R}^w$, $i\in [n]$, define the layer $i$ weight of $f$ on $y$ as:
\[
\mathsf{W}(f,i,y)=\sum_{(u,v)\in{E_{i-1,i}}}\left|e_u^T \M_{i-1\ldots i}y-e_v^T y\right|
\]
 For every $0\leq l< r\leq n$, the total weight between layer  $l$ and layer $r$ of $f$ on $y$ is defined as:
 \[
 \mathsf{W}(f,l,r,y)=\sum_{i=l+1}^r \mathsf{W}(f,i,\M_{i\ldots r} y).
 \]

\end{definition}

We immediately have the following facts:
\begin{lemma}
    $0\leq l< m<r\leq n$,  $\mathsf{W}(f,l,r,y)=\mathsf{W}(f,l,m,\M_{m\ldots r} y)+\mathsf{W}(f,m,r,y)$
\end{lemma}

\begin{lemma}
    $\mathsf{W}(f)=\mathsf{W}(f,0,n,y_{acc})$, where $y_{acc}$ is the accept vertices vector of $f$.
\end{lemma}

\begin{proof}
\begin{align*}
    \mathsf{W}(f,0,n,y_{acc})&=\sum_{i=1}^n \mathsf{W}(f,i,\M_{i\ldots n} y_{acc})=\sum_{i\in [n]}\sum_{(u,v)\in{E_{i-1,i}}}\left|e_u^T \M_{i-1\ldots n}y_{acc}-e_v^T \M_{i\ldots n} y_{acc}\right|\\
    &=\sum_{(u,v)\in E}\left|\E[f^{u\to}-f^{v\to}]\right|=\mathsf{W}(f)
\end{align*}
    
\end{proof}

We need a main result of \cite{bravermanPseudorandomGeneratorsRegular2014}.

\begin{lemma}[\cite{bravermanPseudorandomGeneratorsRegular2014}]\label{lem:prg for bounded weight}
    For every $n,s,w\in\mathbb{N}$ and $\varepsilon>0$, there exists a PRG $G:\{0,1\}^d\to(\{0,1\}^s)^w$ such that for every $f\in\mathcal{B}_{n,s,w}$ and every $y\in \mathbb{R}^w$:
    \[
        \left\|\left(\E_{x\in\{0,1\}^d}f^{[l,r]}(G(x))\right)y-\M_{l\ldots r}y\right\|_\infty\leq \varepsilon \mathsf{W}(f,l,r,y).
    \]
    Furthermore, $d=s+O(\log n(\log\log n+\log(w/\varepsilon))$
\end{lemma}

We use the error reduction polynomials in \cite{chattopadhyayRecursiveErrorReduction2023}. Without loss of generality, assume that $n$ is a power of $2$ and define the set $\BS_n$ as
\begin{align*}
    \BS_n = \{(l,r) \in [n]^2 \mid \exists i,k\in \mathbf{N}\cup \{0\}, l=i\cdot 2^k, r=l+2^k, 0\leq l< r \leq n\}.
\end{align*}
Given a set of substochastic matrices $\{\M_1,\ldots,\M_n\}$, suppose for every $(l,r)\in \BS_n$, we have a matrix $\M^{(0)}_{l\ldots r}$ that coarsely approximates the product $\M_{l+1}\cdots \M_r$. They define the following matrices $\M^{(k)}_{l\ldots r}$ as
\begin{align}\label{eq:binary-splitting}
    \M^{(k)}_{l\ldots r} =
    \left\{
    \begin{array}{ll}
        \M_{r} & \text{if } r=l+1, \\
        \sum_{i+j=k} \M^{(i)}_{l\ldots m} \M^{(j)}_{m\ldots r} - \sum_{i+j=k-1} \M^{(i)}_{l\ldots m} \M^{(j)}_{m\ldots r} & \text{otherwise, where } m = (l+r)/2.
    \end{array}
    \right.
\end{align}

The following result of \cite{chattopadhyayRecursiveErrorReduction2023} shows that $\M^{(k)}_{l\ldots r}$ is a good approximation as long as $k$ is large enough.

\begin{theorem}[Lemma 3.6 of \cite{chattopadhyayRecursiveErrorReduction2023}]\label{thm:regular error reduct}
    For any $0<\varepsilon <\frac{1}{30\log n}$ Assume that for every  $(l,r)\in \BS_n$
    \[
    \forall y\in \mathbb{ R}^w,\left\|\left(\M^{(0)}_{l\ldots r} -\M_{l\ldots r}\right)y\right\|_\infty\leq \varepsilon\frac{\mathsf{W}(f,l,r,y)}{\mathsf{W}(f)}
    \]
    Then for every  $(l,r)\in \BS_n$ and $k\in \mathbb{N}$
    \[
    \forall y\in \mathbb{ R}^w,\left\|\left(\M^{(k)}_{l\ldots r} -\M_{l\ldots r}\right)y\right\|_\infty\leq(30 \varepsilon\cdot \log n)^k \frac{\mathsf{W}(f,l,r,y)}{\mathsf{W}(f)}
    \]
\end{theorem}

Now we prove \cref{thm:CHLTW for regular}

\begin{proof}[Proof of \cref{thm:CHLTW for regular}]

Set $k=O(\sqrt{\log (1/\varepsilon)})$. Set $\tau=\frac{\varepsilon^{1/k}}{30\log n\cdot \mathsf{W}(f))}$. Let $\PRG:\{0,1\}^d\to (\{0,1\}^s)^n$ be the $\tau$-PRG given by \cref{lem:prg for bounded weight} and let $(\PRG(\cdot))_{m}$ denote its first $m$ symbols over $\{0,1\}^s$. we define $\M^{(0)}_{l\ldots r}:=\E_{x\in\{0,1\}^d}f^{[l,r]}((\PRG(x))_{r-l})$ and expand $\M^{(k)}_{0\ldots n}$ as a weighted sum of products of $\M^{(0)}_{l\ldots r}$, we get the following sum:
\[
 \M^{(k)}_{0\ldots n}=\frac{1}{K}\sum_{i\in K}\gamma_i\cdot \M^{(0)}_{m_{i,0}\ldots m_{i,1}}\M^{(0)}_{m_{i,1}\ldots m_{i,2}}\ldots \M^{(0)}_{m_{i,t-1}\ldots m_{i,t}},
\]
where $\gamma_i\in\{-K,K\}$ are weights and $0=m_{i,0}\leq m_{i,1}\leq \cdots\leq m_{i,t}=n$ are partitions of $[n]$.
We set the $(\log K,K,\varepsilon)$-reduction $(\reduct{R},\sigma)$ from $\mathcal{R}_{n,s,w}$ to $\mathcal{B}_{t,d,w}$ as
\begin{align*}
\reduct{R}(x_1,x_2,\cdots,x_t,i)&=((\PRG(x_1))_{m_{i,1}-m_{i,0}},(\PRG(x_2))_{m_{i,2}-m_{i,1}},\cdots,(\PRG(x_t))_{m_{i,t}-m_{i,t-1}}),\\
\sigma(i)&=\gamma_i.
\end{align*}

By \cref{lem:prg for bounded weight}, $\forall y\in \mathbb{ R}^w,\left\|\left(\M^{(0)}_{l\ldots r} -\M_{l\ldots r}\right)y\right\|_\infty\leq \frac{\varepsilon^{1/k}}{30\log n}\cdot \frac{\mathsf{W}(f,l,r,y)}{\mathsf{W}(f)}$. Then by \cref{thm:regular error reduct}, $\left\|\left(\M^{(k)}_{0\ldots n} -\M_{0\ldots n}\right)y_{acc}\right\|_\infty\leq  \varepsilon\cdot \frac{\mathsf{W}(f,0,n,y_{acc})}{\mathsf{W}(f)}= \varepsilon$, which indicates that our reduction achieves the target precision.

Given the construction of the error reduction polynomial, one can find $K=n^{O(k)}=2^{O(\log n\sqrt{\log(1/\varepsilon)})}$ and $t=O(k\log n)=O(\log n\sqrt{\log(1/\varepsilon)})$. Also, \cref{lem:prg for bounded weight} indicates $d=s+O(\log n(\log\log n+\log w+\log(\mathsf{W}(f))+\sqrt{\log(1/\varepsilon)})$. That completes the proof.

\end{proof}

\section{A revised WPRG construction for standard ROBPs} 
\label{sec:reducing revised}

We show the following revised WPRG for standard ROBPs, whose seed length attains optimal dependence on the alphabet bit-length, matching that of Chen,
Cohen, Doron, Khaskelberg, and
Ta-Shma~\cite{chenImprovedErrorReduction2026}.

\begin{theorem}
\label{thm:final WPRG revised}
    For all integer $n,s,w$, there exists an explicit construction of a $\varepsilon$-WPRG for $\mathcal{B}_{(n,s,w)}$ with seed length
    $$s+O\left(\frac{\log n\log (nw)}{\max\{1,\log\log w-\log\log n\}}+\log w \left(\log\log\min\{n,\log w\}-\log\log\max\left\{\frac{\log w}{\log \frac{n}{\varepsilon} }, 2\right\}\right)+\log(1/\varepsilon)\right),$$
    and weight $\poly (nw/\varepsilon)$.
\end{theorem}
The revised construction does not use the sampler-based techniques of \cite{hozaBetterPseudodistributionsDerandomization2021}.
It follows a three-stage weighted pseudorandom reductions. 
In the first stage we apply a single reduction to reduce the
        length from \(n\) to $n_0=\min\{n,\log(n/\varepsilon)/\log n\}$.
        Then we consider two cases. Case 1 is that $ w> n/\varepsilon$, and for this case we mainly follow our previous construction but replacing the alphabet reduction by using the PRG of \cite{chenSimplifyingArmonisPRG2025}.  
        Case 2 is that $w\le n/\varepsilon$ and this is the part we mainly adjust.
        We can further reduce the length to
        \[
           n'=  O\!\left(\frac{\log^2(n/\varepsilon)}{\log^2 w}\right),
        \]
        by using a stage of iterative reductions.
        If this is smaller than a constant then we will finish by using the PRG of \cite{chenSimplifyingArmonisPRG2025}.
        So for the remaining (the 3rd stage) we only consider this length to be a super constant.
        Notice that at the beginning of the 3rd stage, we have a collection of ROBPs of length $n'$,
        each of which needs to be fooled with error \(\varepsilon/\poly(nw)\). 
        Notice that $\log(n/\varepsilon)\geq \Theta(\sqrt{n'}\log (w))\geq \Theta(\log^2n'\log(n'w))$, by the definition of $n'$.
        We then set
        \[
            \varepsilon_0
            =
            2^{-\frac{\log(n/\varepsilon)}{\log^2 n'}}.
        \]
        Therefore, an application of the
        PRG of  \cite{chenSimplifyingArmonisPRG2025} with error \(\varepsilon_0\) requires
        only
        \[
            O(\log(n'w/\varepsilon_0)\log n')
            =
            O\!\left(\frac{\log(n/\varepsilon)}{\log n'}\right)
        \]
        bits of extra randomness. By inserting this PRG into the error reduction framework, we can shorten the length to \(O(\log^2 n')\) at the
        cost of increasing the alphabet bit length by
        \[
           + O\!\left(\frac{\log(n/\varepsilon)}{\log n'}\right).
        \]
       This procedure can be repeated until the length is
        reduced to \(O(1)\) and then we use the PRG of  \cite{chenSimplifyingArmonisPRG2025}.
        The total amount of extra randomness used accumulates
        to at most
        \[
            O(\log(n/\varepsilon)).
        \]

        \subsection{Three weighted pseudorandom reductions}
        We use the Chen-Ta-Shma generator \cite{chenSimplifyingArmonisPRG2025} which achieves the optimal seed length dependency on the alphabet size.
        \begin{theorem}[Chen-Ta-Shma generator \cite{chenSimplifyingArmonisPRG2025}]
        \label{thm:ct-improved-inw}
        For every length \(n\), width \(w\), alphabet \(\Sigma\), and error
        \(0<\varepsilon<1/2\), there is an explicit PRG that
        \(\varepsilon\)-fools every length-\(n\), width-\(w\),
        alphabet-\(\Sigma\) ordered branching program, with seed length
        \[
        r
        =
        \log|\Sigma|
        +
        O\!\left(
            \frac{
                \log(wn/\varepsilon)\cdot \log n
            }{
                \log\!\left(
                    2+\frac{\log w}{\log(n/\varepsilon)}
                \right)
            }
        \right).
        \]
        \end{theorem}
        Instantiating our length reduction framework~\cref{lem:framework} with this generator, we have the following reduction lemmas, which now optimally depend on the alphabet. 
        The first two lemmas are variants of \cref{lem:length reduction} and \cref{lem:armoni reduction}, adapted to the setting of arbitrarily small error $\varepsilon$.
        
        \begin{lemma}[Length reduction 1]\label{lem:ben armoni reduction}
            For all integers $n,s,w$ and $\varepsilon>0$, there exists an explicit $(\log K,K, \varepsilon)$-weighted pseudorandom reduction from $\mathcal{B}_{(n,s,w)}$ to
            $\mathcal{B}_{\left(k, s+O\left(\frac{\log n\log (nw)}{\log\log w-\log\log n}\right),w\right)}$, where  $K=(8n)^{k+1}$, $k = \frac{\log(n/\varepsilon)}{\log n} $.

        \end{lemma}

        \begin{proof}
        We use \cref{thm:ct-improved-inw} for $\mathcal{B}_{(n,s,w)}$  with error  $\varepsilon_0=\frac{2^{-4\log n}}{(n+1)^2}$. 
        By \cref{thm:ct-improved-inw}, the seed length is $s+O\left(\frac{\log n\log (nw)}{\log\log w-\log\log n}\right)$. 
        Then we can set  $k=\frac{\log (n/\varepsilon)}{\log n}\geq \frac{2\log((n+1)/\varepsilon)}{\log(1/((n+1)^2\varepsilon_0))}$ in Lemma~\ref{lem:framework} and invoke it. This gives a reduction of error $(\varepsilon_0(n+1)^2)^\frac{k+1}{2}\cdot(n+1)\leq \varepsilon$.
        \end{proof}

        \begin{lemma}[Length reduction 2]\label{lem:ben length reduction}
            For any constants $c\in \mathbb{N}$, for all integers $n,w$ and $\varepsilon>0$, if $n\le \log^c w  $, then there exists an explicit $(\log K,K,\varepsilon)$-weighted pseudorandom reduction from $\mathcal{B}_{(n,s,w)}$ to $\mathcal{B}_{(k, s+C'\log w,w)}$, where $K\leq (8n)^{k+1}, k =  n^{1/c}\cdot \frac{\log(n/\varepsilon)}{\log w}$, $C'$ is a constant depending on $c$.
        \end{lemma}

        \begin{proof}
        Fix $c$ to be a constant. We use \cref{thm:ct-improved-inw}
        for $\mathcal{B}_{(n,s,w)}$ with error $\varepsilon_0=\frac{2^{-\frac{4\log w}{n^{1/c}}}}{(n+1)^2}$.
        By \cref{thm:ct-improved-inw}, the seed length is $s+O\left(\log(nw/\varepsilon_0)\cdot\frac{\log n}{\log\log w-\log\log(n/\varepsilon_0)}\right)= s+O\left(\log w\cdot\frac{\log n}{\log n^{1/c}}\right)=s+C' \log w $, where $C'$ depends on $c$.
        Then we can set $k=n^{1/c}\cdot \frac{\log(n/\varepsilon)}{\log w} \ge \frac{\log(1/\varepsilon)+\log (n+1)}{1/2\cdot\log(1/((n+1)^2\varepsilon_0))}$ in Lemma~\ref{lem:framework} and invoke it.  This gives a reduction of error $(\varepsilon_0(n+1)^2)^\frac{k+1}{2}\cdot(n+1)\leq \varepsilon$.
        \end{proof}



        Our third reduction, which resembles \cref{cor:permutation-reduction-2}, but instead instantiated with the Chen-Ta-Shma generator.

        \begin{lemma}[length reduction 3]\label{lem:ben small error reduction}
            For all integers $n,s,w$ and $\varepsilon>0$ satisfying $\varepsilon \leq 2^{-2\log^2n\log (nw)}$, there exists an explicit $(\log K,K, \varepsilon)$-weighted pseudorandom reduction from $\mathcal{B}_{(n,s,w)}$ to $\mathcal{B}_{\left(k,s+O\left(\frac{\log (1/\varepsilon)}{\log n}\right),w\right)}$, where  $K=(8n)^{k+1}$, $k = 10\log^2 n$.
            
        \end{lemma}

        \begin{proof}
        We use \cref{thm:ct-improved-inw} for $\mathcal{B}_{(n,s,w)}$  with error  $\varepsilon_0=\frac{2^{-\frac{\log((n+1)/\varepsilon)}{\log^2 n}}}{(n+1)^2}$. The condition $\varepsilon \leq 2^{-2\log^2n\log (nw)}$ guarantees that $\log(nw/\varepsilon_0)\leq 2\log(1/\varepsilon_0)$.

        By \cref{thm:ct-improved-inw}, the seed length is $s+O(\left(\log(nw/\varepsilon_0)\cdot \log n\right))=s+O\left(\frac{\log (1/\varepsilon)}{\log n}\right)$.
        Then we can set $ k=10\log^2 n\geq \frac{\log((n+1)/\varepsilon)}{1/2\cdot \log(1/(n+1)^2\varepsilon_0)}$ in Lemma~\ref{lem:framework} and invoke it. This gives a reduction of error $(\varepsilon_0(n+1)^2)^\frac{k+1}{2}\cdot(n+1)\leq \varepsilon$.
        \end{proof}

        \subsection{Proof of \cref{thm:final WPRG revised}}

        \begin{proof}

             
             We do three stages of reductions: In the first stage, we reduce the length from $n$ to $n_0=\min\{n,\log(n/\varepsilon)/\log n\}$. Then we split the algorithm into two cases: if $w> n/\varepsilon$, we reduce the length to $\frac{\log w}{\log n/\varepsilon}$, then fool it with PRG; if $w\leq  n/\varepsilon$, we reduce the length to $\left(\frac{10\log(n/\varepsilon)}{\log w}\right)^2$, then we further reduce the length to $O(1)$ using a different parameter scheme, and fool it with PRG.

             For the first stage, if $n\leq \frac{\log(n/\varepsilon)}{\log n}$, then the goal is already achieved, and we set $n_0=n$. Otherwise, we use \cref{lem:ben armoni reduction} to reduce the length from $n$ to $\frac{\log(n/\varepsilon)}{\log n}$ via a $(\log K,K, \varepsilon/2)$-weighted pseudorandom reduction, where  $K\leq n^2/\varepsilon^2$. 
            Recall that the number of reduced ROBPs is $K$.
            The reduced ROBPs have length $n_0=\frac{\log(n/\varepsilon)}{\log n}$, width $w$, and alphabet bit-length
            \[
                s_0=s+O\left(\frac{\log n\log (nw)}{\log\log w-\log\log n}\right).
            \]

            We set $\varepsilon'=\varepsilon/2K^2$ which will serve as the error parameter for the remaining stages.

            \begin{enumerate}
                \item 
            \textbf{Case $w>n/\varepsilon$:} We iteratively use \cref{lem:ben length reduction} for $l$ times ($l$ will be defined later): In iteration $0\leq i\leq l$, we reduce the length from $n_i$ to $n_{i+1}=n_i^{1/4}\frac{\log(n_iw/\varepsilon')}{\log w}\leq 4n_i^{1/4}$, decomposing one ROBP into $K_i$ shorter ones and increasing the alphabet bit-length from $s_i$ to $s_{i+1}$. Parameter calculation shows
            \[
                s_{i+1}=s_i+C\log w,
                \qquad
                K_i\leq (8n_i)^{\frac{\log(n_i/\varepsilon')}{\log w}n_i^{1/4}+1}\leq
                (n/\varepsilon)^{\frac{2n_i^{1/4}\log n_i}{\log w}}\leq 
                (n/\varepsilon)^{1/\log^{1/4} w},
            \]
            where the first inequality is directly due to \cref{lem:ben length reduction} and the second inequality follows from $n_i\leq n_0=\log(1/\varepsilon)/\log n\leq \log^2w $.

            The iteration goes until $n_l\leq \frac{\log w}{\log n/\varepsilon'}$, which continues for at most $l=\log\log n_l-\log\log n'\leq \log\log\min\{n,\log w\}-\log\log\frac{\log w}{\log n/\varepsilon}$ rounds. When the iteration stops, 
            the reduced ROBPs of length $n_l$ and alphabet bit-length $s_l$ could be $\varepsilon'$-fooled by Chen-Ta-Shma PRG with seed-length  $s_l+O\left(\log(nw/\varepsilon')\cdot \frac{\log n_l}{\log\log w-\log\log(n_l/\varepsilon')}\right)=s_l+O(\log(nw/\varepsilon)$.

            To account for the total error introduced, we first bound the number of reductions we use in the second stage. We have $l$ rounds of reductions, and in each round we have at most $K\cdot \prod_{i=0}^{l} K_i$ parallel reductions. Therefore, the total amount of reductions we use is at most $l\cdot K\cdot \prod_{i=0}^{l} K_i \leq K\cdot (n/\varepsilon)^{\frac{2\log\log w}{\log^{1/4} w}}\leq K^2$. The total error introduced in the second stages is at most $K^2\cdot \varepsilon'=\varepsilon/2$. Together with the error $\varepsilon/2$ introduced in the first stage, we have that the total error introduced is at most $\varepsilon$.

            Then we bound the randomness we use. The randomness used in selecting the index of reductions is at most $\log(K\cdot \prod_{i=0}^{l} K_i)\leq O(\log (nw/\varepsilon))$. The randomness used in fooling the final reduced ROBPs equals $s_{l}+O(\log nw/\varepsilon')$. By the constructions, we introduce $s_0-s=O\left(\frac{\log n\log (nw)}{\log\log w-\log\log n}\right)$ bits in the first stage, $s_l-s_0=O(l\cdot \log w)$ bits, which is $O\left(\log w \cdot \left(\log\log\min\{n,\log w\}-\log\log\frac{\log w}{\log n/\varepsilon}\right)\right)$ bits in the second stage. Therefore, the total randomness we use is at most
       \[
            s+O\left(\frac{\log n\log (nw)}{\log\log w-\log\log n}+\log w (\log\log\min\{n,\log w\}-\log\log\frac{\log w}{\log n/\varepsilon})+\log (1/\varepsilon)\right)
            \]
            as we claimed.

            \item
            \textbf{Case $w\leq n/\varepsilon$:} We will reduce the length to $n'=\left(\frac{10\log(n/\varepsilon)}{\log w}\right)^2$. If $n_0\leq n'$, the goal is already achieved and we directly go to the 3rd stage.
            
            Now consider $n' < n_0$. 
            One can deduce that $\log(1/\varepsilon)\leq \log^2 w\log n$, since $n_0 \leq    \log(1/\varepsilon)/\log n$.
            We iteratively use \cref{lem:ben length reduction} for $l$ times ($l$ will be defined later): In iteration $0\leq i\leq l$, we reduce the length from $n_i$ to $n_{i+1}=n_i^{1/4}\frac{\log(n_iw/\varepsilon')}{\log w}$, decomposing one ROBP into $K_i$ shorter ones and increasing the alphabet bit-length from $s_i$ to $s_{i+1}$. Notice that $n_{i+1}< n_{i}^{3/4}$ when $n_i\geq \left(\frac{\log(n_iw/\varepsilon')}{\log w}\right)^2$, which is always true when $n_i\geq n'$.
            Parameter calculation shows
            \[
                s_{i+1}=s_i+C\log w,
                \qquad
                K_i\leq (8n_i)^{\frac{\log(n_iw/\varepsilon')}{\log w}n_i^{1/4}+1}\leq
                (nw/\varepsilon)^{\frac{2n_i^{1/4}\log n_i}{\log w}}\leq 
                (nw/\varepsilon)^{1/\log^{1/4} w},
            \]
            where the first inequality is directly due to \cref{lem:ben length reduction} and the second inequality follows from $n_i\leq n_0=\log(1/\varepsilon)/\log n\leq \log^2w $.
            The iteration goes until $n_l\leq n'$, which continues for at most $l=\log\log n_l-\log\log n'\leq \log\log\min\{n,\log w\}$ rounds.  
            
            Now we enter the third stage and we have that $n'\leq \left(\frac{\log(1/\varepsilon')}{\log w}\right)^2$. Thus, $\log(1/\varepsilon')\geq \log w \cdot \sqrt{n'}\geq \log(n'w)\cdot \log^2 n'$, enabling us to apply \cref{lem:ben small error reduction}. We use an iterative approach: We set $n'_0=n_l<n', s'_0=s_l$, and in iteration $0\leq i\leq l'$, we reduce the length from $n'_i$ to $n'_{i+1}=10\log^2 n'_i$, decomposing one ROBP into $K'_i$ shorter ones and increasing the alphabet bit-length from $s'_i$ to $s'_{i+1}$. Parameter calculation shows
            \begin{align*}
                s'_{i+1}&=s'_i+O\left(\frac{\log (1/\varepsilon')}{\log n'_i}\right),\\
                K'_i&\leq (8n'_i)^{10\log^2 n'_i+1}\\
                &\leq  (8n')^{10\log^2 n' + 1}  \\
                &\leq  2^{100\log^3 n' }  \\
                &\leq  2^{100\log^3 \log(1/\varepsilon')}\\
                &\leq  2^{{100\log^3 \log(n/\varepsilon)}}\\
                &\leq  2^{\frac{\log (n/\varepsilon)}{\log^{1/2}w}}\\
                &\leq  (n/\varepsilon)^{1/\log^{1/2} w}.
            \end{align*}
            We stress that the sequence $\log n_i$ decreases exponentially, making the sum $\sum_{i=0}^{l'} O\left(\frac{\log (1/\varepsilon')}{\log n'_i}\right)$ dominated by a geometric series (i.e. $\sum_{i=0}^{l'} O\left(\frac{\log (1/\varepsilon')}{2^{l'-i}\log n'_{l'}}\right)$), and therefore the total increase of alphabet bit-length is at most $O(\log (1/\varepsilon'))$.
            The iteration continues for at most $l'=\log\log n'_l\leq \log\log \log (n/\varepsilon)\leq \log\log w$ rounds, until $n'_{l'}\leq O(1)$ and we fool the final reduced ROBPs with Chen-Ta-Shma PRG with seed-length $s'_{l'}+O(\log nw/\varepsilon')$.
            
            Overall, for this case, the total amount of reductions we use is at most $(l+l')\cdot K\cdot \prod_{i=0}^{l} K_i \cdot \prod_{i=0}^{l'} K'_i\leq K\cdot (n/\varepsilon)^{\frac{2\log\log w}{\log^{1/4} w}}\leq K^2$. The total error introduced in the second and third stages is at most $K^2\cdot \varepsilon'=\varepsilon/2$. Together with the error $\varepsilon/2$ introduced in the first stage, we have that the total error introduced is at most $\varepsilon$.


            Then we bound the randomness we use. The randomness used in selecting the index of reductions is at most $\log(K\cdot \prod_{i=0}^{l} K_i \cdot \prod_{i=0}^{l'} K'_i)\leq O(\log (nw/\varepsilon))$. The randomness used in fooling the final reduced ROBPs equals $s'_{l'}+O(\log nw/\varepsilon')$. By the constructions, we introduce $s_0-s=O\left(\frac{\log n\log (nw)}{\log\log w-\log\log n}\right)$ bits in the first stage, and $s_l-s_0=O(l\cdot \log w)$ bits in the 2nd stage, which is $O\left(\log w \cdot \log\log\min\{n,\log w\}\right)$. In the third stage, we introduce $s'_{l'}-s_l=O(\log (1/\varepsilon'))=O(\log (n /\varepsilon))$ bits. Therefore, the total randomness we use is at most
        \[
            s+O\left(\frac{\log n\log (nw)}{\log\log w-\log\log n}+\log w \cdot\log\log\min\{n,\log w\}+\log (n /\varepsilon)\right)
            \]
            as we claimed.

            \end{enumerate}

        \end{proof}

\end{document}